%% file: main.tex
\documentclass[%
superscriptaddress,
preprint,
nofootinbib,
 amsmath,amssymb,
 aps,
]{revtex4-2}

\usepackage[utf8]{inputenc}
\usepackage{xcolor}
\usepackage{graphicx, epsfig, array, subfigure, epstopdf}
\usepackage[colorlinks=true, citecolor=blue, linkcolor=blue, urlcolor=blue]{hyperref}
\usepackage{amssymb}
\usepackage{enumitem}
\usepackage[version=4]{mhchem}
\usepackage{bm}
\usepackage{cleveref}[2012/02/15]
\usepackage{ upgreek }
\crefformat{footnote}{#2\footnotemark[#1]#3}
\usepackage{fancyhdr}
\makeatletter
\def\l@subsection#1#2{}
\def\l@subsubsection#1#2{}
\makeatother

\renewcommand{\paragraph}[1]{\noindent{\bf #1}}
\newcommand{\fermilab}{Fermi National Accelerator Laboratory, Batavia, IL, 60510, USA}
\newcommand{\cern}{CERN, Esplande des Particules, 1211 Geneva 23, Switzerland}
\newcommand{\UKentucky}{University of Kentucky, KY, 40506, USA}
\newcommand{\uvic}{Department of Physics and Astronomy, University of Victoria, Victoria, BC V8P 5C2, Canada}

\usepackage{lineno}
\begin{document}
\preprint{ FERMILAB-FN-1145, LA-UR-22-21987}

\title{Physics Opportunities for the Fermilab Booster Replacement}


\author{John Arrington}
\affiliation{Lawrence Berkeley National Laboratory, Berkeley, California 94720, USA}
\author{Joshua Barrow}
\affiliation{The Massachusetts Institute of Technology, Cambridge, MA, USA}
\affiliation{Tel Aviv University, Tel Aviv, Israel}
\author{Brian Batell}
\affiliation{University of Pittsburgh, Pittsburgh, PA 15260}
\author{Robert Bernstein}
\affiliation{\fermilab}
\author{Nikita Blinov}
\affiliation{\uvic}
\author{S. J. Brice}
\affiliation{\fermilab}
\author{Ray Culbertson}
\affiliation{\fermilab}
\author{Patrick deNiverville}
\affiliation{Los Alamos National Laboratory, Los Alamos, NM 87545}
\author{Vito Di Benedetto}
\affiliation{\fermilab}%
\author{Jeff Eldred}
\affiliation{\fermilab}%
\author{Angela Fava}
\affiliation{\fermilab}%
\author{Laura Fields}
\affiliation{Department of Physics, University of Notre Dame, Notre Dame, Indiana 46556, USA}
\author{Alex Friedland}
\affiliation{SLAC National Accelerator Laboratory, 2575 Sand Hill Road, Menlo Park, CA 94025, USA}
\author{Andrei Gaponenko}
\affiliation{\fermilab}
\author{Corrado Gatto}
\affiliation{Istituto Nazionale di Fisica Nucleare Sezione di Napoli (Italy)}
\affiliation{Northern Illinois University, DeKalb, IL 60115 (USA)}
\author{Stefania Gori}
\affiliation{Santa Cruz Institute for Particle Physics and Department of Physics, University of California, Santa Cruz,
1156 High Street, Santa Cruz, CA 95064
}
\author{Roni Harnik}
\email{roni@fnal.gov}
\affiliation{\fermilab}
\author{Richard J. Hill}
\affiliation{\fermilab}
\affiliation{\UKentucky}
\author{Daniel M. Kaplan}
\affiliation{Illinois Institute of Technology, Chicago, IL 60616, USA}
\author{Kevin J. Kelly}
\affiliation{\fermilab}
\affiliation{\cern}
\author{Mandy Kiburg}
\affiliation{\fermilab}
\author{Tom Kobilarcik}
\affiliation{\fermilab}
\author{Gordan Krnjaic}
\affiliation{\fermilab}
\author{Gabriel Lee}
\affiliation{Department of Physics, LEPP, Cornell University, Ithaca, NY 14853, USA}
\affiliation{Department of Physics, Korea University, Seoul 136-713, Korea}
\affiliation{Department of Physics, University of Toronto, Toronto, Ontario, Canada M5S 1A7}
\author{B. R. Littlejohn}
\affiliation{Illinois Institute of Technology, Chicago, IL 60616, USA}
\author{W. C. Louis}
\affiliation{Los Alamos National Laboratory, Los Alamos, NM 87545}
\author{Pedro Machado}
\affiliation{\fermilab}%
\author{Anna Mazzacane}
\affiliation{\fermilab}
\author{Petra Merkel}
\affiliation{\fermilab}
\author{William M. Morse}
\affiliation{Brookhaven National Laboratory, Upton, NY 11973}
\author{David Neuffer}
\affiliation{\fermilab}
\author{Evan Niner}
\affiliation{\fermilab}
\author{Zarko Pavlovic}
\affiliation{\fermilab}%
\author{William Pellico}
\affiliation{\fermilab}%
\author{Ryan Plestid}
\affiliation{\fermilab}
\affiliation{\UKentucky}
\author{Maxim Pospelov}
\affiliation{School of Physics and Astronomy, University of Minnesota, Minneapolis, MN 55455, USA}
\author{Eric Prebys}
\affiliation{University of California Davis, One Shields Avenue, Davis, CA 95616}
\author{Yannis K. Semertzidis}
\affiliation{Center for Axion and Precision Physics Research, IBS, Daejeon 34051, Republic of Korea}
\affiliation{Department of Physics, KAIST, Daejeon 34141, Republic of Korea}
\author{M. H. Shaevitz}
\affiliation{Columbia University, New York, NY 10027}
\author{P.  Snopok}
\affiliation{Illinois Institute of Technology, Chicago, IL 60616, USA}
\author{M.J. Syphers}
\affiliation{Northern Illinois University, DeKalb, IL 60115, USA}
\author{Rex Tayloe}
\affiliation{Indiana University, Bloomington, IN 47405}
\author{R. T. Thornton}
\affiliation{Los Alamos National Laboratory, Los Alamos, NM 87545}
\author{Oleksandr Tomalak}
\affiliation{\fermilab}
\affiliation{Los Alamos National Laboratory, Los Alamos, NM 87545}
\affiliation{\UKentucky}
\author{M. Toups}
\affiliation{\fermilab}
\author{Nhan Tran}
\affiliation{\fermilab}
\author{Yu-Dai Tsai}
\affiliation{\fermilab}
\affiliation{Department of Physics and Astronomy, University of California, Irvine, CA 92697-4575, USA}
\author{Richard Van de Water}
\affiliation{Los Alamos National Laboratory, Los Alamos, NM 87545}
\author{Katsuya Yonehara}
\affiliation{\fermilab}
\author{Jacob Zettlemoyer}
\affiliation{\fermilab}
\author{Yi-Ming Zhong}
\affiliation{Kavli Institute for Cosmological Physics, University of Chicago, Chicago, IL 60637, USA}
\author{Robert Zwaska}
\affiliation{\fermilab}

\date{\today}
\begin{abstract}
    This white paper presents opportunities afforded by the Fermilab Booster Replacement and its various options. Its goal is to inform the design process of the Booster Replacement about the accelerator needs of the various options, allowing the design to be versatile and enable, or leave the door open to, as many options as possible. The physics themes covered by the paper include searches for dark sectors and new opportunities with muons. 
\end{abstract}
\maketitle

\newpage

\setcounter{tocdepth}{1}
\tableofcontents 

\pagestyle{fancy}

\section{Introduction - The Fermilab Booster Replacement}

The motivation to replace the Fermilab Booster accelerator ring stems from the needs of the long-baseline neutrino program. 
The P5 recommendation is for 2.4MW to be delivered for DUNE~\cite{ParticlePhysicsProjectPrioritizationPanel(P5):2014pwa}.
A power of 2.4 MW requires $1.5\times10^{14}$ particles from the Main Injector (MI) accelerator ring at 120 GeV, assuming a 1.2 s cycle time.
The existing Fermilab Booster is not capable of accelerating the $2.5\times 10^{13}$ particles required for each of the six batches needed to fill the MI irrespective of the injection energy, or how it is upgraded. Achieving more than 2 MW will thus require replacement of the Booster and possible upgrades of the MI. 

As past experience at Fermilab and around the world shows, a versatile accelerator complex opens the door to important physics opportunities at low and intermediate energies. The Booster Replacement should enable the following capabilities:
\begin{itemize}
    \item Deliver 2.4 MW @ 60-120 GeV from the Main Injector to the LBNF beamline in support of the DUNE experiment
\item Deliver up to 80 kW @ around 8 GeV to support g-2, Mu2e, and short-baseline neutrino experiments.
\item Deliver ~100 kW continuous wave (CW) @ 800 MeV or a higher energy to support a second generation of the Mu2e experiment. 
\item Exploit the capabilities of CW superconducting RF (SRF) PIP-II linac and the full accelerator sequence to enable other physics opportunities, including searches for dark sectors. 
\end{itemize}

An optimal accelerator design will depend on the physics opportunities that may be enabled and pursued.
This white paper grew out of the need to inform such designs. It identifies physics opportunities that may be enabled or brought closer to realization by the PIP-II linac, by the Booster Replacement, and by the enhanced power in the main injector. The goal here is \emph{not} to prioritize the potential experiments, nor is it to assess the cost of various options. Rather the goal is to inform the accelerator design about what experiments may be proposed in the years ahead, and what are the special requirements of the design needed to pursue them. In allowing for an informed accelerator design, we hope that many doors will remain open to pursue exciting physics goals, such as searches for dark sectors, a slew of charged lepton flavor violation searches, and precision measurements, as well as R\&D towards a muon accelerator at the energy frontier. 

In collecting possible physics opportunities we have striven to identify concrete options that are feasible in the short term, but also to take a long view. We remind the reader that the existing Fermilab Booster was designed over 50 years ago. The uses and utility of that machine have certainly exceeded the expectations of its designers. While thinking about physics opportunities enabled by its replacement, we should remember that it will also likely serve the HEP community for decades. It will hopefully will be versatile enough to enable not only the experiments that are most motivated in the near term, but also those of the of the next generation. We note that our exercise is not the first of its kind. Notably the Project~X study~\cite{Kronfeld:2013uoa}, has comprehensively explored the physics opportunities afforded by an 8~GeV Linac extension.   

In the following sections we present the various physics opportunities put forth by the community, beside the long-baseline neutrino program of LBNF and DUNE. The topic was discussed in an open remote workshop on May 19th 2020 and input was collected from the community in the following weeks. Every section contains a brief motivation and physics case, and a description of the experimental setup. Since the goal of this compilation is to inform the accelerator design, each section also contains a subsection that focuses the accelerator needs, such as the type of beam, the beam energy and intensity, the needed time structure, etc. For future prioritization, it is of importance to understand the status of the various subfields globally, a topic which is also addressed throughout the following sections. A preliminary version of this report was presented to accelerator experts at Fermilab, and a companion paper with options for upgrades to the Fermilab complex have been presented in a companion white paper~\cite{Ainsworth:2021ahm}. 

The physics topics represent a broad array, pursuing goals in dark sector physics, neutrino physics, charged lepton flavor violation (CLFV), precision tests, as well as R\&D facilities, both for detector development, and to explore new directions for HEP. The list of presented topics is shown in Table~\ref{tab:summary} and labeled in these broad categories. 
\begin{table}[t]
    \centering
\includegraphics[width=14cm]{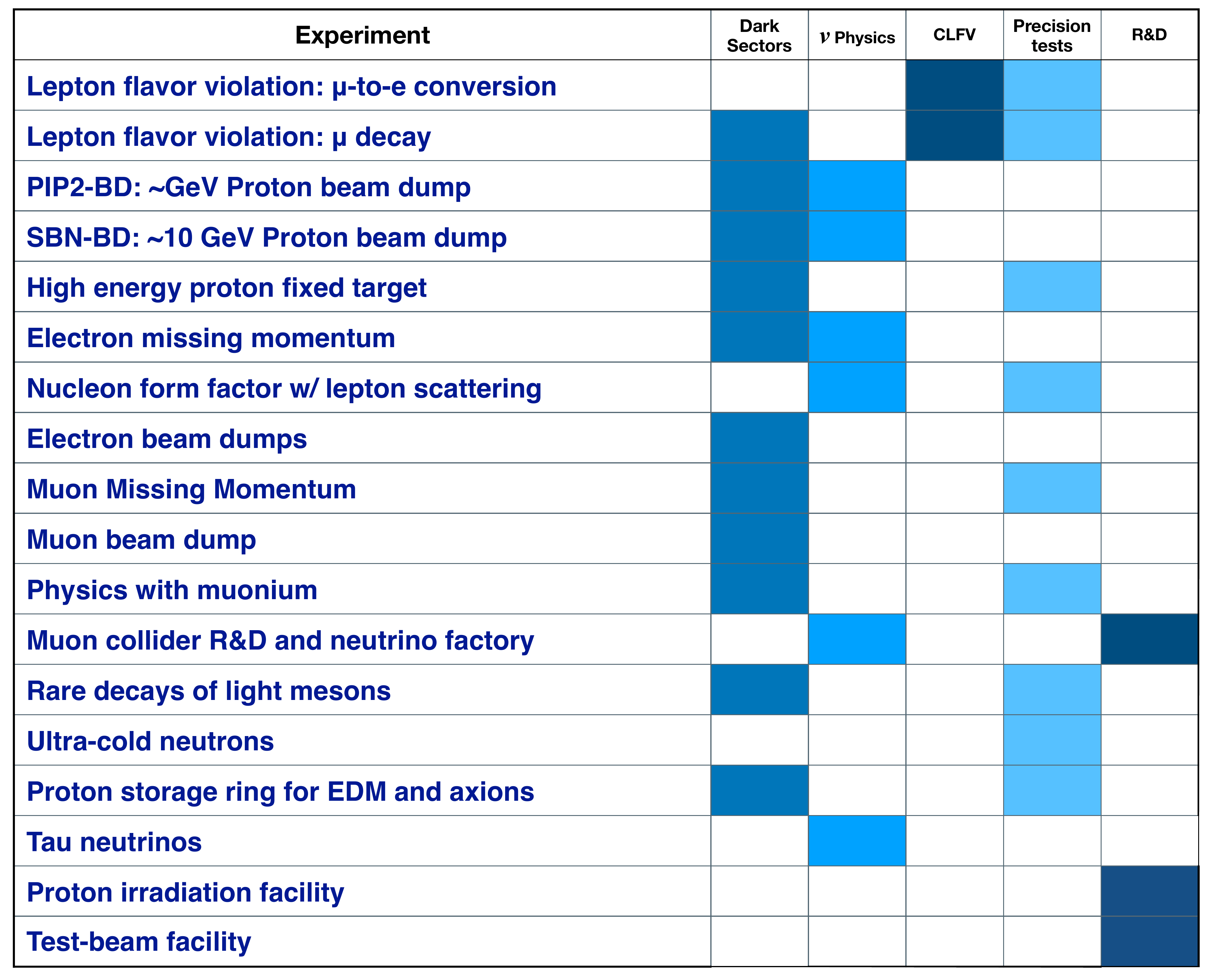}
    \caption{A summary of the physics opportunities presented in this document, categorized by areas of physics they pursue. Each table entry is epresented by a section of this document in order.}
    \label{tab:summary}
\end{table}
%
%
It should be noted that several physics opportunities cover more than one area. 
Before presenting the various opportunities and their accelerator requirements, we will briefly motivate some of the central themes that emerge from this compilation, other than neutrino experiments. These include searches for dark sectors, and opportunities with muons.  

\subsection{Theme: Dark Sectors}
In the past decade interest in probing the possible existence of dark sectors at intensity frontier facilities has grown significantly. In light of this, a significant number of the physics opportunities presented here focus on this direction. Given the prominent place dark sector searches are taking, we review their motivation briefly. More in-depth reviews and community papers are available, e.g.~\cite{Essig:2013lka,Alexander:2016aln, BRN}.

The Standard Model (SM) of particle physics is a rich and complex structure, including several gauge groups and several copies of matter particles. It is easily conceivable that additional dynamics is present in this spirit, in a new sector of the theory. The interactions of such a new sector would have to respect the symmetries of the SM and those of the new sector. It is thus convenient to categorize these interactions in terms of the SM portals which they couple to, as reviewed in~\cite{Alexander:2016aln}.
For example, a new dark photon, $A'$, can couple to the SM photon through the kinetic mixing portal $F^{\mu\nu}F'_{\mu\nu}$, with $F$'s being the usual field strengths of the two photons. The coefficient of this interaction $\epsilon$ can be naturally small. 
In any process that produces photons, such as meson decays, a dark photon can be emitted instead with a small probability if kinematically allowed.
The dark photon may decay either visibly or invisibly. The visible decay will invariably need to go through the (weak) portal interaction, and may thus be displaced, enabling interesting search channels.
The dark photon may also be a mediator into the dark sector. For example, if the dark sector contains a new light state charged under the dark $U(1)$ interaction, perhaps dark matter, the dark photon will decay into pairs of the dark states with a high branching ratio. In this way, a fixed target setup can be a source of dark matter, or of mediators, which can be searched for.

Other portals are also well motivated. A singlet fermion in the dark sector may couple to the so-called neutrino portal, $HLN$, and thus be produced  in the decay of charged mesons. Heavy neutral leptons such as $N$ also have a variety of search channels. Light pseudo-scalars, also dubbed axion-like particles, can also couple to standard model mesons and be produced in decays among them, e.g. $K\to \pi a$. Another important dark extension of the standard model introduces millicharged particles, particles with small electric charges, which like dark photons, can be emitted in any process in which photons are produced, such as meson decays.

Dark matter provides another important motivation for the search for dark sector physics. In particular, light dark matter (i.e., with masses smaller than $\sim m_p$) has been understood to be a major blind spot for traditional searches, like standard direct detection and high energy colliders~\cite{Battaglieri:2017aum}. Here theoretical understanding of the DM production in the early universe strongly motivates the existence of mediator particles, like the dark photons, heavy neutrinos and axion-like particles mentioned above. For example, DM can be produced through a freeze-out mechanism akin to the WIMP paradigm; for DM masses below the GeV scale, however, the annihilations of DM to SM particles, $\chi \chi \leftrightarrow\mathrm{SM}$,  cannot occur via any known force carriers~\cite{Lee:1977ua}; this necessitates the existence of a new mediator with a mass well below the weak scale~\cite{Boehm:2003hm}. Additional restrictions on the mediator mass relative to DM, as well as the Lorentz structure of its interactions can be obtained from astrophysics and cosmology (e.g., by requiring that residual DM annihilations do not modify the observed Cosmic Microwave Background~\cite{Slatyer:2009yq}).
In many cases requiring a given model reproduce the observed relic abundance of DM, singles out a preferred region of parameter space, thereby providing a concrete experimental target -- see Ref.~\cite{Berlin:2018bsc} for some examples. Two main classes of experiments can target such predictive scenarios: searches for missing momentum (where SM collisions produce a mediator that decays to DM, or DM directly) or via scattering (where DM produced in beam-target collisions deposits energy in a downstream detector via scattering or de-excitation). 

The minimal DM models described above necessitate only two dark sector particles: the DM itself and the mediator. It has recently become apparent that freeze-out is but one of a wide range of predictive DM production mechanisms. These other mechanisms may require a much richer dark sector, more akin to the SM. For example, instead of $2\to 2$ annihilations (like $\chi \chi \leftrightarrow\mathrm{SM}\;\mathrm{SM}$) being responsible for DM production, $3\leftrightarrow 2$ reactions like $\chi \chi \chi \to \chi \chi$ can play the dominant role (as in the $2\to2$ case such reactions can reduce the DM number density until a desired relic abundance is obtained)~\cite{Hochberg:2014dra,Fitzpatrick:2020vba}. Such interactions require the coalescence of 3 particles at a point, a somewhat rare event, even in the dense early universe plasma; as a result, sufficiently large rates are naturally obtained in strongly interacting models where $\chi$'s are mesons of a new non-Abelian gauge group~\cite{Hochberg:2014kqa}. In other words, this class of DM production mechanisms is naturally realized when the dark sector contains something like a dark chromodynamics, complete with its rich spectrum of bound states. Remarkably even these complex models can feature specific targets that can be strived for with the intense beams available at FNAL~\cite{Berlin:2018tvf}. The typical signals of these models include the cascade decays of the mediator particle into other visibly-decaying and stable dark sector states.

In summary, dark sector and dark matter present interesting and well motivated challenges for high energy physics at intensity facilities. As this document shows, these searches can be launched off of a variety of beams, in the full range of energies that Fermilab's future facilities will span, from $\sim 1$~GeV to $\sim 120$ GeV, as well as intermediate energies.
\\

\subsection{Theme: Muon-based searches}

Muon experiments are playing a prominent role in Fermilab's current and near future program. The muon $g-2$ experiment has recently confirmed the  longstanding muon magnetic moment anomaly ~\cite{Abi2021} and will further test this tantalizing hint of new physics. The Mu2e collaboration is slated to improve the reach of muon to electron conversion by four orders of magnitude, with interesting reach for beyond the standard model physics~\cite{Kuno:1999jp,Raidal:2008jk,deGouvea:2013zba,Calibbi:2017uvl}. Given the prominence of muon physics, both in Fermilab's near term future and as a motivated search avenue for new physics, one of the notable themes of this document is an opportunity for Fermilab to build on its strength. 
In addition to the lab's existing muon portfolio, it may also be possible
for Fermilab to host a future muon beam fixed target experiments
 to discover possible new physics responsible for the $g-2$ anomaly and other
 connections to dark matter and hidden forces \cite{Kahn:2018cqs}. 
These topics are covered in Section~\ref{clfv-conversion} (on muon conversion), Section~\ref{clfv-decays}, and Section~\ref{sec:muon-missing-momentum} (on muon missing momentum). It is also notable that some explanations of the muon g-2 deviation may be tested in a test version of the muon missing momentum concept, even before the Booster replacement, with largely existing accelerator infrastructure~\cite{Kahn:2018cqs}.

\subsection{Theme: Muon collider R\&D}

Over the past several years, there has been rekindled interest in the physics potential of future muon colliders at the few 100 GeV and multi-TeV scale (for a review of recent progress see Refs. \cite{ali2021muon,Neuffer:2018yof,Delahaye:2019omf}).  Unlike protons or ions, muons are elementary particles whose full center-of-mass energy can be converted to heavy new states in $\mu^+ \mu^-$ annihilation events, which enhances the physics reach of such facilities and  yields a cleaner collision environment. Unlike electrons as beam particles, muons are heavier and therefore suffer lower radiative losses due to synchrotron radiation, which allows for a more compact accelerator design that can exploit tunnels at existing collider facilities \cite{Neuffer:2018yof}. Thus, muon colliders offer the possibility of both high CM energy and a clean environment for precision measurements of SM and possible BSM physics.  It has been shown that TeV scale muon colliders with 1/ab luminosities can decisively probe models of WIMP dark matter using mono-X searches and greatly improve sensitivity to Higgsino-like WIMPs with disappearing tracks and displaced vertex searches \cite{Han:2020uak}. If the muon g-2 anomaly is confirmed as an indirect discovery of new physics, a staged muon collider program will directly produce the new states responsible through a variety of observables \cite{Rodolfo2022,Buttazzo:2020ibd}.  It has also been shown that such facilities can test new muon-philic physics responsible for lepton flavor universality breaking  anomalies currently observed by LHCb at $>4$ sigma significance \cite{RK-2022}. Relative to other future collider concepts (e.g. FCC and CLIC), muon colliders may also have competitive sensitivity to Standard Model Higgs-fermion couplings and offer precision measurements of its trilinear self interaction \cite{franceschini2021higgs}. Intriguingly, such machines also benefit from large vector boson fusion enhancements to various new physics processes \cite{maltoni-VBF}.

Muon collider R\&D has paused in the US during the previous P5 process. To some extent, work is continuing elsewhere. However, a versetile Booster Replacement facility may be an opportunity to resume this well motivated activity and potentially open the door to bringing the energy frontier back to the US in the future.
\\

\paragraph{Appendix:} The Booster Replacement plan revolved around the goal of bringing the Main Injector to 2.4 MW and beyond. As part of the exercises we have also considered the future of the Fermilab facility beyond the Main Injector. In an appendix we present the results of a study that asks what the needs of the DUNE long baseline neutrino program would be in terms of proton beam energy, if the constraints of the main injector are relaxed, allowing for a lower energy beam. If this avenue were pursued, new challenges would emerge, particularly in the area of targetry.

\newpage
\include{clfv-conversion}

\newpage
\include{clfv-decays}

\newpage
\include{stopped_pions}

\newpage
\include{protons-medium-energy}


\newpage
\include{high-energy-dump}

\newpage
\include{electron-missing-momentum}

\newpage
\include{lepton-nucleon}

\newpage
\include{electron-dump}

\newpage
\include{muon-missing-momentum}

\newpage
\include{muon-beam-dump}

\newpage
\include{muonium}

\newpage
\include{NuFact-MuCol}

\newpage
\include{REDTOP_v3}

\newpage
\include{NNbar}

\newpage
\include{proton-storage-ring}

\newpage
\include{tau-neutrinos}

\newpage
\include{proton-irradiation}

\newpage
\include{test-beam}

\newpage
\include{DUNE-proton-momentum}

\section*{Acknowledgements}
Fermilab is operated by Fermi Research Alliance, LLC under Contract No. DE-AC02-07CH11359 with the United States Department of Energy. The work is supported by the US Department of Energy through the Los Alamos National Laboratory. Los Alamos National Laboratory is operated by Triad National Security, LLC, for the National Nuclear Security Administration of U.S. Department of Energy (Contract No. 89233218CNA000001).

\newpage
\pagestyle{plain}
\bibliography{refs,refs-clfv,refs-redtop,bibliography,DUNEMomentum}

\end{document}

%% file: clfv-conversion.tex
\section{Charged lepton flavor violation in muon to electron conversion}\label{clfv-conversion}

\paragraph{Authors and proponents:}
R.\ Bernstein, R.\ Culbertson, J.\ Eldred, A.\ Gaponenko, D.\ Neuffer, B.\ Pellico (Fermilab), E.\ Prebys\footnote{rhbob@fnal.gov, rlc@fnal.gov, gandr@fnal.gov, neuffer@fnal.gov, eprebys@ucdavis.edu} (UC Davis), Mu2e-II Collaboration

\paragraph{Related sections:}
Charged lepton flavor violation with muon decays, Sec.~\ref{clfv-decays}.

\subsection{Physics Goals, Motivation, and Setup}
The potential of charged lepton flavor violation searches to probe
physics beyond the Standard Model is well established
\cite{Kuno:1999jp,Raidal:2008jk,deGouvea:2013zba,Calibbi:2017uvl}.
The ongoing Mu2e and COMET experiment aim to reach sensitivities
in the $10^{-16}$--$10^{-17}$ range.  Mu2e will use about $5\times10^{20}$
8 GeV protons and produce and stop in the detector about $10^{18}$ muons
to reach its goal.  The average proton beam power in Mu2e is 8 kW.
A future generation of conversion experiments will need need to
stop more than an order of magnitude more muons to either extend the sensitivity
reach, or characterize an observed CLFV signal by exploiting different
stopping target materials.

Two types of future muon conversion experiments are being developed.
One, ``Mu2e-like'', uses a beamline made of superconducting solenoids
to capture negative pions produced when the proton beam strikes a production target
and deliver muons resulting from their decay directly to the experiment
stopping target \cite{Abusalma:2018xem}.   This type  of experiments requires a pulsed proton
beam with a high level of ``extinction'', that means the fraction
of protons outside of nominal beam pulses must be below a $10^{-10}$ level.
Times around muon beam arrival to the stopping target must be excluded
from the signal search window, because surviving negative pions are a source
of background.   Only muons that remain stopped in the experiment target
by the end of the muon beam flash contribute to the physics sensitivity.
Those muons have a stopping target material dependent life time that varies
from an order a microsecond (e.g. 846 ns for aluminum) to only dozens of nanoseconds
for heavy elements (e.g. 72 ns for gold).   The optimal beam pulse
repetition rate should match the stopped muon lifetime.
This type of experiments is also subject to background from antiprotons
if any are produced by the proton beam.   Therefore proton
beam energy for future experiments must be below the kinematic
threshold of antiproton production.

The other type of future experiments, ``PRISM/PRIME like'',
plans on using a muon storage ring, where after a few turns
pions decay to a negligible level.
Another important function of the muon storage ring is
delivering muon beam with a low momentum spread.
Muons are initially produced with a broad momentum spectrum,
therefore thick targets (e.g. a total of 3.4 mm of aluminum in Mu2e)
have to be used to efficiently stop the muons.
The PRISM idea is to use very short proton pulses,
then phase rotate the narrow-in-time-broad-in-momentum
muon distribution to make it narrow-in-momentum, at the price
of time broadening.
With pure, low energy, muon beam, it is unnecessary to exclude the
muon pulse arrival time from
signal search window \cite{Barlow:2011zza}.  
This makes the beam pulse repetition rate not critical
and allows to freely choose stopping target material.
The most important parameter of the accelerator system
is the amount of muons that can be delivered on a macroscopic
time scale.

\textcolor{blue}{}
\subsection{Accelerator Requirements}

\paragraph{Accelerated particles:}
Protons

\paragraph{Beam Energy:}
The range of 1 to 3 GeV seems to be optimal for a Mu2e-based
experiment. It is important to keep proton energy below
the antiproton production threshold.

\paragraph{Beam intensity:}
100 kW (Mu2e-like, \cite{Abusalma:2018xem}) to 2 MW (PRISM-like, \cite{Alekou:2013eta}).

\paragraph{Beam time structure:}
Mu2e-like searches: narrow proton pulses (tens of ns or better) separated by 200--2000~ns. Flexible
timing is needed for different muon stopping targets.

PRISM-like: narrow (15 ns) proton pulses at repetition rate about 1 kHz.

\paragraph{Target requirements:}
Thick target optimized for muon production

\paragraph{Other requirements:}
Flexible time structure and minimal pulse-to-pulse variation for Mu2e-like searches.

\paragraph{Timescales, R\&D needs, and similar facilities:}

R\&D needs:
\begin{itemize}
    \item High power, thick muon production target.
        (PSI uses a thin target that utilizes only a fraction
        of the available  proton beam power to produce muons and delivers
        the rest of the power in the ``spent'' beam to a spallation neutron facility.)

    \item Radiation Shielding of the production solenoid superconductor

    \item PRISM-like muon storage/manipulation ring

    \item Proton bunch compression for PRISM-like scheme

    \item Alternative ways to improve $\mu^{+}$ stopping density.  There are ideas like using an induction linac
        to slow down muons, or insert wedges in dispersive regions of the muon beamline.

\end{itemize}

Similar facilities: the COMET experiment in Japan is a competitor to Fermilab's present Mu2e.
The author of the  PRISM concept proposed constructing it in Japan \cite{Kuno:1997dr}, \cite{Kuno:2000kd},
and an LOI has been submitted in 2003 \cite{2003-PRISM-LOI}.   However it is not an officially
approved project, and, for example, the 2019 roadmap report \cite{2019-KEK-roadmap} does not mention it.
Fermilab is in a unique position to develop world leading muon physics program, which will
include an evolution of Mu2e with the booster beam into  Mu2e-II with PIP-II beam into a future
muon conversion experiment using infrastructure of the booster replacement accelerators.

%% file: clfv-decays.tex
\section{Charged lepton flavor violation with muon decays}\label{clfv-decays}

\paragraph{Authors and proponents:}
R.\ Bernstein, R.\ Culbertson, A.\ Gaponenko, D.\ Neuffer\footnote{rhbob@fnal.gov, rlc@fnal.gov, gandr@fnal.gov, neuffer@fnal.gov} (Fermilab)

\paragraph{Related sections:}
Charged lepton flavor violation with muon conversion, Sec.~\ref{clfv-conversion}.

\subsection{Physics Goals, Motivation, and Setup}
Searches for  $\mu^{+}\to{}e^{+}\gamma$ decay with their null
results played an important role in
establishing the Standard Model \cite{Calibbi:2017uvl}.
A rate of this charged lepton flavor violating decay  that is observable in current or foreseeable future
experiments is predicted by a broad range of New Physics models
\cite{Kuno:1999jp,Raidal:2008jk,deGouvea:2013zba,Calibbi:2017uvl}.
A $\mu^{+}\to{}e^{+}e^{-}e^{+}$ decay can be related to $\mu^{+}\to{}e^{+}\gamma$
via internal conversion of the photon, but is additionally sensitive to tree
level lepton-lepton couplings.
There is an active program at PSI to search for both of these decays,
the MEG-II \cite{Baldini:2013ke} and Mu3e \cite{Blondel:2013ia} experiments.
The MEG-II experiment plans to run with the muon beam intensity of $7\times10^{7}$
to reach a sensitivity of $6\times10^{-14}$ \cite{Baldini:2013ke}.
It has been pointed out that ``If a muon beam rate exceeding $10^9$ muons per second is available,
the much cheaper photon conversion option would be recommended'' for the detector
to reach a sensitivity of a few $10^{-15}$  \cite{Cavoto:2017kub}.

The Mu3e experiment plans to reach a $10^{-15}$ sensitivity using
up to $10^8$ muons/s beam flux in phase I.   Going beyond that
to $10^{-16}$ in phase II will require a $10^{10}$ muons/s beam
that is not currently available at PSI, but is achievable
with the proposed  High Intensity Muon Beam upgrade.

A Fermilab linac (PIP-II or its extension) can deliver beam
that is continuous on the muon life time scale, as is required
for $\mu\to{}e\gamma$ and $\mu\to{}eee$ searches.
The Mu2e solenoidal beamline under construction will be capable of delivering beam fluxes
above $10^{11}$ muons/s, although the ``out of the box'' Mu2e design
will produce $\mu^{+}$ beam with a larger momentum spread and lower
relative muon stopping density than what is provided by
PSI surface muon beamline.   The stopping density
issue can likely be mitigated with some R\&D, which
will make  the Fermilab muon campus an attractive venue
for the next generation CLFV searches in muon decays.

\subsection{Accelerator Requirements}

\paragraph{Accelerated particles:}
Protons

\paragraph{Beam Energy:}
Not critical, 0.8 to a few GeV.  A few GeV beam would be
easier to deliver to the pion production target in the current Mu2e solenoid.

\paragraph{Beam intensity:}
0.1 MW or more; probably limited by the production target and/or production solenoid rad hardness

\paragraph{Beam time structure:}
Continuous beam, on the time scale of free muon lifetime.  That is, proton pulses that are separated by a microsecond or less.
The more "continuous" the better.

\paragraph{Target requirements:}
Thick target to efficiently use the proton beam.  

\paragraph{Other requirements:}

\paragraph{Timescales, R\&D needs, and similar facilities:}
Necessary detector technologies are/will be available in the near term.
While some detector R\&D will be necessary, it is ongoing elsewhere.
R\&D that is specific to Fermilab is needed on:
\begin{itemize}
    \item Thick muon production target.
        (PSI uses a thin target that utilizes only a fraction
        of the available  proton beam power to produce muons.)

    \item Radiation Shielding of the production solenoid superconductor

    \item Develop ways to improve $\mu^{+}$ stopping density.  There are ideas like using an induction linac
        to slow down muons, or insert wedges in dispersive regions of the muon beamline. Explore the surface muon option.
\end{itemize}

Similar facilities: PSI conducts and active program of searches for
LFV muon decays, and an upgrade of the muon beamline has been proposed
\cite{Iwai:2020jye}.  The upgrade plans to achieve the surface muon
rate of the order of $10^{10}$ muons per second, by optimizing the
pion production target and muon capture and delivery.  but using the
existing proton beam.  The production target must be kept thin because
the passing proton beam serves a spallation neutron facility.  Muon
production at Fermilab is free from such constraint, and higher rates
should be achievable.  Pursuing a program of LFV searches with muon
decays at Fermilab in addition to the muon conversion searches will
exploit synergies between positive and negative muon beams and grow the
experimenter community.

%% file: stopped_pions.tex
\section{PIP2-BD: GeV Proton Beam Dump 
}\label{sec:1GeV}

\paragraph{Authors and proponents:} 
Matt Toups\footnote{\label{contacts}toups@fnal.gov, vdwater@lanl.gov} (FNAL), Richard Van de Water\cref{contacts} (LANL), Brian Batell (U. of Pittsburgh), S.J.~Brice (FNAL), Patrick deNiverville (LANL), Jeff Eldred (FNAL), Kevin J. Kelly (FNAL/CERN), Tom Kobilarcik (FNAL), Gordan Krnjaic (FNAL), B. R. Littlejohn (IIT), Bill Louis (LANL), Pedro A.~N. Machado (FNAL), Z.~Pavlovic (FNAL), Bill Pellico (FNAL), Michael Shaevitz (Columbia University), P. Snopok (IIT), Rex Tayloe (Indiana University), R. T. Thornton (LANL), Jacob Zettlemoyer (FNAL), Robert Zwaska (FNAL)

\subsection{Physics Goals, Motivation, and Setup}

Two recent developments in particle physics clearly establish the need for a GeV-scale high energy physics (HEP) beam dump facility. First, theoretical work has highlighted not only the viability of sub-GeV dark sectors models to explain the cosmological dark matter abundance but also that a broad class of these models can be tested with accelerator-based, fixed-target experiments, which complement growing activity in sub-GeV direct dark matter detection~\cite{Alexander:2016aln, Battaglieri:2017aum, BRN}. Second, the observation of coherent elastic neutrino-nucleus scattering (CEvNS)~\cite{Freedman:1973yd, Kopeliovich:1974mv} by the COHERENT experiment~\cite{Akimov:2017ade} provides a novel experimental tool that can now be utilized to search for physics beyond the standard model in new ways, including in searches for light dark matter~\cite{deNiverville:2015mwa} and active-to-sterile neutrino oscillations~\cite{Anderson:2012pn}, which would provide smoking-gun evidence for the existence of sterile neutrinos.

The completion of the PIP-II superconducting LINAC at Fermilab as a proton driver for DUNE/LBNF in the late 2020s creates an attractive opportunity to build such a dedicated beam dump facility at Fermilab.  A unique feature of this Fermilab beam dump facility is that it can be optimized from the ground up for HEP. Thus, relative to spallation neutron facilities dedicated to neutron physics and optimized for neutron production operating at a similar proton beam power, a HEP-dedicated beam dump facility would allow for better sensitivity to dark sector models, sterile neutrinos, and CEvNS-based NSI searches, and more precise measurements of neutrino interaction cross sections relevant for supernova neutrino detection. For example, the Fermilab facility could be designed to suppress rather than maximize neutron production and implement a beam dump made from a lighter target such as carbon, which can have a pion-to-proton production ratio up to $\sim$2 times larger than the heavier Hg or W targets used at spallation neutron sources. The facility could also accommodate multiple, 100-ton-scale high energy physics experiments located at different distances from the beam dump and at different angles with respect to the incoming proton beam. This flexibility would allow for sensitive dark sector and sterile neutrino searches, which can constrain uncertainties in expected signal and background rates by making relative measurements at different distances and angles.

The continuous wave capable PIP-II LINAC at Fermilab can simultaneously provide sufficient protons to drive megawatt-class $\mathcal{O}$(GeV) proton beams as well as the multi-megawatt LBNF/DUNE beamline. By coupling the PIP-II LINAC to a permanent magnet or DC-powered accumulator ring, the protons can be compressed into pulses suitable for a proton beam dump facility with a rich physics program. The accumulator ring could be located in a new or existing beam enclosure and be designed to operate at 800~MeV and initially provide 100~kW of beam power.  One such storage ring,  which could be implemented on the timescale of the completion of PIP-II, has been proposed in~\cite{Pellico}.

One variant of this accumulator ring would be a $\sim$100 m circumference ring operating at 1.2~GeV with a pulse width of 20~ns and a duty factor of $\mathcal{O}(10^{-6})$, which would greatly reduce steady-state backgrounds.  Another is a accumulator ring coupled to a new rapid cycling synchrotron replacing the Fermilab Booster with an increased proton energy of 2~GeV and an increased beam power of 1.3 MW~\cite{Ainsworth:2021ahm}.

\subsection{Dark Sector Searches}

\begin{figure}
\centering
  \includegraphics[width=0.48\textwidth]{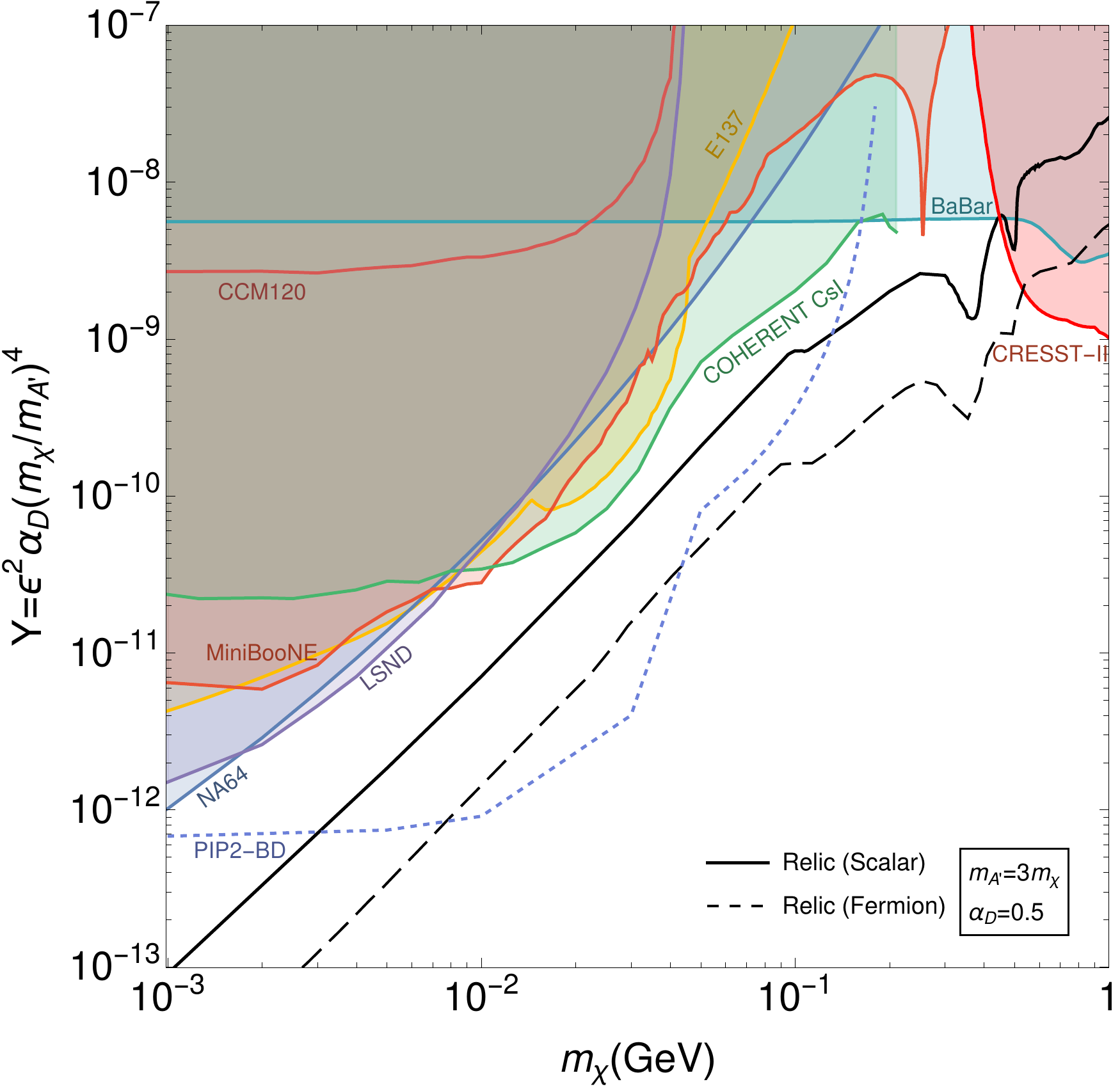}
  \caption{PIP2-BD vector portal dark matter sensitivity curve for $5.65\times10^{22}$ protons on target compared to thermal relic density targets and existing 90\% exclusion limits as a function of the dimensionless scaling variable $Y=\epsilon^2\alpha(m_\chi/m_{A'})^4$, assuming $\alpha=0.5$ and $m_A=3m_{\chi}$.}
  \label{fig:darksector}
\end{figure}

Proton beam dump experiments are potentially sensitive to any dark sector models that produce light dark matter directly through hadronic interactions or through the subsequent decay of light mesons. This includes, for example, both standard vector portal dark matter models that can be probed with beam proton and electron beams as well as other models, such as leptophobic or hadrophilic dark matter models, for which proton beams provide unique sensitivity~\cite{Batell:2018fqo}. Some of the best limits on vector portal dark matter for dark matter masses in the 10--100~MeV range come from reinterpretations~\cite{Batell:2009di,deNiverville:2011it,Kahn:2014sra} of $\nu_e$-electron elastic scattering measurements made by the high power, 800-MeV proton beam dump experiment LSND~\cite{Auerbach:2001wg}. While an LSND-like experiment optimized for dark matter searches at the Fermilab GeV beam dump facility could likely improve on existing bounds~\cite{Jordan:2018gcd}, CEvNS provides an additional channel with which to search for dark matter~\cite{deNiverville:2015mwa}.

The COHERENT collaboration recently reported the first detection of CEvNS on argon using an $\mathcal{O}$(10 kg) liquid argon scintillation detector achieving a 20 keV recoil energy threshold~\cite{Akimov:2020pdx}. Studies of the sensitivity of an upgraded 750-kg liquid argon scintillation detector to scalar light dark matter models indicate the importance of larger mass detectors, utilizing the angular dependence of the dark matter flux, and reduced flux uncertainty (which can be addressed with relative measurements at different angles using identical or moveable detectors), to expand the reach of these searches~\cite{Akimov:2019xdj}. We consider here a 100-ton LAr detector placed on-axis, 18 m downstream from a carbon proton beam dump with a 50 keV recoil energy threshold (to suppress neutrino backgrounds) and an efficiency of 70\%.  Assuming a 5-year run of an upgraded 1.2 GeV proton accumulator ring with 20~ns pulses and a 75\% uptime, we generate argon recoils from dark matter scattering using the BdNMC~\cite{deNiverville:2016rqh} simulation.  We then pass these, as well as signals from beam-related and steady-state backgrounds, through a Geant4-based simulation of the detector response to obtain the argon recoil sensitivity curves shown in Fig.~\ref{fig:darksector}.  We probe thermal relic density targets for both fermion and scalar dark matter.

\subsection{Sterile Neutrino Searches}

\begin{figure}
\centering
  \includegraphics[width=0.3\textwidth]{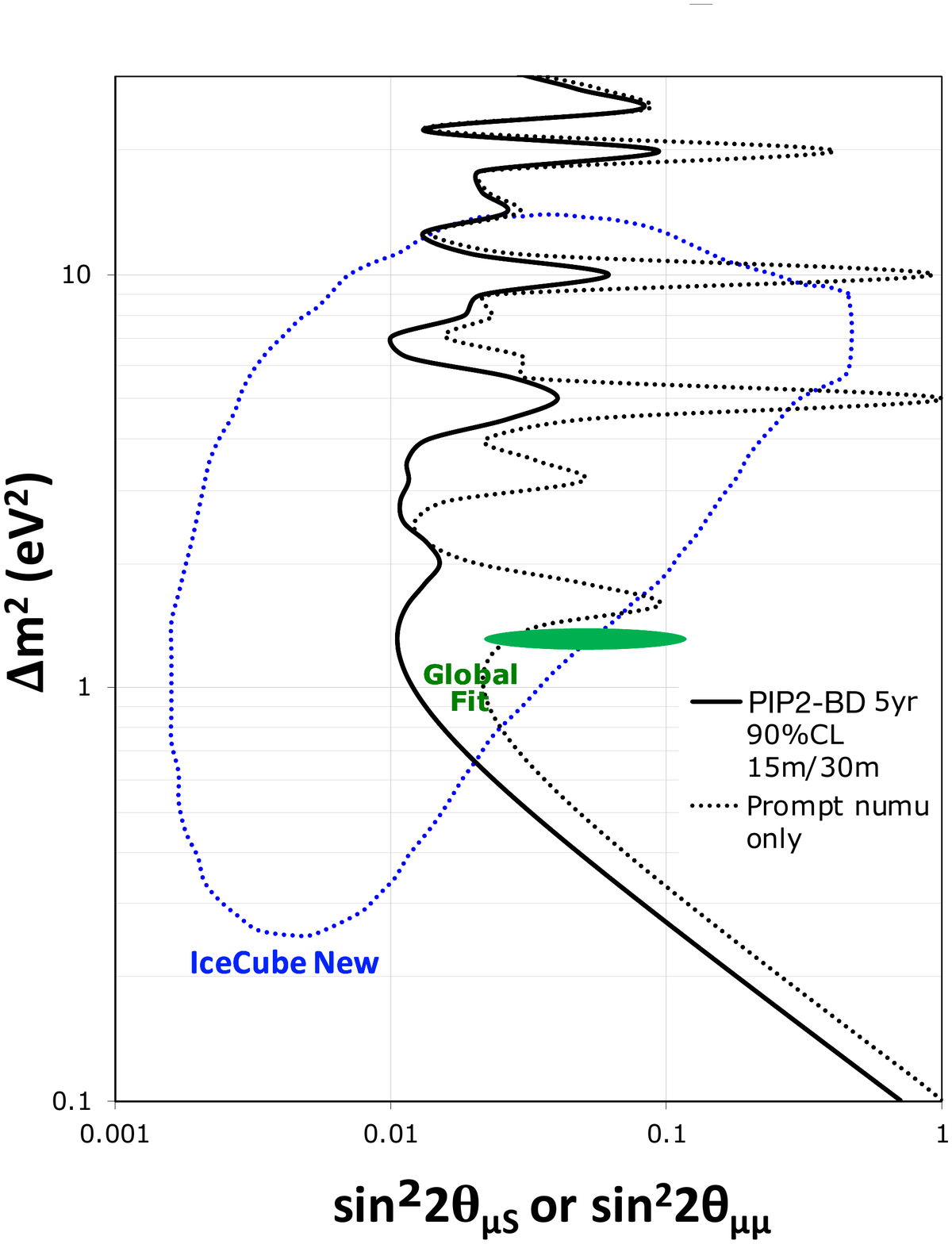}  \includegraphics[width=0.3\textwidth]{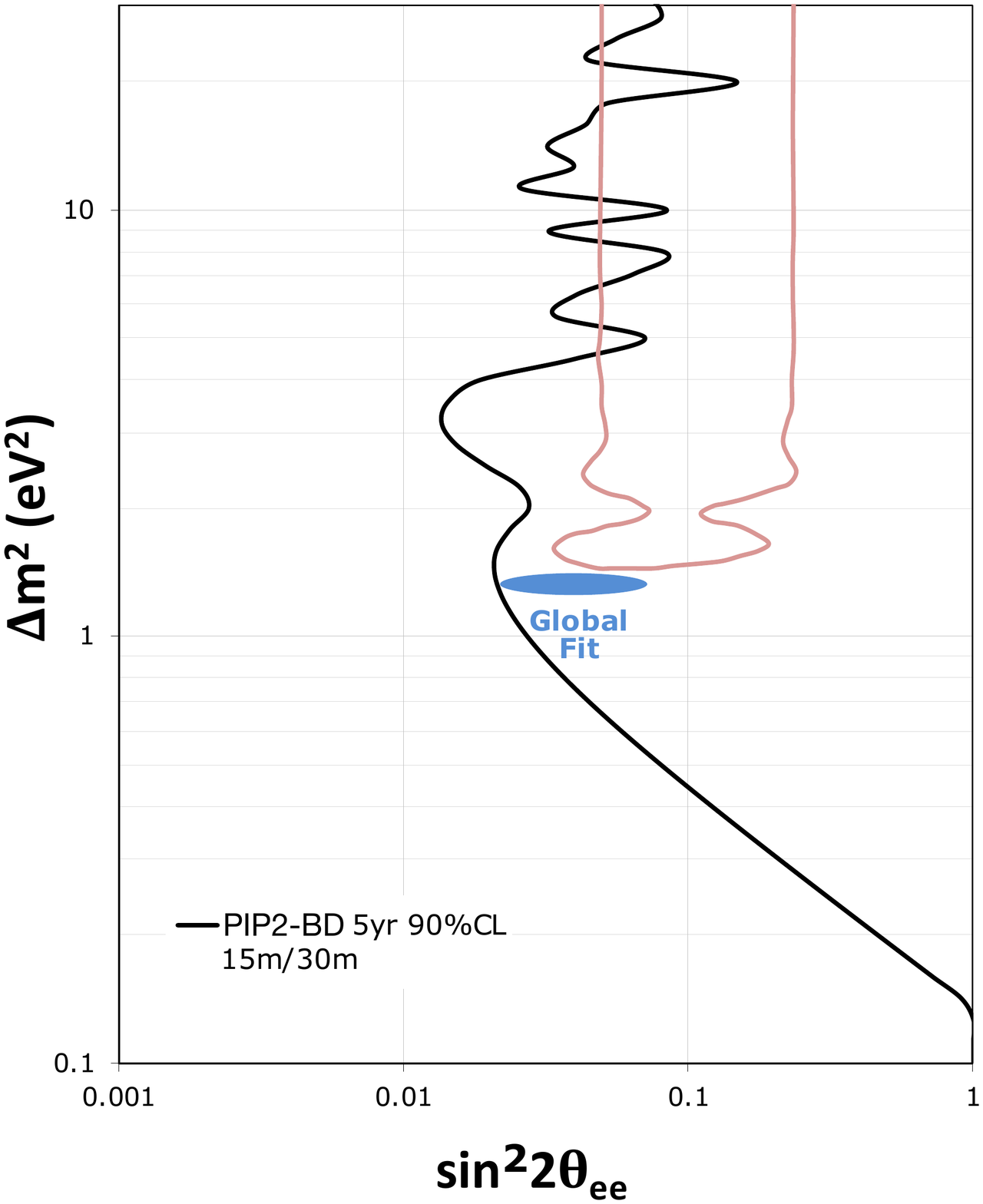}  \includegraphics[width=0.3\textwidth]{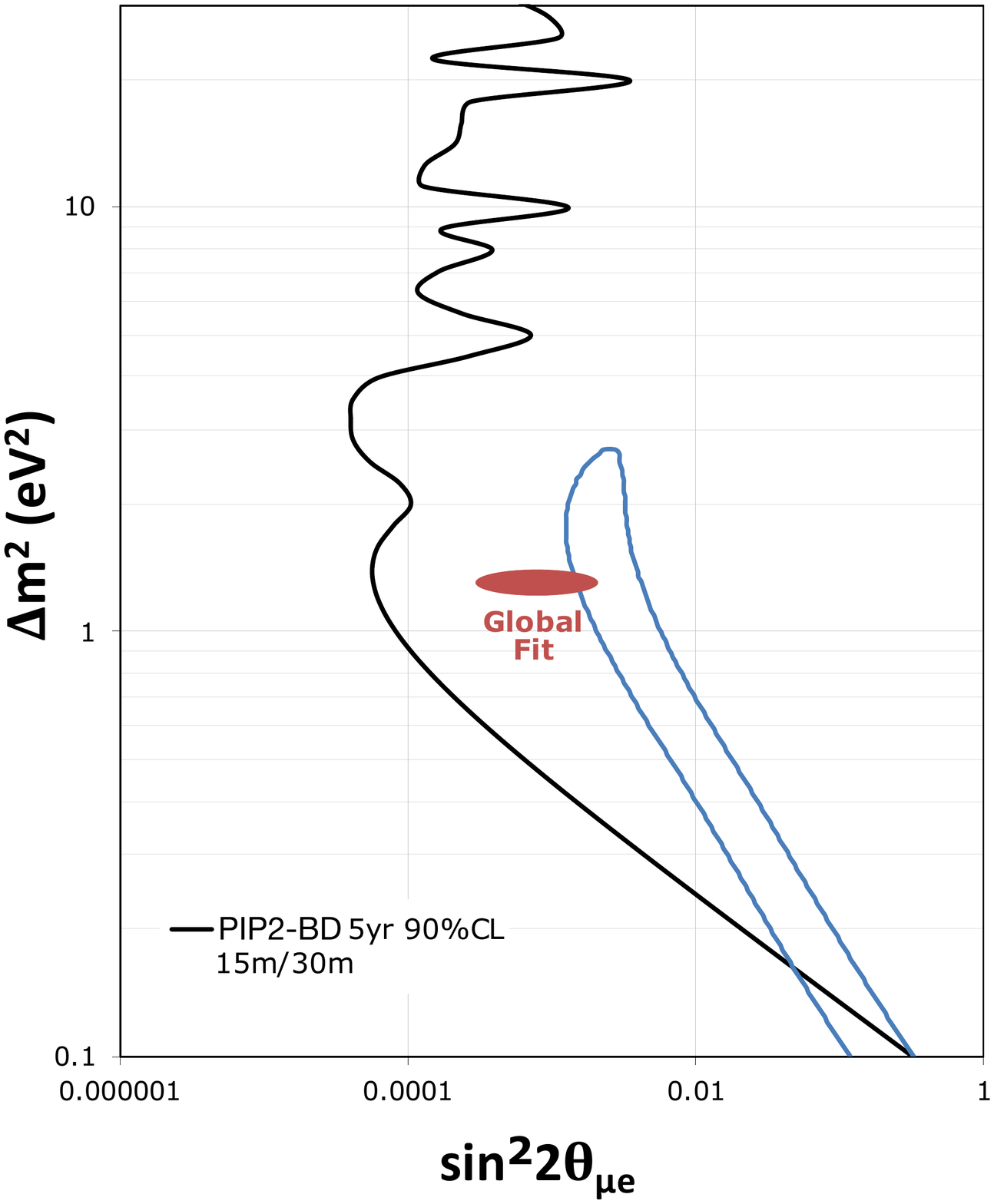}
  \caption{PIP2-BD 90\% confidence limits on active-to-sterile neutrino mixing compared to existing $\nu_\mu$ disappearance limits from IceCube~\cite{Aartsen:2020iky} and a recent global fit~\cite{Diaz:2019fwt}, assuming a 5~year run (left). Also shown are the 90\% confidence limits for $\nu_\mu$ disappearance (left), $\nu_e$ disappearance (middle), and $\nu_e$ appearance (right), assuming the $\bar\nu_\mu$ and $\nu_e$ can be detected with similar assumptions as for the $\nu_\mu$. }
  \label{fig:sterilenu}
\end{figure}

Decay-at-rest neutrinos from a stopped pion beam dump provide an excellent source of $\nu_\mu$, $\bar\nu_\mu$, and $\nu_e$ with a time structure that can separate $\nu_\mu$ from $\bar\nu_\mu$ and $\nu_e$. Using a lightly-doped oil Cerenkov detector, the LSND experiment found evidence for an excess of $\bar\nu_e$ 30~m downstream from a high-powered, 800~MeV proton beam dump, which can be interpreted as evidence for short-baseline $\bar\nu_\mu \rightarrow \bar\nu_e$ oscillations driven by a light sterile neutrino with a $\Delta m^2 \sim 1$~eV$^2$ mass-squared splitting~\cite{Aguilar:2001ty}. A larger, follow-up experiment could be mounted at the Fermilab beam dump facility as a direct test of the LSND anomaly, using the same technology as LSND but located far off-axis and taking advantage of the low duty factor~\cite{Elnimr:2013wfa}. On the other hand, CEvNS provides a unique tool to definitively establish the existence of sterile neutrinos through active-to-sterile neutrino oscillations~\cite{Anderson:2012pn}.

Using CEvNS, we can explore both mono-energetic $\nu_\mu$ disappearance with $E_\nu=$ 30 MeV and the summed disappearance of $\nu_\mu$, $\bar\nu_\mu$, and $\nu_e$ to $\nu_S$, which can also put constraints on $\nu_\mu \rightarrow \nu_e$ oscillation parameters in a 3+1 sterile neutrino model. We consider here a setup consisting of identical 100-ton LAr scintillation detectors, located 15~m and 30~m away from a carbon proton beam dump with a 20 keV recoil energy threshold and an efficiency of 70\%.  We assume the neutron background in this dedicated facility could be suppressed to a negligible level for this experiment and that the signal-to-noise ratio for the remaining steady-state backgrounds is 1:1. In Fig.~\ref{fig:sterilenu}, we calculate the 90\% confidence limits on the $\nu_\mu \rightarrow \nu_S$ mixing parameter $\sin^22\theta_{\mu S}$ for a 5-year run of an upgraded 1.2 GeV proton accumulator ring operating with a pulse width of 20~ns, a duty factor of $\mathcal{O}(10^{-6})$, and a 75\% uptime, assuming a 9\% normalization systematic uncertainty correlated between the two detectors and a 36~cm path length smearing. Also shown are the 90\% confidence limits for $\nu_\mu$ disappearance, $\nu_e$ disappearance, and $\nu_e$ appearance, assuming the $\bar\nu_\mu$ and $\nu_e$ can be detected with similar assumptions as for the $\nu_\mu$.

\subsection{Accelerator Requirements}

\paragraph{Accelerated particles:}
Protons

\paragraph{Beam Energy:}
$\mathcal{O}$(1~GeV) beam energy.  Above 1.5 GeV, more of the beam energy goes into producing hadrons other than $\pi^+$.  Further optimization of the beam energy can be done, as increasing beam energy closer to 2-3~GeV provides access to kaon-decay-at-rest (KDAR) physics at the expense of increased decay-in-flight neutrino backgrounds.

\paragraph{Beam intensity:}
$\mathcal{O}$(0.1--1~MW)-class beam intensity. Another important parameter here is the protons-on-target (POT)/year of beam operation. A value of $\geq10^{22}$~POT/year would be competitive with other currently operating stopped-pion sources.  

\paragraph{Beam time structure:}
As the dominant backgrounds for the headline physics measurements described above are steady-state, non-beam-related backgrounds, a pulsed beam with a low duty factor, defined as the repetition rate times the proton pulse duration, is necessary to achieve adequate background rejection.  There are multiple ranges of interest for the duration of the beam pulse, which can provide different physics handles:
\begin{itemize}
    \item $\leq\mathcal{O}(1~\mu$s): Allows separation of neutrinos produced from charged pion decay from neutrinos produced from muon decay
    \item $\leq30$~ns: Allows separation of light dark matter produced from neutral pion decay from beam-related backgrounds (including neutrons).
\end{itemize}

\noindent A duty factor of $\mathcal{O}(10^{-5})$ or better allows for the maximum physics reach. Part of the physics program, particularly measurements relying on reconstructing higher-energy, MeV-scale deposits in the detector (which notably does not include either of the searches described above), can also be accomplished with a higher duty factor proton beam.

\paragraph{Target requirements:}
A lighter target material such as graphite is preferred as it will produce fewer neutron backgrounds relative to a higher-Z target material.  This is a thick target more appropriately characterized as a beam absorber.  The beam can be painted over a relative large phase space as it impinges on the target (current simulations assume a Gaussian beam profile with a standard deviation of 4.5 cm).

\paragraph{Other requirements:}

\paragraph{Timescales, R\&D needs, and similar facilities:}
While previous and current experiments such as Coherent CAPTAIN-MILLS and the COHERENT collaboration's CENNS-10 experiment have demonstrated that the technology is sufficiently mature to execute this experimental program, R\&D related to improved timing (for background rejection and $\nu_\mu$ versus $\bar\nu_\mu/\nu_e$ separation), energy reconstruction, lower thresholds, PID, and Cherenkov light reconstruction in single phase liquid argon scintillation detectors would provide additional reach to the physics program. In addition, the program will benefit from R\&D associated with maintaining a high light yield while scaling up to larger detector masses (purity, optical properties, and Xenon-doping, for example).  Finally, the logistics of obtaining large quantities of isotopically pure argon will be needed for achieving low backgrounds.


There are 4 facilities that can support physics programs with some overlap with the program outlined here.  On a similar timescale envisioned for this program (late 2020s), the Spallation Neutron Source at Oak Ridge National Lab could achieve a $>2$~MW, 1.3 GeV proton beam, followed by a a second target station in the 2030s.  Although the existing HEP experimental program in “Neutrino Alley” would not have sensitivities competitive with the Fermilab program laid out above, a scaled-up HEP program with dedicated space at the second target station would probe some of the same physics goals.  The Lujan Center at Los Alamos National Lab, the J-PARC Material and Life Science Experimental Facility, and the European Spallation Source (ESS) are three additional spallation neutron sources providing 100 kW, 800 MeV proton beams, 1 MW, 3 GeV proton beams, and 5 MW, 2 GeV proton beams, respectively.  While the ESS has a relatively large duty factor of 4\%, a proposed upgrade to the ESS would add a proton accumulator ring and provide 5 MW of 2 GeV protons for a decay-in-flight neutrino oscillation program with a much lower duty factor and has also been studied for its capability to support a beam dump physics program.

%% file: protons-medium-energy.tex






\section{SBN-BD: 10 GeV Proton Beam Dump 
}\label{sec:10GeV}

\bigskip
\paragraph{Contacts: Matt Toups (FNAL) [toups@fnal.gov], R.G. Van de Water (LANL) [vdwater@lanl.gov]}
\medskip

\paragraph{Authors and Proponents: Brian Batell (University of Pittsburg), S.J. Brice (FNAL), Patrick deNiverville (LANL), Jeff Eldred (FNAL), A. Fava (FNAL), Kevin J. Kelly (FNAL/CERN), Tom Kobilarcik (FNAL), W.C. Louis (LANL), Pedro A.~N. Machado (FNAL), Bill Pellico (FNAL), Rex Tayloe (Indiana University), R. T. Thornton (LANL),  Z. Pavlovic (FNAL), J. Zettlemoyer (FNAL)}

%

\subsection{Physics Goals and Motivation}

Proton beam dumps are prolific sources of mesons enabling powerful techniques to search for vector mediator coupling of dark matter to the neutral pion and higher mass meson decays.  In the next five years, the PIP-II linac will be delivering up to 1 MW of proton power to the FNAL campus.  This includes a significant increase of power to the Booster Neutrino Beamline (BNB), which delivers 8 GeV protons to the Short Baseline Neutrino (SBN) detectors.  By building a new dedicated beam dump target station and using the SBN detectors, a greater than an order of magnitude increase in search sensitivity for dark matter relative to the recent MiniBooNE beam dump search can be achieved.  This modest cost upgrade to the BNB would begin testing models of the highly motivated relic density limit predictions.

Recent theoretical work has highlighted the motivation for sub-GeV dark matter candidates that interact with ordinary matter through new light mediator particles~\cite{Alexander:2016aln, Battaglieri:2017aum,BRN}. 
These scenarios are both cosmologically and 
phenomenologically viable and capable of accounting for the dark matter of the universe. Such sub-GeV (or light) dark matter particles are difficult to 
probe using traditional methods of dark matter detection, but can be 
copiously produced and then detected with neutrino beam experiments 
such as MiniBooNE, the Short Baseline Neutrino (SBN) Program, NOvA, and DUNE~\cite{deNiverville:2011it}. 
This represents a new experimental approach to search for dark matter and  
is highly complementary to other approaches such as underground direct 
detection experiments, cosmic and gamma ray satellite and balloon 
experiments, neutrino telescopes, and high energy collider 
experiments~\cite{Alexander:2016aln, Battaglieri:2017aum, BRN}. 
 Furthermore, searches for light dark matter provide an 
additional important physics motivation for the current and future 
experimental particle physics research program at the Fermi National Accelerator Laboratory (FNAL).  

The MiniBooNE experiment that ran at the FNAL Booster Neutrino Beamline (BNB) was originally designed for neutrino
oscillation and cross section measurements.  In 2014 a special beam dump run was carried out which suppressed neutrino produced backgrounds while enhancing the search for sub-GeV dark matter via neutral current scattering, resulting in new significant sub-GeV dark matter limits~\cite{Aguilar-Arevalo:2018wea}. 
 The result clearly demonstrated the unique and powerful ability to search for dark matter with a beam dump neutrino experiment.  

\subsection{A New BNB Beam Dump Target Station and Running in the PIP-II Era}

Leveraging the pioneering work of MiniBooNE's dark matter search, it has become clear that a significantly improved sub-GeV dark matter search can be performed with a new dedicated BNB beam dump target station optimized to stop charged pions which produce neutrino backgrounds to a dark matter search.  The new beam dump target can be constructed within 100~m of the SBN Near Detector (SBND) that is currently under construction~\cite{Machado:2019oxb}. In the PIP-II era 8 GeV protons with higher power can be delivered to the BNB, up to 15 Hz and 115 kW, which is a significant increase from the current 5 Hz and 35 kW. In a five year run this would result in $6\times10^{21}$ Proton on Target (POT) delivered to a new dedicated beam dumped while still delivering maximum levels of protons to the rest of the Fermilab neutrino program. The five year sensitivity would be greater than an order of magnitude better than current MiniBooNE dark matter sensitivity due to the reduced neutrino background from the dedicated beam dump, the detector's close proximity to the beam dump, and higher protons on target.

\begin{figure}[ht]
 \centerline{ 
 \includegraphics[width=0.45\textwidth]{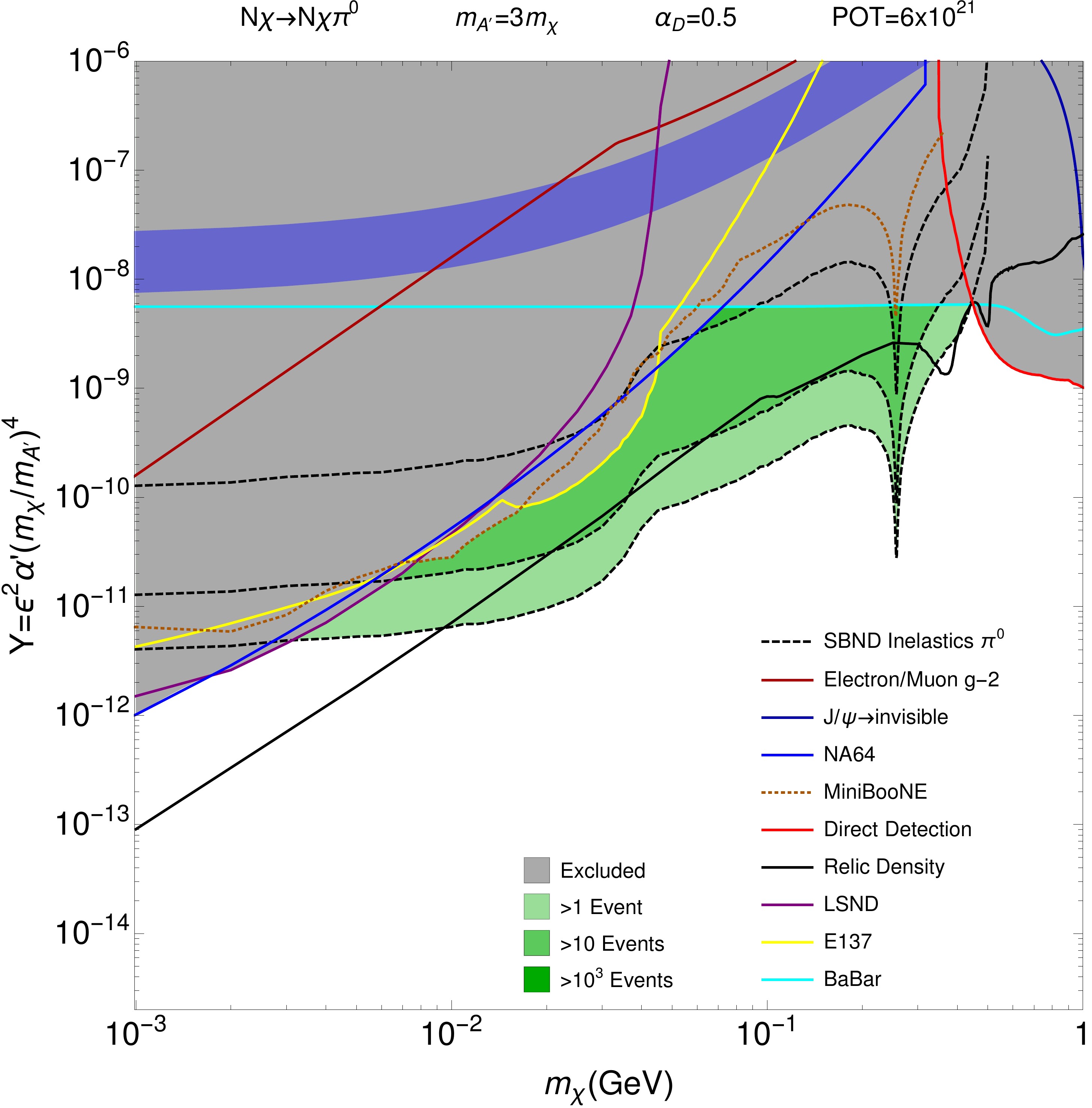}\hspace*{0.3cm}
 \includegraphics[width=0.45\textwidth]{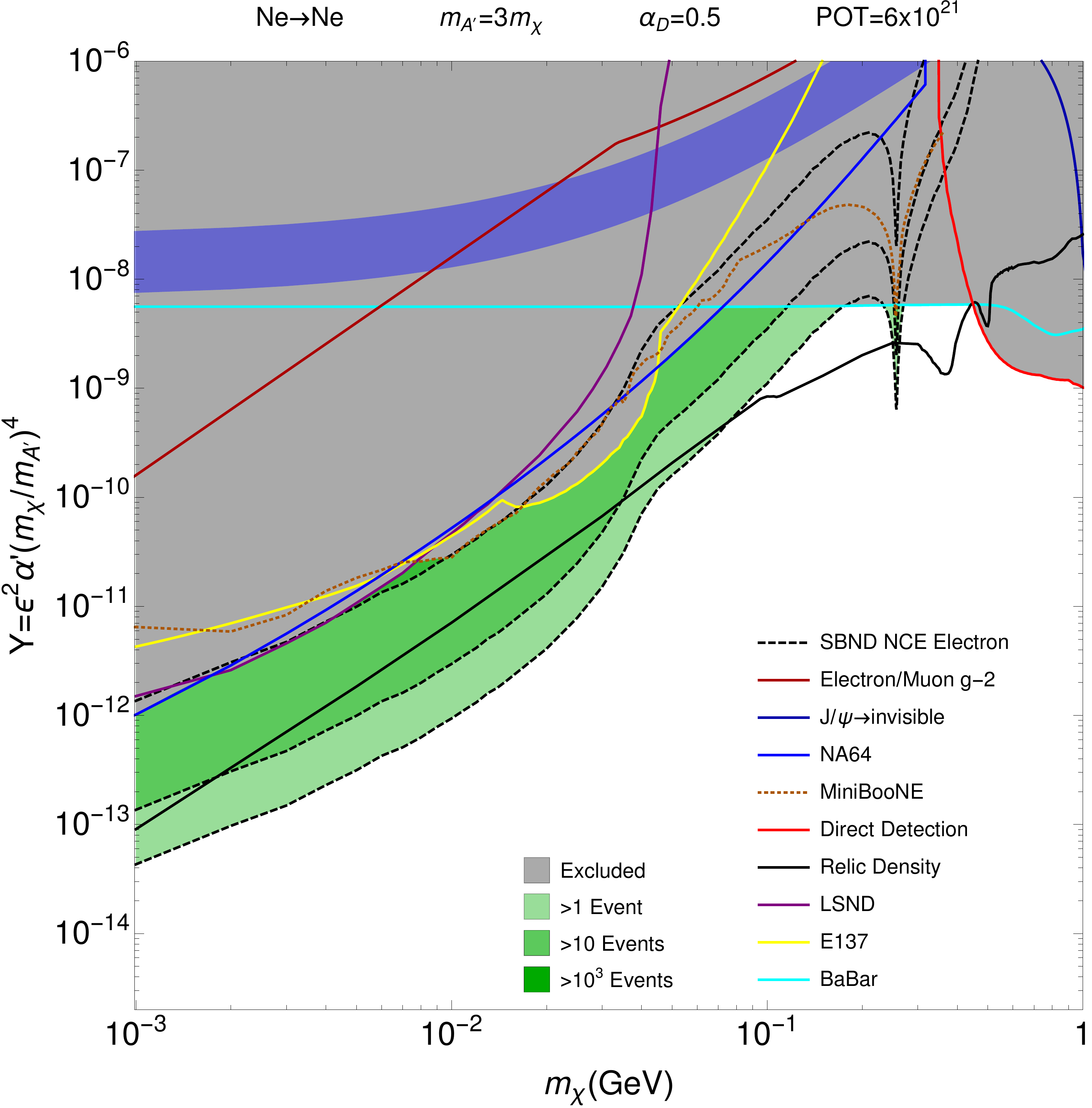}\hspace*{0.3cm}
 }
\caption{\footnotesize Event rate curves in terms of the relic abundance parameter (mixing
  strength) $Y$ vs. dark matter mass $m_\chi$ for $6\times10^{21}$ POT that could be
  achieved in a five year run with dedicated proton beam dump medium energy running in the PIP-II era. Left is the signal sensitivity for NC$\pi^0$ and right for NC-electron scattering with the SBND detector at 100 m from the dedicated beam dump.  Both panels show regions where we expect 1--10 (light green), 10--1000 (green), and more than 1000 (dark green) scattering events. The solid black line is the scalar relic density line describing the points in the parameter space for which the model reproduces the observed dark matter density in the universe.}
\label{ExpSensitivity1}
\end{figure}

\subsubsection{Required Infrastructure}
The Fermilab accelerator complex driven by the new PIP-II linac will be able to deliver 
80 kW of power to a dedicated beam dump on the BNB, while simultaneously delivering 1.2~MW of 120~GeV protons to LBNF/DUNE. A new target station fed by the BNB and on axis with the existing SBN neutrino detectors could be built relatively quickly and at a modest cost. Such a facility could be run concurrently with the SBN neutrino program, only using protons beyond the 35 kW limit. Events would be trivially separated on a pulse by pulse basis based on the target to which the beam is being delivered. The facility will require a Fe target about 2 m in length and 1 m in width to absorb the protons and resulting charged pions. Shielding and cooling requirements up to 80 kW are straightforward. Such a target would reduce the neutrino backgrounds by another three orders of magnitude relative to the regular neutrino running (see next section for details). Besides the higher power, the reduced neutrino flux background enables a significantly more sensitive search for dark matter relative to the MiniBooNE beam off target run.

\subsubsection{Neutrino Flux Reduction with Improved Beam Dump}

\begin{figure}[h]
 \centerline{ 
 \includegraphics[width=0.5\textwidth]{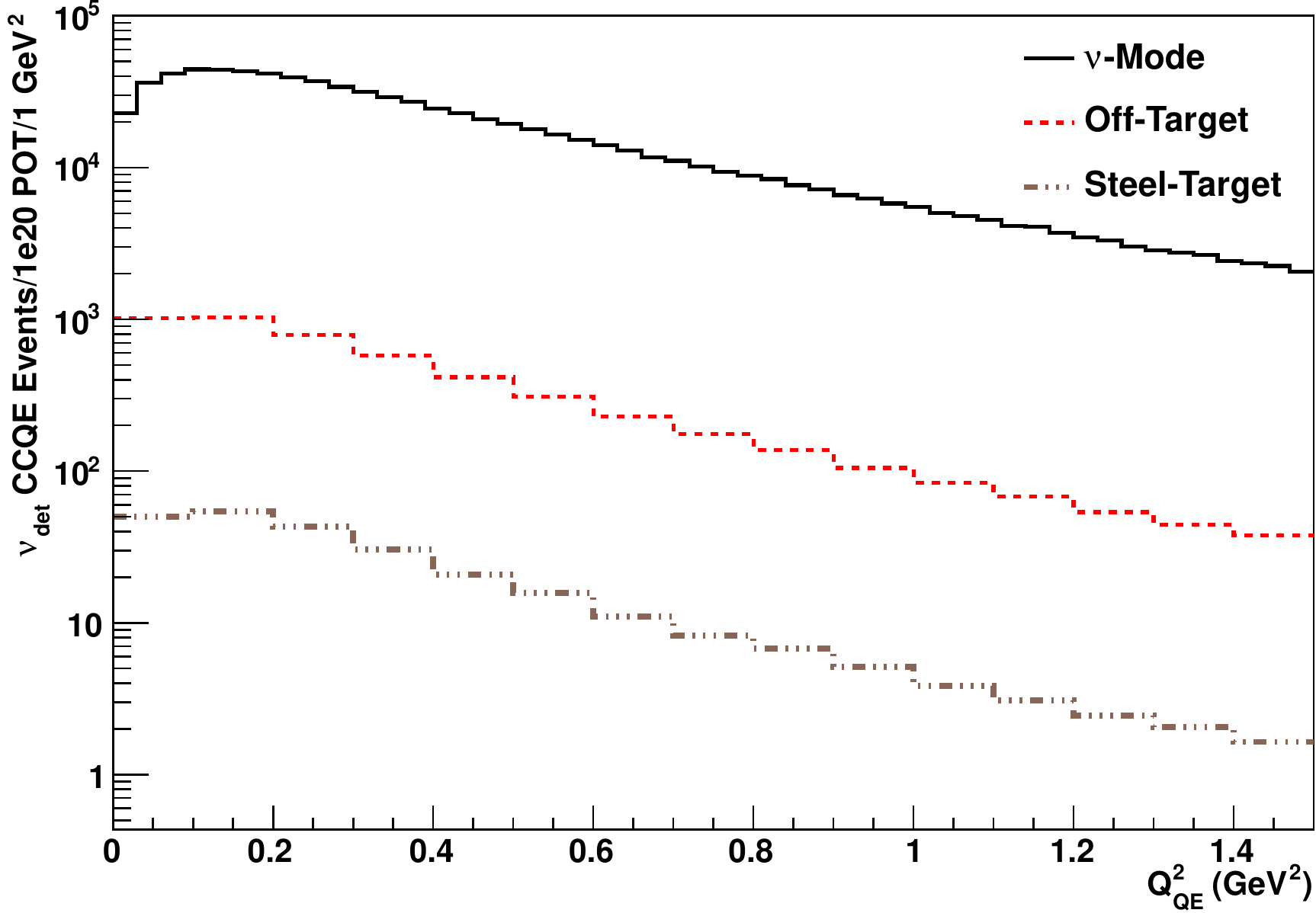}\hspace*{0.3cm}}
\caption{\footnotesize Detailed neutrino flux estimation for neutrino running 
  (solid black line), beam-off-target 50 m absorber running (dotted red line), 
  and a dedicated new BNB beam dump target station (dotted brown line).   In this final mode, the neutrino 
  flux reduction is a factor of 1000, or about 20 times better than 
  50 m absorber running. }
\label{NuFlux}
\end{figure}

To leverage the increased signal rate production, a corresponding reduction in neutrino-induced backgrounds is required.
The MiniBooNE-DM beam-off-target run steered the protons past the Be
target/horn and onto the 50 m absorber.  This reduces the
neutrino-induced background rate by a factor of $\sim$50, but there
was still significant production of neutrinos from proton interactions
in the 50 m of decay pipe air and the beam halo scraping of the target.
Further reduction of neutrino production occurs by directing the proton beam directly onto a dense beam stop absorber made of Fe or W.  This puts the end of the proton beam pipe directly onto the dump with no air gap.  Detailed BNB dump beamline simulations, which have been verified by data \cite{Aguilar-Arevalo:2018wea}, demonstrate that this would reduce neutrino-induced backgrounds by a factor of 1000 over Be-target neutrino running, which is a factor of twenty better than the 50 m
absorber as demonstrated in Figure \ref{NuFlux}. 

\subsection{Accelerator Requirements}

This proposal requires a new dedicated beam dump either reusing the the BNB or a separate new beam line.  The best and cheapest option is to build a dedicated dump feed by the BNB (kicker magnet just upstream of the current neutrino target/horn) and 100 m from the SBND detector.  

\paragraph{Accelerated particles:}
Protons

\paragraph{Beam Energy:}
On the order of 10 GeV.  Sensitivity studies have been performed for 8 GeV BNB line, but slightly lower and higher energies would have similar sensitivities.  

\paragraph{Beam intensity:}

Currently the BNB runs at a maximum power of 35 kW due to constraints on the horn and target.  Having a separate beam dump target station would allow up to 115 kW of beam power, limited by radiation beamline losses and target cooling requirements.

\paragraph{Beam time structure:}

Beam spills less than a few $\mu$s with separation between spills greater than 50 $\mu$s.  Shorter beam spills reduce cosmogenic and other random backgrounds. RF structure on the scale of $\sim$1 ns enable dark matter time of flight measurements increasing sensitivity. Requires detectors with similar timing capability.

\paragraph{Target requirements:}

Thick ($\sim$meters) dense targets (Fe, W, U, etc) are ideal for ranging out charged pions and stopping them before decaying into neutrinos, which form the biggest background for beam dark matter searches.

\paragraph{Other requirements:}
Building a dedicated beam dump and maintaining the current neutrino target allow flexible physics goals (sterile neutrino and dark sector searches), maximum use of POT, and utilization of shared resources like the BNB beam line.

\paragraph{Timescales, R\&D needs, and similar facilities:}

The timescale is similar to the construction of PIP-II and the expected upgrade in protons once online. The SBN detectors are expected to run for at least 10 years. The new dedicated beam dump could be built sooner and begin running using the SBN detectors at a lower rate until the PIP-II upgrade is complete. Such a facility could be built in as little as 1-2 years, and at modest cost below \$5M.  

There are no similar facilities in the world currently or planned in the next five years that can directly probe for dark matter masses up to 1 GeV with a medium energy proton beam.  This is a unique opportunity for FNAL to leverage existing SBN resources to lead the dark matter search and to directly probe relic density limits at the sub-GeV mass scale.






  



%% file: high-energy-dump.tex
\section{High Energy Proton Fixed Target}\label{sec:proton-beam-dumps-high}

\paragraph{Authors and proponents:} 
SpinQuest, DarkQuest, LongQuest\\
Stefania Gori, Nhan Tran, Ming Liu, Yu-Dai Tsai, Nikita Blinov

\paragraph{Related sections:}
For similar concepts at intermediate proton energy, see the previous section.

\subsection{Physics Goals, Motivation, and Setup}
High energy ($\mathcal O(100~{\rm{GeV}}$)) proton fixed target beam-dump experiments can be used to search for $\sim 10$ GeV scale and below dark sector particles. In general, high-intensity and high-energy proton beam dumps provide the largest production rates of dark sector particles, compared to lower energy proton or lepton fixed target experiments.

In a spectrometer-based experiment such as SeaQuest and its proposed upgrade (the DarkQuest experiment), weakly coupled long-lived dark sector states, once produced, can be detected through their displaced decays to visible SM particles. The high-intensity 120 GeV proton beam hitting the target can produce these dark sector particles.  
 The $\sqrt s=15$ GeV center of mass energy offers kinematic access to heavier dark particles in the mass range of $\sim 10$ GeV and below. Furthermore, the relatively short baseline (or shield thickness) of the order of 5 m, will allow accessing dark particle life-times of $\mathcal O(10{\rm{cm}}-1{\rm{m}})$.
 
 
 Depending on the specific detector setup, signatures containing electrons, muons, charged pions, and photons could be recorded and identified. This would result in the reach of large regions of unexplored parameter space of many dark sector models: (1) minimal models containing a ``mediator'' particle decaying resonantly to two SM particles (e.g. a dark photon decaying into an electron-positron pair), see \cite{Gardner:2015wea,Berlin:2018pwi,Tsai:2019buq,Batell:2020vqn,Blinov:2021say} for the study of minimal dark sector models at DarkQuest; (2) non minimal models containing, for example, an excited DM state that decays to the DM candidate together with visible particles, as it happens e.g. in strongly interacting massive particle (SIMP) dark matter \cite{Berlin:2018tvf} or in models of inelastic DM \cite{Berlin:2018pwi}. 
 
\begin{figure}[tbh!]
\centering
\includegraphics[width=0.47\textwidth]{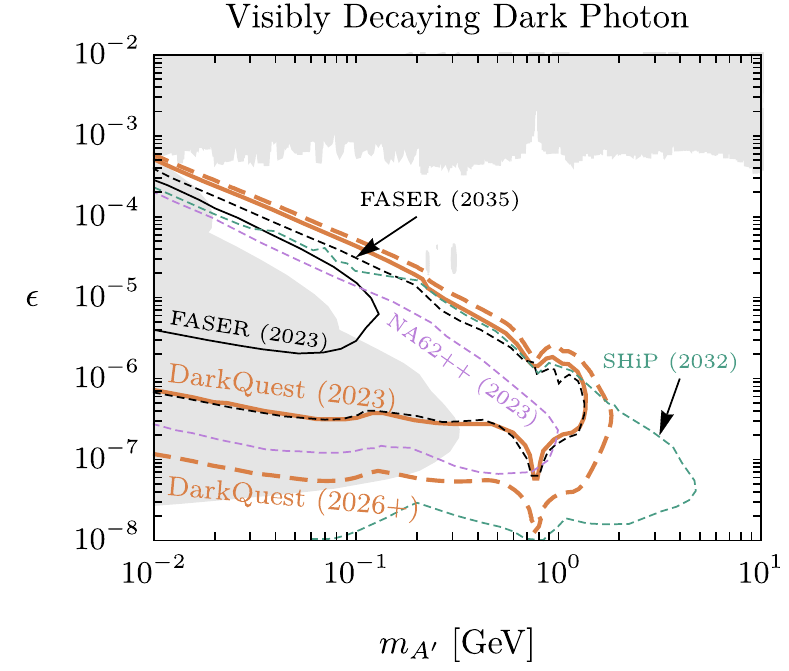}
\includegraphics[width=0.47\textwidth]{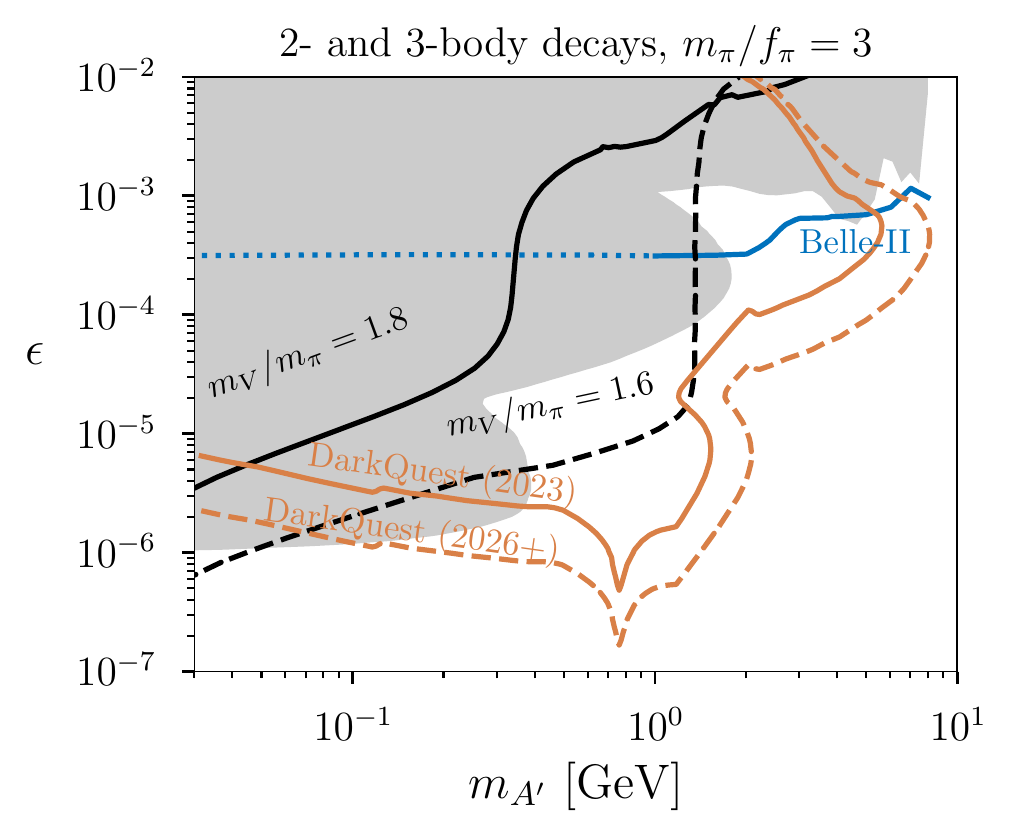}
\caption{Existing constraints (gray regions) and sensitivity of DarkQuest and of future (proposed or presently running) experiments. The reach of DarkQuest is shown for two different luminosities: $10^{18}$ POTs (solid lines denoted ``DarkQuest (2023)''), and $10^{20}$ POTs (dashed lines denoted ``DarkQuest (2026+)''). Left panel: the sensitivity of DarkQuest to displaced decays of dark photons (plot updated from \cite{Berlin:2018pwi}). Right panel: the sensitivity of DarkQuest to signals of strongly interacting dark sectors (SIMPs). Dark sector pions make up all of the dark matter along the black contours (from \cite{Berlin:2018tvf}).}
\label{fig:sesitivity_comparison}
\end{figure}

 In Fig. \ref{fig:sesitivity_comparison}, we present the reach on the parameter space of a minimal dark photon model (left panel) and of a SIMP model (right panel). In gray, we show the regions of parameter space already probed by past experiments. As we can observe, the main strength of the experiment will be to probe relatively high masses and an intermediate range of life-times that were not accessible to past experiments. Particularly, already with $10^{18}$ POTs, DarkQuest can extend the reach of past proton beam dumps that were using a longer baseline.
 
\subsection{Accelerator Requirements}
We assume that it has similar capabilities to what is currently being delivered to SpinQuest via Switchyard.  

\paragraph{Accelerated particles:}
Protons.

\paragraph{Beam Energy:}
120~GeV.

\paragraph{Beam intensity:}
approximately $10^{12}$ protons/second.

\paragraph{Beam time structure:}
CW (via resonance extraction).

\paragraph{Target requirements:}
Currently targeting 1e12 PoT/s, but higher is better.

\paragraph{Other requirements:}
Currently, the switchyard gets beam 4s per minute but could be increased by an order of magnitude if more data is needed.

\paragraph{Timescales, R\&D needs, and similar facilities:}
We could start in 2023, since much of the detector is already in place, aside from the EMCal upgrade.  However, given its size, it would be not easy to move. New detectors would be required for other future concepts, but they would all use existing technology.

\subsection{Global Context}

Dark sector searches with the DarkQuest apparatus provide a unique opportunity, particularly in the near term.  In general, the very short beam dump baseline ($\sim$5\,m) probes a challenging region of the lifetime/coupling phase space.  The primary effort which is similar to DarkQuest is the NA62 experiment at CERN.  However, the baseline of the experiment ($\sim$400\,m) for a similar beam energy means that the two efforts are complementary -- probing complementary regions of phase space.  FASER at the LHC is another similar effort, though being at a colliding beam requires a longer time to integrate the necessary statistics (over the life of the HL-LHC) and, therefore, to reach the planned sensitivity.

%% file: electron-missing-momentum.tex
\section{Electron missing momentum}
\label{sec:electron-missing-momentum}\label{sec:LDMX}

\paragraph{Authors and proponents:} 
LDMX-like, Nhan Tran, Gordan Krnjaic

\paragraph{Related sections:}
This detector concept is closely related to muon missing momentum, see Sec.~\ref{sec:muon-missing-momentum}. For connections to neutrino-nucleaus cross section measurments see Sec.~\ref{sec:sasha+ryan}.

\subsection{Physics Goals, Motivation, and Setup}

Searches for such light hidden sectors can be sensitive to the mediators as well as to the dark matter itself. Fixed target, accelerator-based searches employing the missing-momentum technique provide a particularly comprehensive probe of hidden sector physics. In the simplest and most predictive scenarios, those with dark matter annihilating through a light vector mediator, a missing momentum search such as the Light Dark Matter eXperiment (LDMX) can definitively explore thermal relic dark matter over most of the MeV to GeV range, and simultaneously test a broad range of other dark matter models and light, weakly coupled physics beyond dark matter. This unique strength is largely the result of probing dark matter interactions in the relativistic limit. Competing techniques that probe non-relativistic dark matter, such as direct detection, would require a 1e20 improvement in sensitivity over current constraints to have comparable physics reach.

Missing momentum experiments with electrons are particularly powerful as electron beams are high intensity, pure, and precise \cite{Akesson:2018vlm}. The missing momentum technique has a distinct advantage of experiments which detect rescattering of dark matter in a downstream detector \cite{Izaguirre:2014bca}.  Rescattering experiments scale as $\epsilon^4$ while missing momentum techniques scale as $\epsilon^2$.  Furthermore, missing momentum experiments are also sensitive to a wide range of more complex dark sector scenarios~\cite{Berlin:2018bsc}.  The detector can also be used to make important measurements of electron-nucleon scattering processes for the DUNE program~\cite{Ankowski_2020}.

LDMX requires a low current, high-repetition-rate electron beam to achieve high statistics with an energy in the few-GeV range. The dark force carrier is produced in interactions of the electron beam with a thin target via dark bremsstrahlung. The experimental signature is a soft wide-angle scattered electron and missing energy. The detector is composed of a tracker surrounding the target to measure each incoming and outgoing electron individually and a fast hermetic calorimeter system capable of sustaining a O(100) MHz rate while vetoing low-multiplicity SM reactions that can mimic the DM signal. LDMX leverages mature and developing detector technologies and expertise from existing programs like the Tevatron and LHC to achieve the necessary detector performance.

\subsection{Accelerator Requirements}
A low current, high-repetition-rate electron beam -- bunch multiplicities should be of O(1).  

\paragraph{Accelerated particles:}
Electrons

\paragraph{Beam Energy:}
$\sim$3~GeV to $\sim$20~GeV, O(10 GeV) is sort of the sweet spot as we can have a compact detector (better for hermiticity). Going too low creates real challenges for understanding hadronic backgrounds -- needs to be confirmed.  

\paragraph{Beam intensity:}
Mentioned above but O(1) electron per RF bucket

\paragraph{Beam time structure:}
CW-ish -- an electron per RF bucket at 53 Mhz

\paragraph{Target requirements:}
None (part of experiment)

\paragraph{Other requirements:}
N/A

\paragraph{Timescales, R\&D needs, and similar facilities:}
All the technology currently exists on the detector side.  

\subsection{Global Context}

There are no planned missing momentum/energy/mass searches in the world that have the sensitivity that LDMX has planned. NA64 at CERN is the nearest competitor sensitivity-wise and uses the missing energy technique.  However, it doesn't not reach the sensitivity of LDMX (or an LDMX-like experiment) and will not reach all thermal relic density milestones for sub-GeV dark matter.  The current LDMX collaboration is planning  to move forward with detector design based on being hosted by SLAC off the LCLS-II beamline.

%% file: lepton-nucleon.tex
\section{Nucleon Electromagnetic Form Factors from Lepton Scattering}\label{sec:sasha+ryan}

\paragraph{Authors and proponents:} 

John Arrington (Lawrence Berkeley National Lab), Richard J.\ Hill (University of Kentucky and Fermilab), Gabriel Lee (Cornell University, Korea University, and University of Toronto), Ryan Plestid  (University of Kentucky and Fermilab)\footnote{rpl225@uky.edu}, Oleksandr Tomalak (University of Kentucky, Fermilab and Los Alamos National Laboratory)\footnote{tomalak@lanl.gov}

\paragraph{Related sections:}
Related to Sections~\ref{sec:LDMX} and ~\ref{sec:muon-missing-momentum}.

\subsection{Physics Goals, Motivation, and Setup}

Nucleon electromagnetic form factors are ubiquitous in any description of nucleon interactions at GeV energies and below. If known with sufficient precision,  they provide predictive input for a number of ``frontier'' physics problems. Two such prominent examples are charged-current elastic neutrino-nucleon scattering, and structure-dependent shifts of atomic energy levels.

Current data sets on nucleon electromagnetic form factors have some notable deficiencies. First, there is significant tension in the determination of the magnetic form factor of the proton using either the most recent and most precise (A1@MAMI) data set or the pre-existing global data set. Second, the precision and kinematic range of the existing data for the neutron form factor is lacking. In addition to the experimental limitations, most information on form factors come from electron scattering measurements, where large radiative corrections, including model-dependent two-photon exchange contributions, must be understood and removed to isolate the form factors needed in neutrino scattering and in atomic physics measurements.
Recently, polarization experiments have allowed for measurements with reduced radiative corrections for specific observables, but muon scattering can also improve our understanding of the form factors due to the significantly reduced Bremsstrahlung contributions. 

The limitations in existing data discussed above have considerable impact in the kinematic range relevant for neutrino-nucleon charged-current elastic scattering, namely for $Q^2 < 1~\mathrm{GeV}^2$ (above this value, resonance and deep inelastic scattering dominate). Currently, calculations of the neutrino-nucleon charged-current elastic cross section differ at the 3--5\% level when using recent A1@MAMI~\cite{Bernauer:2013tpr} form factor data as opposed to previous world data. In addition, questions have been raised~\cite{Lee:2015jqa,Sick:2012zz} about the corrections and systematic uncertainties in the A1 extractions. This means that there is a significant uncertainty in even the proton form factors based on high-statistics measurements, which will significantly limit the ability to make measurements aiming for percent-level precision. 
New high-precision lepton-scattering measurements of the proton magnetic form factor with independent systematics can potentially resolve this discrepancy.  

The isovector vector form factors that enter into the charged-current elastic process are given by differences of proton and neutron electromagnetic form factors.
Therefore, similar measurements using a deuteron target, with common kinematics and experimental conditions, will improve on the precision of the isovector form factors and scrutinize nuclear models (with one of the simplest nuclei) used for the analysis of the experimental data.
Moreover, the robust treatment and experimental verification of nuclear physics corrections in electron-deuteron scattering will improve the extraction of the axial form factor from the neutrino scattering data. If sub-percent level precision is achieved, this will effectively fix the isovector vector form factors and allow future experiments~\cite{Snowmass2021LoInuHD} with neutrino beams to precisely determine the axial structure of nucleons. In particular, since the charged-current elastic process is an important input for $\nu_e A$ and $\nu_\mu A$ scattering at $Q^2 < 1~\mathrm{GeV}^2$, it is essential to have robust data coverage up to this threshold (a slightly higher threshold is required for $\nu_\tau A$ scattering $Q^2 < 3~\mathrm{GeV}^2$).

Extractions of the proton's root-mean-square charge radius from electron-proton scattering data give contradictory results. The muon-proton scattering experiment MUSE@PSI~\cite{Gilman:2017hdr} will yield a complementary data set in the kinematic range $Q^2 \lesssim 0.08~\mathrm{GeV}^2$ and provide a consistency check between electron and muon scattering in the low-$Q^2$ region. Scattering experiments with muons are subject to smaller radiative corrections, but different systematic errors. 
However, MUSE is designed to make precise comparisons of electron to muon scattering. While it will provide an important consistency check of the radii extracted with electrons and muons, it will not provide a precise absolute measurement of the proton charge radius. To extract the charge radius, complementary data over a wide kinematics range, in particular at higher $Q^2$, is needed. The same data will help resolve the proton magnetic form factor problem.

Nucleon form factors also play a central role in the spectroscopy of light nuclei. The current understanding of the proton magnetic form factor limits our knowledge of the proton Zemach radius (not to be confused with the charge radius), an important quantity in atomic physics. This radius provides the leading structure-dependent correction to the hyperfine splitting in hydrogen. In the next decade, three experiments will extract the proton Zemach radius by performing the first precise measurements of the ground-state hyperfine splitting in muonic hydrogen~\cite{Pohl:2016tqq,Dupays:2003zz,Adamczak:2016pdb,Ma:2016etb}. The Zemach radius is sensitive to form factors below $Q^2 < 1~\mathrm{GeV}^2$~\cite{Tomalak:2017npu}, and precise knowledge in this regime will allow us to better understand the agreement or tension with Standard Model predictions. In the same vein, the uncertainties of the proton form factors (both electric and magnetic) also introduce theoretical uncertainties in the calculation of the structure-dependent two-photon exchange corrections to S levels in muonic hydrogen~\cite{Pohl:2010zza,Antognini:1900ns,Carlson:2011zd}, and to 1S--2S splitting in ordinary hydrogen~\cite{Parthey:2010aya,Parthey:2011lfa,Tomalak:2018uhr};  this theoretical uncertainty is of the order of the experimental precision. Measurements of the 1S--2S line have the smallest uncertainty in hydrogen spectroscopy and serve as the main input in the determination of the Rydberg constant.

The A1@MAMI data set remains the single most precise high-statistics data set; however, it is in significant tension with previous world data~\cite{Bernauer:2013tpr,Lee:2015jqa,Ye:2017gyb,Borah:2020gte}. This tension demands further study and, as we have outlined above, will impact neutrino-nucleon scattering and spectroscopy measurements. Motivated by the above considerations, we propose the use of hydrogen and deuterium targets in a high-intensity electron (or muon) beam to improve on the precision of proton and neutron form factors. As emphasized above, to evaluate cross sections with muon and electron neutrinos, form factors should be measured at $Q^2 < 1~\mathrm{GeV}^2$ (for tau neutrinos, $Q^2 < 3~\mathrm{GeV}^2$).\footnote{Existing measurements of nucleon form factors are described in Refs.~\cite{Punjabi:2015bba,Ye:2017gyb}.} The kinematic range, together with the angular resolution of the detector, determine the minimum energy of the beam. We note that the current leading precision from A1@MAMI is \emph{systematics} limited, so issues such as angular resolution and beam stability are likely to be more important than maximizing statistics.

\subsection{Accelerator Requirements}

\paragraph{Accelerated particles:}

Electrons or muons: an electron beam has larger statistics and is easier to produce, whereas a muon beam is subject to different systematics, but furnishes a novel probe of particular radiative corrections.

\paragraph{Beam Energy:}

Assuming angular coverage of the detector up to $\sim 150$ degrees, energies above $850$--$900~\mathrm{MeV}$~$[1.8$--$2~\mathrm{GeV}]$ are required to probe the $Q^2 < 1~\mathrm{GeV}^2$~$[Q^2 < 3~\mathrm{GeV}^2]$ range.

For a measurement using a very forward-angle detector, similar to the PRAD experiment at Jefferson Lab, larger energies (1--3~GeV) would be desired.   

\paragraph{Beam intensity:}

For electron beams, beam currents of roughly 1~nA are sufficient for forward-angle measurements, and up to 1--10~$\mu$A for coverage of a wider range of kinematics. For muon measurements, 10$^7$--10$^8$ muons per second would permit forward-angle measurements and, depending on target and detector configuration, would begin to open up the desired kinematic range. Note that for the forward detection, it will be important to have a well-characterized muon beam with a small emittance. 

\paragraph{Beam time structure:}

For most measurements, a continuous or pulsed structure (ideally with a duty factor of $\sim$1\% or larger) should be sufficient. It will be important to have reliable measurements of the beam position and beam current that maintain precision over the full range of beam intensities, including very low beam currents. 

\paragraph{Target requirements:}

Hydrogen and deuterium targets. 

\paragraph{Other requirements:}

Unpolarized scattering measurements are the traditional way to access form factors. A double-polarization experiment would also work, requiring a polarized beam and either a recoil polarimeter or polarized target, the latter being required for very low $Q^2$ measurements.


\paragraph{Timescales, R\&D needs, and similar facilities:}

Technology exists to pursue the simplest unpolarized electron-proton scattering measurement. To perform a complementary experiment with other systematics, one can exploit muon beams or the double-polarization transfer technique. These experiments would be novel and R\&D will be needed. Alternatively, nucleon form factors can be also constrained by measuring the production of lepton pairs~\cite{Pauk:2015oaa,Carlson:2018ksu}.


Data should be analyzed before the analysis of future neutrino scattering experiments on elementary targets. 

There are no current or planned experiments dedicated to measurements of $G_M^p$ in the range $Q^2 < 1~\mathrm{GeV}^2$.

Measurements of the proton magnetic form factor using the polarization transfer technique at $Q^2 < 0.05~\mathrm{GeV}^2$ will be performed by MAGIX@MESA~\cite{Grieser:2018qyq,Caiazza:2020sda}.

Future $ed$ scattering data will come from experiments at MAMI~\cite{Mainz_neutron}.

The A1@MAMI experiment~\cite{Bernauer:2013tpr} covered almost the whole kinematic region of interest. To avoid background from scattering from walls, a new measurement with a gas-jet target is being planned~\cite{Mainz_neutron}.

Recent and planned experiments with unpolarized electron beams tuned for the low-$Q^2$ region include:
PRAD@JLab~\cite{Xiong:2019umf}, ProRAD@PRAE~\cite{Faus-Golfe:2019lxi}, Ultra-low $Q^2$@Tohoku, and experiments in Mainz. 

Recent experiments with unpolarized electron and positron beams dedicated to two-photon exchange measurements include:
VEPP-3~\cite{Rachek:2014fam}, CLAS@JLab~\cite{Adikaram:2014ykv,Rimal:2016toz}, OLYMPUS@DESY~\cite{Henderson:2016dea}. 

Current facilities with naturally polarized muon beams tuned for low-$Q^2$ region: MUSE@PSI~\cite{Gilman:2017hdr} and COMPASS@CERN~\cite{Denisov:2018unj}.

The JLab, MAMI, and MIT-Bates polarization transfer measurements at low $Q^2$ can be found in Ref.~\cite{Zhan:2011ji}. Future JLab measurements are tuned to a higher $Q^2$ region or extremely low $Q^2$ (mainly below 0.01~GeV$^2$) for extraction of the proton charge radius.

%% file: electron-dump.tex
\section{Electron beam dumps}
\label{sec:ebeamdumps}

\paragraph{Authors and proponents:} 
Gordan Krnjaic\footnote{krnjaicg@fnal.gov} (Fermilab), BDX Collaboration

\paragraph{Related sections:}
Electron beam missing momentum (LDMX) from Sec. \ref{sec:electron-missing-momentum}, muon
beam missing momentum (M$^3$) from Sec. \ref{sec:muon-missing-momentum}, and proton beam dumps (DUNE,DarkQuest) Sec. \ref{sec:proton-beam-dumps}.

\subsection{Physics Goals, Motivation, and Setup}

\begin{figure}
    \centering
    \includegraphics{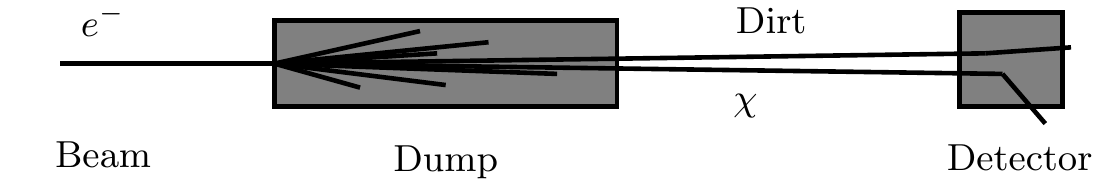}
    \caption{Schematic cartoon of an electron beam dump experiment to search for dark matter. A high intensity, multi-GeV electron beam impinges on a beam dump in which new particles $\chi$ can be produced. A detector is placed downstream of the target $\chi$ can deposit energy  by either scattering off detector targets (as shown here) or by decaying as they decay to visible matter upstream of the detector}
    \label{fig:ebeamdump}
\end{figure}

As shown schematically in Fig. \ref{fig:ebeamdump}, the basic setup of an electron beam dump experiment consists
of a high intensity multi-GeV electron beam impinging on a 
thick target and a downstream detector to register anomalous energy depositions; between the target and detector, there is typically also additional shielding to prevent Standard Model particles from reaching the detector and faking new physics signals. Unlike proton beam dump experiments, in which mesons are copiously produced in the target and yield large fluxes of neutrinos, electron beam dump experiments can only produce neutrinos through electroweak processes or inelastic electron-nucleus interactions, which emit pions that produce secondary neutrinos.

Such experiments are powerful tools 
for probing light weakly coupled particles below the GeV-scale. A key motivation
of these efforts is to probe dark matter candidates that achieve their relic density via thermal freeze out into standard model particles. If the annihilation rate proceeds via DM DM $\to$ SM SM processes, the same combination of couplings responsible for beam dump production and detection is in direct correspondence with the couplings responsible for setting the dark matter abundance in the early universe. Furthermore, this setup can also probe long lived particles that are produced in the target and decay upstream of the detector (e.g. visibly decaying dark photon or axion like particles)

Unlike traditional detectors designed for non-relativistic dark matter scattreing searches (e.g. XENON1T), here the signal consists of a large energy deposition inside the downstream detector from the boosted dark matter particles produced in the target. Thus, the detector requirements are fairly modest; a world leading beam dump experiment could consist of CSi crystals recycled from the BaBar experiment (as with the BDX) or mineral oil (as with the MiniBooNE experiment). As long as the detector can reigstier $\gtrsim$ 100 MeV energy deposits, detection efficiencies will be fairly high; detector performance is rarely a limiting factor for these searches. 

\subsection{Accelerator Requirements}
Electron beam dumps perform best with $\sim$ few GeV+ electron beams that can deliver at least $\sim 10^{20}$ electrons on target over an experimental runtime. Since the main backgrounds for these searches are induced by cosmic rays (e.g. secondary neutrons from cosmic ray showers in the atmosphere), it is beneficial to run with a pulsed beam and a small duty factor (beam on/off ratio) and only trigger on events within a certain time window with respect to beam pulses.  

\paragraph{Accelerated particles:} Electrons

\paragraph{Beam Energy:} At least $\sim$ GeV, but lower energies
can be useful if very high statistics can be achieved $\gtrsim 10^{22}$ electrons delivered to the target over an experimental lifetime

\paragraph{Beam intensity:} This depends somewhat on beam energy, as the available parameter space for interesting models depends on the masses of particles involved; probing lighter particles $\lesssim 10$ MeV benefits from greater statistics more than beam energy, but higher energies can probe heavier particles whose couplings are poorly constrained by comparison

\paragraph{Beam time structure:}
Can work with either CW or pulsed beams, but pulsed is always better for optimal cosmic background rejection strategies.

\paragraph{Target requirements:}
 Thick target, but no specific requirements so long as Standard Model particles can efficiently be stopped in the target or with additional downstream shielding

\paragraph{Timescales, R\&D needs, and similar facilities:}
 Technology is currently available and fairly inexpensive; similar experiments (e.g. E137 \cite{Batell:2014mga}) have been conducted in the past, but under conditions that optimized for long lived axion-like particles over a $\sim$ few 100 meter baseline with $10^{20}$ electrons on target; there is considerable room for improvement for dark matter detection, specifically. Potentially useful electron beams also currently exist in the United States at Jefferson Lab CEBAF at 12 GeV 
 \cite{Dudek:2012vr} and the SLAC LCLS-II at 4-8 GeV (depending on upgrade) \cite{Dalesandro:2017rtx,Raubenheimer:2018mwt}.
At CERN the secondary 100 GeV electron beam at the SPS
is currently being utilized for dark photon searches 
by the NA64 experiment \cite{NA64:2019imj} and 
there are lower energy options currently available: the Mainz Mircotron MAMI-C \cite{Achenbach:2008uf}
 and the DA$\Phi$NE test beam facility at INFN Frascati \cite{Raggi_2014} currently operate relativistic electron beams at few 100 MeV-GeV scale energies.  In principle any of these could be
 adapted for various dark sector beam dump searches

 Although there are proposals to utilize some of these facilities for dark sector experiments in the early 2020s -- particularly the LDMX missing momentum experiment at SLAC \cite{Akesson:2018vlm}, the BDX beam dump experiment at JLab \cite{Battaglieri:2016ggd}, and the PADME missing mass search at INFN Frascati \cite{frankenthal2019searching},
only NA64 at the CERN SPS  is is currently running a dark sector search experiment \cite{NA64:2019imj}.
 Possible future facilities include the a JLab polarized positron source \cite{accardi2020ejlab}, a possible electron beam dump at the future BNL electron ion collider \cite{Accardi:2012qut}, 
 the ILC beam dump \cite{Sakaki:2020mqb}.
 For a detailed list of many facilities and 
 their see Table II of Ref. \cite{Battaglieri:2017aum}.

%% file: muon-missing-momentum.tex
\section{Muon Missing Momentum}
\label{sec:muon-missing-momentum}

\paragraph{Authors and proponents:} 
Gordan Krnjaic (Fermilab), Nhan Tran (Fermilab)

\paragraph{Related sections:}
Electron beam missing momentum (LDMX) from Sec. \ref{sec:electron-missing-momentum}, electron beam dumps (BDX) from Sec \ref{sec:ebeamdumps} and proton beam dumps (DUNE,DarkQuest) Sec.  and~\ref{sec:proton-beam-dumps-high}. See Section~\ref{sec:sasha+ryan} for nuclear form factor measurements.

\subsection{Physics Goals, Motivation, and Setup}

\begin{figure}
    \centering
    \includegraphics{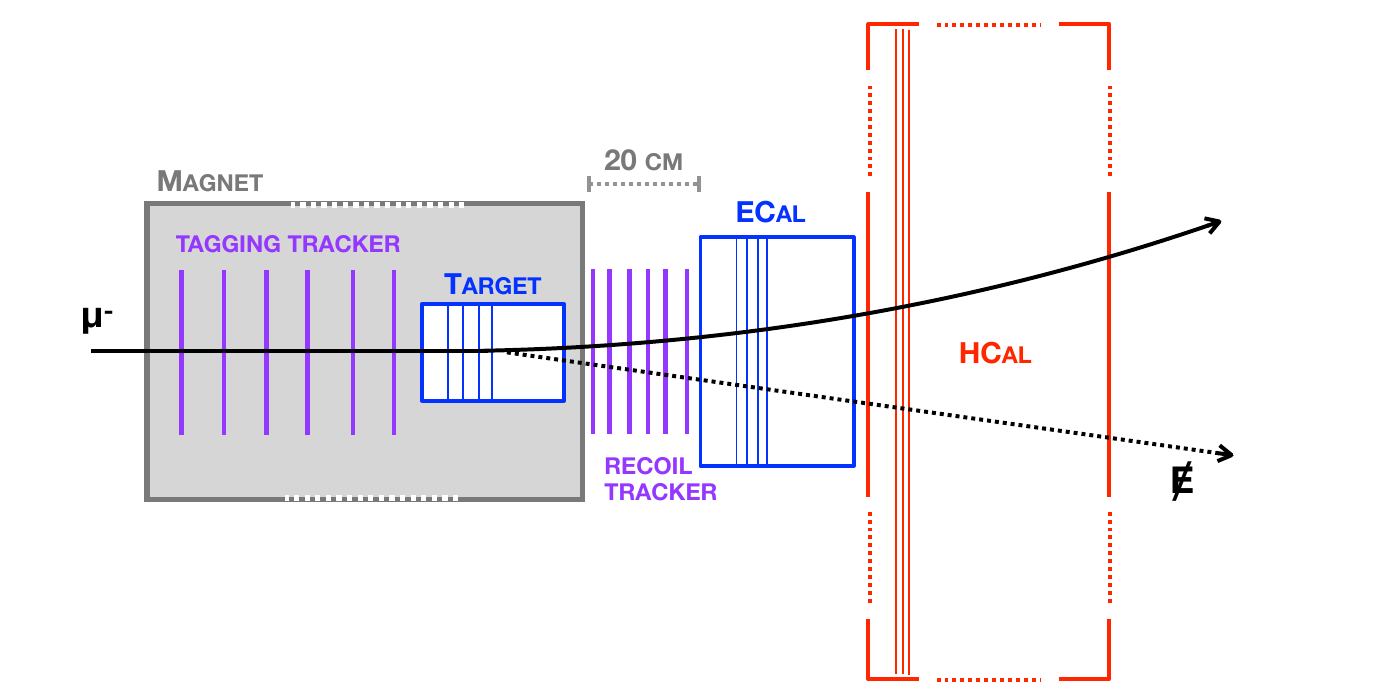}
    \caption{Schematic cartoon of a muon missing mometnum experiment. The incident $\sim$ 10+ GeV muon beam impinges on a {\it thick} target of many electron radiation lengths $\sim 50 X_0$, which is immersed in a $\sim$ few T magnetic field and bracketed by tracking material. Since the beam is relatively low current compared to beam dump experiments, the incident muon's momentum is measured before and after the target; signal events involve muons that have lost most of their energy/momentum due to invisible new particle production inside the target}
    \label{fig:m3}
\end{figure}

The muon missing momentum concept is proposed in~\cite{Kahn:2018cqs}.
As shown in Fig. \ref{fig:m3}, muon missimg momentum experiments involve a low current, multi-GeV muon beam that passes through a thick, active target, in which it can produce well motivated invisibly decaying particles beyond the Standard Model. The incident muon's energy is measured before and after it passes through the target and signal events are defined by  final state has lost a large fraction of its incident energy. Unlike electron beam missing momentum experiments from Sec \ref{sec:electron-missing-momentum}, here the energy of the final state recoiling particle cannot be obtained calorimatrically, so the target region and downstream tracking material must be immersed in a $\sim$ T strength magnetic field which enables a momentum measurement using track curvature. Downstream of the target, the ECAL/HCAL detector vetoes any additional Standard Model particles produced in the target.

Such an experiment is complementary to electron beam efforts because there is unique sensitivity to any new physics that couples preferentially to muons (e.g. $L_\mu-L_\tau$ gauge bosons), which can viably resolve the muon $g-2$ anomaly with light new, weakly coupled particles. This is the last viable scenario for explaining $g-2$ using only SM netural particles below the GeV scale. Furthermore, this setup has unique sensitivity to predictive thermal dark matter models 
in which the relic abundance is set by DM DM $\to \mu^+ \mu^-$ annihilation in the early universe. Finally, since muons are electrically charged, they can also compete with electron beam missing momentum experiments to cover invisibly decaying dark photon models  
 
\subsection{Accelerator Requirements}
Need a multi GeV muon beam particles. though the precise energy is not as important as the total number of muons delivered to the target. Anything above $\sim 10^{10}$ muons on target can cover interesting candidate explanations of the $g-2$ anomaly and $10^{13}$ muons can cover predictive dark matter thermal targets in muon-philic scenarios that cannot be probed in any other way

\paragraph{Accelerated particles:} Muons

\paragraph{Beam Energy:} Since muons cannot be controlled as well as electrons, threre is generically a spread of energies. However, optimal performance is obtained with an average energy of $\sim$ few 10s of GeV muons

\paragraph{Beam intensity:}
Typically the challenge is to produce as many muons as possible while still being able to track them individually. The optimal design for this setup has not yet been identified, but important physics goals can be achieved with $10^{10}$ muons per experimental runtime; 

\paragraph{Beam time structure:}
Pulsed is preferred to reject cosmic backgrounds. However, the currents here are fairly low, so 

\paragraph{Target requirements:}
No specific requirements, but better sensitivity is achieved with higher $Z$ targets. Unlike electron missing momentum (e.g. LDMX), here the target can be much thicker than an electron radiation lenght, so the target can be active and aid in the rejection of SM background events.

\paragraph{Timescales, R\&D needs, and similar facilities:}
Technology still in R\&D phases, need to know whether beam purity can be achieved at desirable levels. Possible precursor experiment at CERN if NA64 runs in muon beam mode.

\subsection{Global Context}

The main experimental effort which is similar is NA64mu and is based at CERN.  There is no official plan yet to run NA64mu.  Subject to beam parameters and running conditions, an experiment based at FNAL like $(M^3)$ and an NA64mu experiment at CERN would have similar sensitivities.

%% file: muon-beam-dump.tex
\section{Muon Beam Dump}

\paragraph{Authors and proponents:} 
Maxim Pospelov (University of Minnesota), Yi-Ming Zhong\footnote{ymzhong@kicp.uchicago.edu} (University of Chicago)

\paragraph{Related sections:}
See section~\ref{sec:ebeamdumps}.

\subsection{Physics Goals, Motivation, and Setup}

New Physics (NP) at low-mass has become an actively pursued topic of the intensity
frontier physics given the abundant evidence for NP in the neutrino and dark matter sectors, coupled with the lack of NP signal at the Large Hadron Collider (LHC).  The current $\sim 3.5 \sigma$ discrepancy between the predicted and observed value of the
muon anomalous magnetic moment has provided a strong motivation for light dark sector searches.  Popular candidates such as dark photons or dark Higgs have been tightly
scrutinized as possible explanations for such anomaly. Nevertheless, other solutions, such as a muonic dark sector with particles that dominantly couples to muons, remain viable and deserve attention.

\begin{figure}[htbp]
   \centering
   \includegraphics[width=0.8\textwidth]{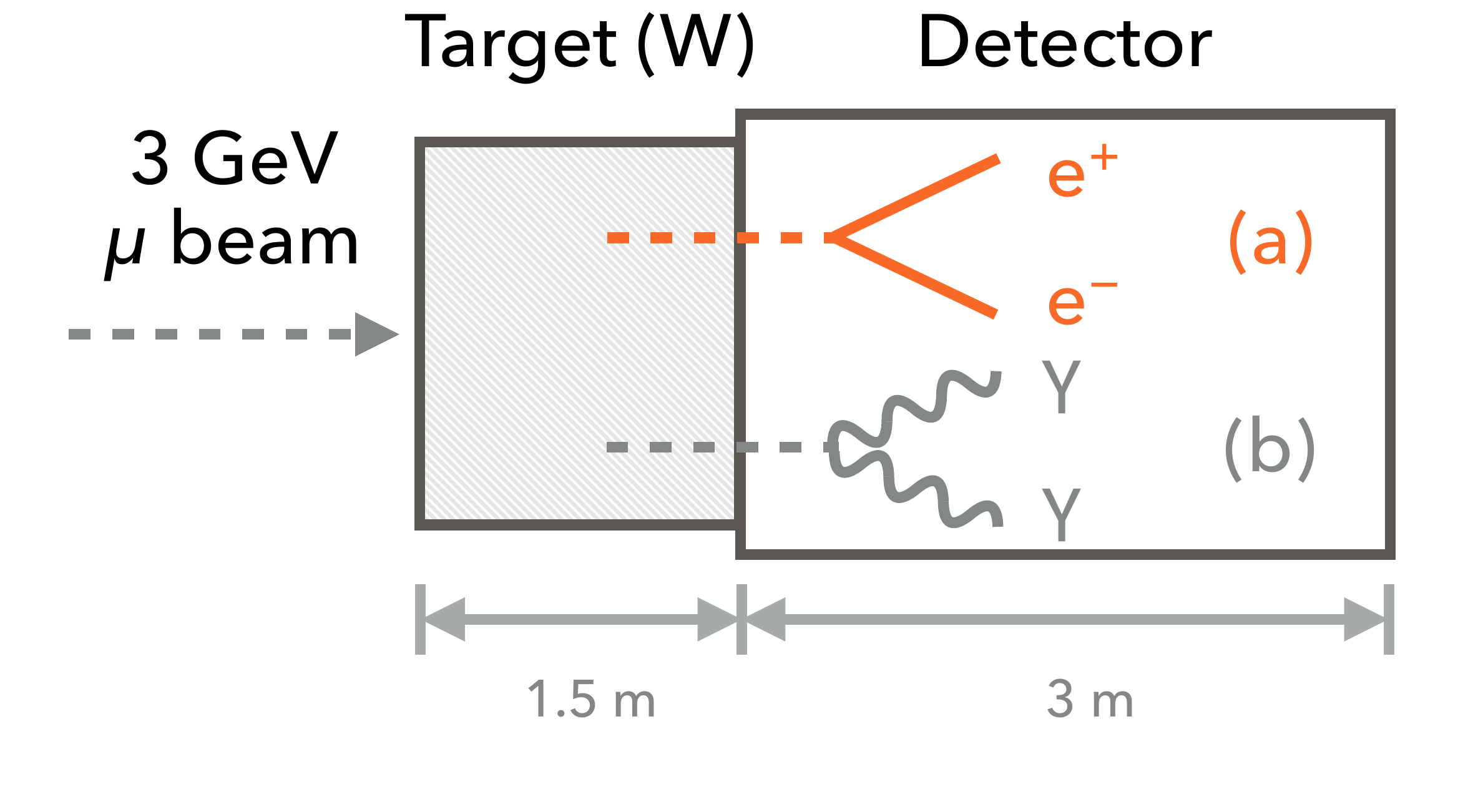} 
   \caption{Setup for muon beam-dump experiments at Fermilab. The muon beam energy is $\sim 3$ GeV and the target material is tungsten. The lengths of the targets and the detectors are shown in the plot. Here we focus on the displaced vertices from the decays of the dark sector particle.}
   \label{fig:setups-Fermi-mu}
\end{figure}

We suggest the simplest muon beam dump experiment using the existing Fermilab muon beam source with the anomalous energy deposition downstream from the dump~\cite{Chen:2017awl}. The experimental setup is shown as Fig.~\ref{fig:setups-Fermi-mu}. The incident muon beam energy we propose for the
experiment is around 3 GeV, as the accelerator complex is already tuned to this energy for the Muon g-2 experiment~\cite{Chapelain:2017syu}. Such a beam will be completely stopped in 1.5 meter thickness tungsten target. Dark sector particles that produced through the muon-nucleon bremsstrahlung interactions can then visibly decay inside the 3-meter detector equipped with electron or photon tracker/calorimeter. The signal signature of the dark sector particle is a displaced vertex that reconstructed from $e^+e^-$ or $\gamma\gamma$ pairs.

\subsection{Accelerator Requirements}

We assume the muon beam source is similar to what is currently being delivered to the Muon g-2 experiment. 

\paragraph{Accelerated particles:}
Muons

\paragraph{Beam Energy:}
Around 3 GeV.

\paragraph{Beam intensity:}
$10^7$ muons per second for 1 year or $3\times 10^{14}$ muons in total on target to reach a sensitivity of $\mathcal{O}(10^{-5})$ for the dark scalar coupling.

\paragraph{Beam time structure:}
CW

\paragraph{Target requirements:}
1.5 meter tungsten target. Other high-$Z$ material should also work if it can efficiently stop the muons.

\paragraph{Other requirements:}
No requirement on the muon beam polarization.

\paragraph{Timescales, R\&D needs, and similar facilities:}
The muon beam source exists and the detector technology is currently available. The projected parameter space can be also probed by the NA64-$\mu$ experiment at CERN, which is expected to start at Run 3 of the LHC. 

%% file: muonium.tex
\section{Physics with Muonium}

\paragraph{Authors and proponents:} 
{
Daniel M. Kaplan (Illinois Institute of Technology),\footnote{kaplan@iit.edu} on behalf of the MAGE Collaboration}

\paragraph{Related sections:}
 Charged lepton flavor violation with muon decays (Ch.~\ref{clfv-decays}).

\subsection{Physics Goals, Motivation, and Setup}

Muonium is an exotic, hydrogen-like atom consisting of an electron bound to an antimuon. Some key elements of the muonium physics case:

\begin{itemize}
    \item 
Muonium has been used to perform precision tests of quantum electrodynamics in a bound state free of  hadronic effects and to search for beyond-the-standard model physics via possible muonium--antimuonium oscillations~\cite{GORRINGE201573} via double charged-lepton flavor-violating processes. 
    \item
It is readily formed by stopping a $\mu^+$ beam in matter. 
    \item
It can be used to produce a high-quality, low-energy $\mu^+$ beam, as proposed for the $g-2$ experiment at J-PARC, in which a muonium beam is formed and then laser-stripped to produce the needed slow muon beam~\cite{KANDA2014212}. 
    \item
Muonium is likely the only avenue to a measurement of gravitational effects on 2$^{\rm nd}$-generation particles, and (along with positronium) one of only two potential avenues for gravitational effects on antileptons. 
    \item
An experiment to study the gravitational interaction of muonium in the field of the Earth is under development by the MAGE Collaboration~\cite{Antognini:2018nhb}. It employs a precision 3-grating interferometer illuminated by a horizontal beam of slow muonium, produced by stopping surface muons in superfluid helium~\cite{TAQQU2011216}.
\end{itemize}


\subsection{Accelerator Requirements}


\paragraph{Accelerated particles:}
Protons (producing surface muons via decay at rest of $\pi^+$ near the exit surface of the pion-production target).

\begin{figure}[tbh]
    \centering
    \includegraphics[width=.8\textwidth]{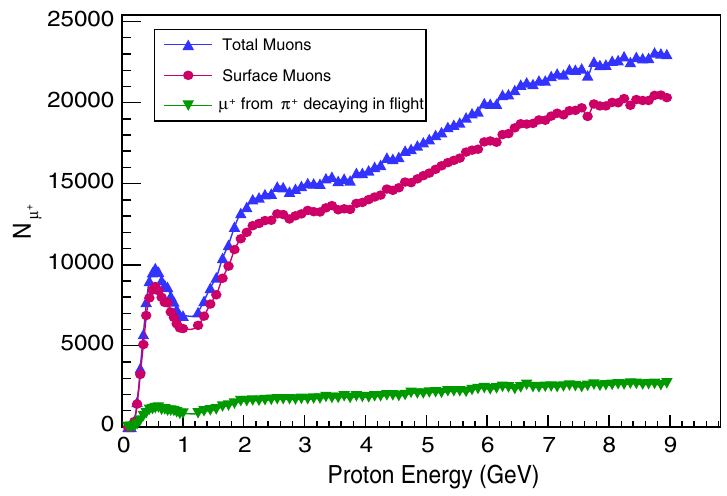}
    \caption{Production of muons, surface muons, and decay-in-flight muons in $p$C interactions vs.\ proton kinetic energy (from \cite{Bungau:2013hd}).}
    \label{fig:Mu-Prod-Bungau}
\end{figure}

\paragraph{Beam Energy:}
800\,MeV (or possibly higher). The production of surface muons by low-energy protons on a graphite target peaks at a proton kinetic energy of about 550\,MeV but is down from the peak by only about 20\% at 800\,MeV (see Fig.~\ref{fig:Mu-Prod-Bungau}).  Thus the PIP-II superconducting linac is an excellent source for the purpose. Above 1.5\,GeV, the production rate is seen to rise above the low-energy peak value, thus $\sim$\,1.5 to 8\,GeV is another possible operating range; the optimal choice of energy could depend on design optimization details. However, the availability of near-CW beam at 800\,MeV is likely to outweigh the higher production rate in the (pulsed) 8\,GeV option.

\paragraph{Beam intensity:} $\sim\,10^{13\pm 1 }$\,Hz of protons on target.
The current world leader in surface-muon intensity is PSI, with (CW) muon rate up to $\sim$\,$10^9$\,Hz. A PSI upgrade is under discussion with the goal of producing $\sim$\,$10^{10}$\,Hz of surface muons. This requires $\sim\,10^{12}$\,Hz of protons on target (depending on beam energy and target configuration), so an intensity in this ballpark is required in order to be competitive. To obtain a more precise estimate will require a study of the trade-offs among beam energy and target material, length, and geometry.

\paragraph{Beam time structure:}
CW (for MAGE, at least). The signature in a muonium experiment is a coincidence of two detected particles: the fast positron from the $\mu^+$ decay, and the slow, now-unbound, electron that is left behind. To suppress possible combinatoric background, it is thus desirable to have as close to a DC beam as possible.

However, the J-PARC $g-2$ approach employs a pulsed beam~\cite{KANDA2014212}.


\paragraph{Target requirements:}
The target thickness will be determined via an optimization study. It should be thick enough to allow a large number of pions to come to rest within it but thin enough to preserve a small beam spot at the exit surface. The target material should be robust against high-power operation. Typically, low-$Z$ materials such as graphite have been used, and graphite remains a good choice. Beryllium could also be investigated.   

\paragraph{Other requirements:}
MAGE does not require polarization, but some other muonium experiments might benefit from the large polarization that is characteristic of surface muon beams. 


\paragraph{Timescales, R\&D needs, and similar facilities:}
The SFHe-produced muonium beam needed for the MAGE experiment is not yet available at any of the world's surface-muon facilities, though R\&D on it is in progress at PSI~\cite{Antognini:2020uyp}. The PSI R\&D program is likely to take a few more years. It is focused on the ``muCool" approach~\cite{PhysRevLett.97.194801}, employing a cooled muon beam in order to minimize the $\mu^+$ stopping distance in liquid helium, at a cost of two to three orders of magnitude in muon intensity. An alternate approach not requiring a cooled beam has been devised~\cite{Antognini:2018nhb} and should ideally be pursued in parallel. Fermilab could be a good venue for this R\&D program.

Existing surface-muon facilities are located at TRIUMF in Canada, J-PARC and MuSIC in Japan, PSI in Switzerland, and ISIS in the U.K. There is discussion of establishing such a facility at Oak Ridge; if this goes ahead it will be the first one in the U.S. A brief further discussion may be found in \cite{Snowmass2021LoILEMu}.


%% file: NuFact-MuCol.tex
\section{Muon Collider R\&D and Neutrino Factory}

\paragraph{Authors and proponents:} 
Katsuya Yonehara (Fermilab), Daniel M. Kaplan (Illinois Institute of Technology) and David Neuffer (Fermilab) 

\subsection{Physics Goals, Motivation, and Setup}
A muon collider has the potential to enable high precision and extremely high energy elementary particle experiments, because it can offer collisions of point-like particles at very high energies, greatly exceeding the energy reach of proposed electron-positron colliders, which are energy-limited by radiative processes. 
Its effective energy can even be competitive with a proton collider of much higher energy, since the muon collision energy is fully available at constituent levels, unlike the case for protons. 

MAP (the Muon Accelerator Program) had the core goal of developing a muon collider that would provide high luminosities at high energies, enabling discoveries and precision physics. 
Since the cross section for s-channel production scales as $\sigma \propto 1/s$, the luminosity goal increases with energy. 
A tentative estimate for the required luminosity is
\begin{equation}
L = \left(\frac{\sqrt{s}}{10 \hspace{0.1cm}\rm{TeV}} \right)^{\!2} \times 10^{35}\,\rm{cm}^{-2} \rm{s}^{-1}.
\end{equation}
Assuming that experiments will have five years of operation, 
a collider energy of 14 TeV and the corresponding luminosity of $4 \times 10^{35} \rm{cm}^{-2}\rm{s}^{-1}$ would have a discovery potential comparable to that of FCC-hh. 
Table \ref{tab:mc} shows the design muon collider parameters,  taken from the MAP study. 
\begin{table}[h!]
\begin{center}
\caption{Key parameters of a muon collider as taken from the MAP study at 1.5, 3.0, and 6.0 TeV (14 TeV values taken from \cite{Neuffer:2018yof})}
    \begin{tabular}{l|c|c|c|c|c}
        CoM Energy & TeV & 1.5 & 3.0 & 6.0 & 14 \\
        \hline
        Avg. Luminosity & $10^{34}\,\rm{cm}^{-2}\rm{s}^{-1}$ & 1.25 & 4.4 & 12 & 33 \\
        Beam Energy Spread & percent & 0.1 & 0.1 & 0.1 & 0.1  \\
        Higgs Production / $10^7$ sec & $10^5$ & 0.375 & 2.0 & 8.2 & 30 \\
        Circumference & km & 2.5 & 4.5 & 6 & 26.7 \\
        Repetition Rate & Hz & 15 & 12 & 6 & 5 \\
        $\beta^{\star}_{x,y}$ & cm & 1 & 0.5 & 0.25 & 0.1 \\
        No. muons/bunch & $10^{12}$ & 2 & 2 & 2 & 2 \\
        Norm. Trans. Emittance, $\varepsilon_{TN}$ & $\mu$m & 25 & 25 & 25 & 25 \\
        Norm. Long. Emittance, $\varepsilon_{LN}$ & mm & 70 & 70 & 70 & 70 \\
        Bunch Length, $\sigma_s$ & cm & 1 & 0.5 & 0.2 & 0.1 \\
        Proton Beam Power & MW & 4 & 4 & 1.6 & 1.3 \\
        Wall Plug Power & MW & 216 & 230 & 270 & 250 \\

    \end{tabular}

\label{tab:mc}
\end{center}
\end{table}

\subsection{Proton Accelerator Requirements}
\begin{figure}[h!]
    \centering
    \includegraphics[width=4.5in]{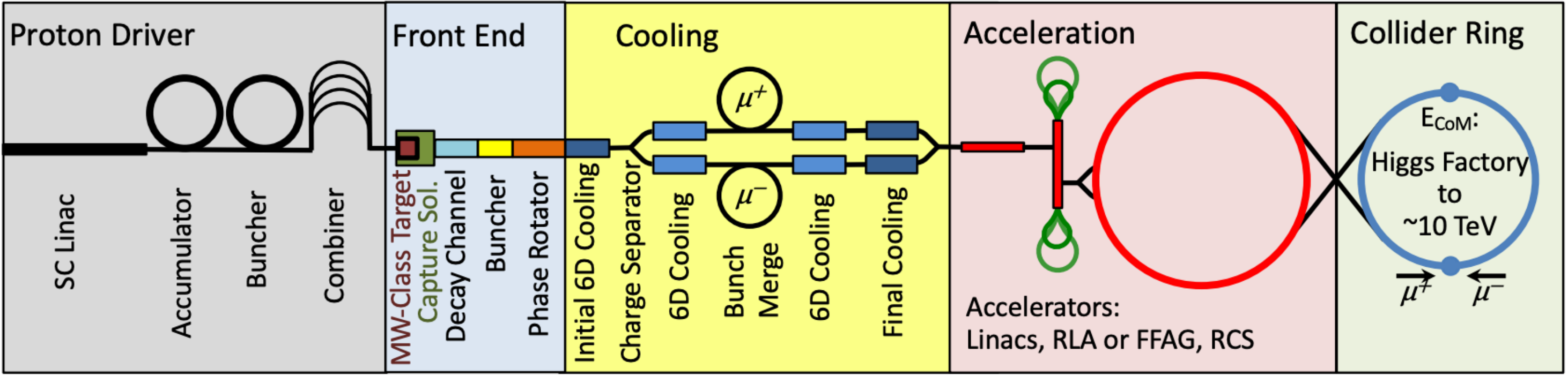}
    \caption{Layout of a muon collider accelerator complex.}
    \label{fig:mcdriver}
\end{figure}

The muon collider requires a high-intensity proton source to produce muons.
Figure \ref{fig:mcdriver} shows the layout of a muon collider accelerator complex. 
Although a superconducting linac with storage rings  is shown in the figure, a rapid-cycling synchrotron (RCS) could also be the primary proton accelerator. 
The accumulator, buncher and compressor are used in the linac scenario to process the proton beam into a small number of short intense bunches. 
A similar, but probably simpler, system would be needed to form the RCS beam into bunches matched to the Front End target. (A full-energy storage ring with bunching rf would probably be sufficient.)  
Muons from pions produced in the target are captured, bunched, and cooled, within the finite lifetime of the muons. 

MAP also considered a neutrino factory exploiting the muon collider front-end channel. 
The required proton beam parameters are slightly different from those of the muon collider scheme: the bunch spacing of a neutrino factory will be between 20 and 70 msec. 

\newpage
\paragraph{Accelerated particles:}
Protons

\paragraph{Beam Energy:}
$5 \sim 30$ GeV

\paragraph{Beam intensity:}
$10^{12} \sim 10^{13}$ protons per bunch

\paragraph{Beam time structure:}
Repetition rate $10 \sim 50$ Hz and  bunch length  $1 \sim 3$~ns

\paragraph{Target requirements:}
A thick target, either a liquid or a solid. 

\paragraph{Other requirements:}
No other specific proton beam parameter is required.

\paragraph{Timescales, R\&D needs, and similar facilities:}
The muon collider scheme still requires a significant amount of R\&D to develop the many elements of a complete collider system. 
After the MAP design study, DOE decided to pause  R\&D activities for the muon collider. 
Now, CERN is planning to host an international muon collider collaboration and continue working on design studies. 
They will study a baseline design and optimize the beam elements, build a test facility to demonstrate the technologies, develop the cost estimates, and publish a CDR in the next decade.

%% file: REDTOP_v3.tex
\section{Rare Decays of Light Mesons to Probe New Physics
}

\paragraph{Authors and proponents:} 
Anna Mazzacane\footnote{mazzacan@fnal.gov},
on behalf of the REDTOP Collaboration\footnote{REDTOP Collaboration \url{https://redtop.fnal.gov/collaboration}.}

\subsection{Physics Goals, Motivation, and Setup}

REDTOP (Rare Eta Decays To Observe Physics
Beyond the Standard Model) is an $\eta/\eta'$ factory which aims at detecting small deviations from the Standard Model by collecting a large event set  from protons impinging on fixed targets. The proposed experiment will produce about  $10^{13}~ \eta$ mesons or $10^{11}~ \eta'$ mesons corresponding to an increase of the existing world sample by four order of magnitude. Decays of the neutral and long-lived $\eta$ and $\eta'$ mesons could shed light on New Physics from a theoretical  and experimental standpoint. All their electromagnetic and strong decays are suppressed at first order and weak decays have branching ratios of order $\leq10^{-11}$. Therefore, they provide an excellent laboratory for precision measurements and a unique window to search for Physics Beyond the Standard Model (BSM) in the MeV-GeV mass  range. The $\eta$ meson has a simple symmetric quark structure which is conducive for fast triggers and online analysis. The detector requires fast and precise timing and should be sensitive only to those particles being produced in the processes of interest. The REDTOP experiment will exploit novel detector technologies aimed at a highly granular, nearly hermetic apparatus, with fast timing $(\sim 30~ps)$ and excellent particle identification. This will offer the opportunity for a broad physics program, exploiting different beam energies and beam configurations along with the world’s  leading  facility to search for physics BSM in flavor-conserving  processes.

\begin{figure}[!]
    \centering
    \includegraphics[width=\textwidth]{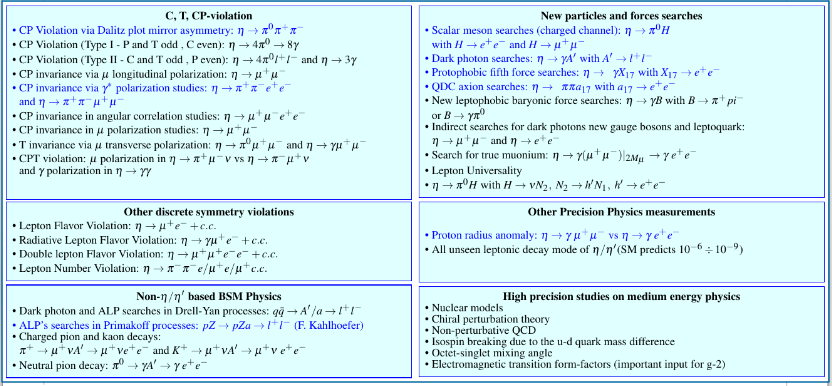}
    \caption{List of the physics processes that REDTOP intends to study.}
    \label{fig:ListProcesses_v3}
\end{figure}

The peculiarity of the $\eta$ and $\eta'$ mesons is that all their quantum numbers are zero. This is a very rare occurrence in nature and strongly constraints the dynamics of those particles. Therefore their decays offer many opportunities for exploring Physics Beyond the Standard Model (BSM). REDTOP will investigate, with large statistics,  violations of discrete symmetries\cite{Gardner_2020, GardnerShi:2020} and  will search for  new weakly-coupled light particles in the MeV-GeV mass scale. They also provide an opportunity to investigate Standard Model predictions with high precision\cite{Gan:2020aco}.
The most important physics processes that REDTOP experiment\cite{Beacham:2019nyx, gatto2019redtop} intends to study are summarized in Figure \ref{fig:ListProcesses_v3}, with the golden modes highlighted in blue. 

The Collaboration has identified three staged physics runs corresponding to different production mechanisms of $\eta$ and $\eta'$ mesons:
\begin{description}[style=nextline] 
\item[Run I - Untagged $\eta/\eta' $ factory]
In this stage, $\eta$ and $\eta'$  mesons are produced from a proton beam scattered on a target made of multiple thin Lithium foils. The production mechanism is based on the formation and decay of intra-nuclear baryonic resonances. Monte Carlo studies have indicated that beam energies of $\sim1.8~ GeV$ ($\sim3.0~GeV$ for the $\eta'$ meson) are optimal. At such energies  the production cross-section is relatively large (of order of several mbarn) and a low-power beam $(30\div 50~W)$ is sufficient to generate the desired statistics.
To cope with large background from hadronic inelastic scattering, a Continuous Wave beam is required.
\item[ Run II - Tagged  $\eta$ factory ]
In this stage, $\eta$ mesons are produced on a gaseous Deuterium target via the nuclear process $p + D \rightarrow \eta + \ce{^3He^++}$. A minimum of $880~MeV$ beam energy is required.  The smaller  $\eta/\eta'$ production cross section can be compensated with a larger beam intensity $(\sim1MW)$.
By tagging the production of the $\eta$ via the detection of the $\ce{^3He^++}$ ion, the combinatorics background from non $\eta$ events is greatly diminished and the sensitivity of the experiment to New Physics is increased. The 4-momentum of the $\ce{^3He^++}$ ion can be measured with the addition of a forward detector. Therefore the kinematics of the reaction is fully closed. Any long-lived, dark particle escaping detection could be identified by measuring the missing 4-momentum. This technique is considerably more powerful than  missing $p_T$ or missing energy and it mirrors an analogous technique adopted by B-factories but with the advantage of $4\times10^4$ larger statistics.
\item[Run III - Tagged  $\eta'$ factory ]
In this stage, $\eta'$ mesons are produced with higher beam energy ($1.7~GeV$) and higher intensity $(> 1MW)$,  using the same target and detector as in Run II.
\end{description}

The expected inelastic scattering rate is of the order of $1~GHz$ (i.e. about 30 times  the rate at LHCb experiment). Detailed MC studies indicate a hadronic background\cite{GENIEHad:2020} with a very low multiplicity ($\le 8$ primary particles).  Most of the physics processes listed in Figure \ref{fig:ListProcesses_v3}  have leptons and $\gamma$s in the final state. Furthermore the detection of displaced vertices (not from $\gamma$ conversion) is  an indication of New Physics. Such an environment poses strict requirements on the detector and it portends to an intense $R\&D$ program:
\begin{itemize}
\item \textbf{Sub-nanosecond timing detector  with $\bm{\sim 30~ps}$ time resolution -} This can be achieved exploiting the prompt nature of the \v{C}erenkov signal combined with the last generation of Si detectors (for example LGAD and 3D digital SiPM).
\item \textbf{Excellent Particle IDentification (PID) -} Several techniques will be exploited to disentangle final state leptons from the slow baryonic background: threshold  \v{C}erenkov, high granularity dual-readout and PFA calorimetry (5D calorimetry), and Time of Flight (ToF) with ${\sim 30~ps}$ time resolution.
\item $\bm{\le100\ \mu m}$ \textbf{resolution vertexing -} The latter will help rejecting the combinatorics from  the $\gamma$ conversion and with secondary vertices reconstruction of long-lived new particles.
\item \textbf{Good energy resolution - } for bump-hunting of decaying particles over a continuous background.
\item  \textbf{Forward detector - } for  tagging and reconstruction of the $\ce{^3He^++}$ in Run II and Run III. 
\end{itemize}
The proposed experimental apparatus includes the following detectors (Figure \ref{fig:Detector}):
\begin{figure}[h]
    \centering
    \includegraphics[width=\textwidth]{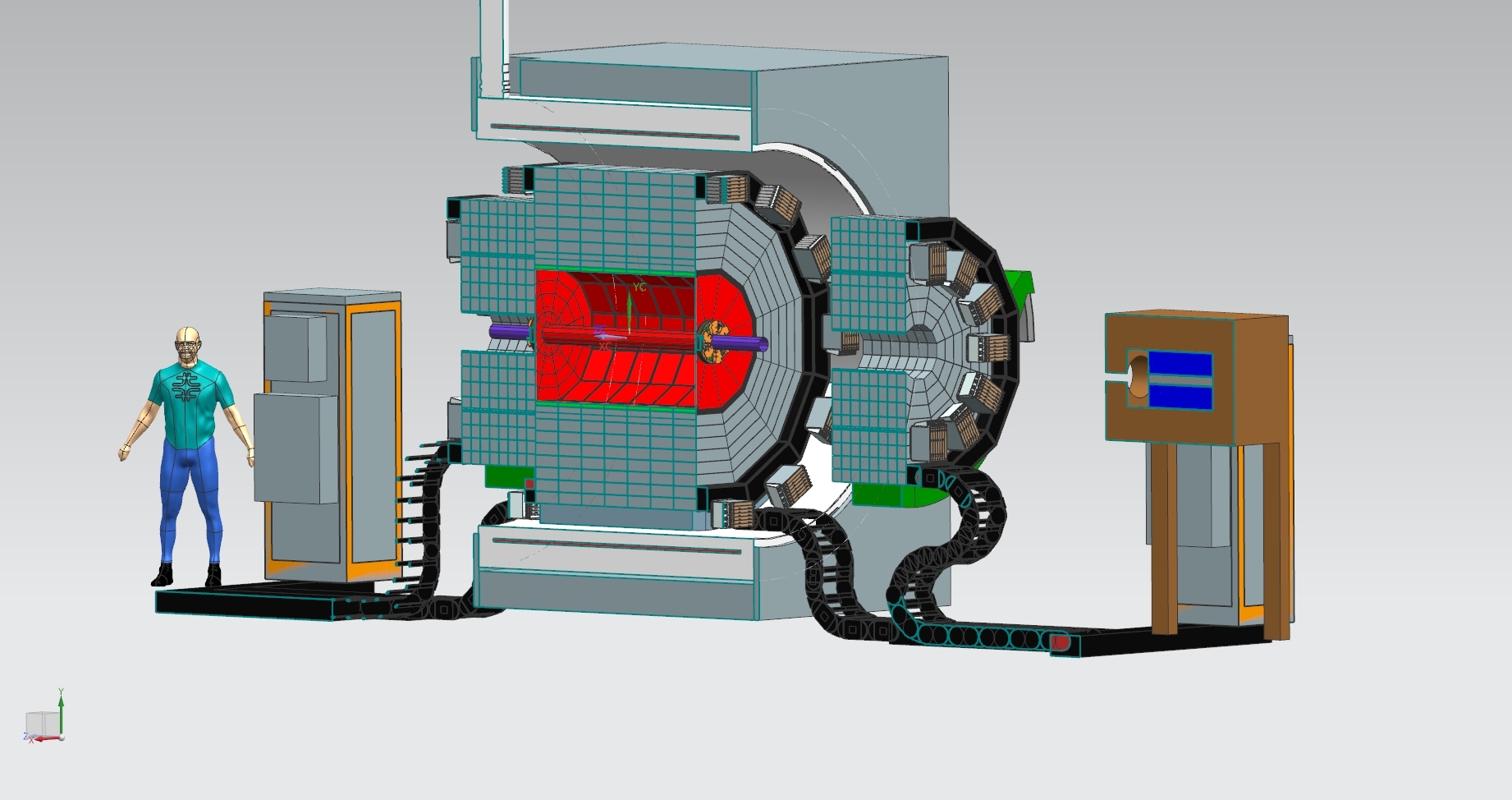}
    \caption{Cross section of the REDTOP detector.}
    \label{fig:Detector}
\end{figure}
\begin{itemize}
\item \textbf{Vertex Fiber Tracker -} 
Its main purpose is to identify displaced vertexes of long lived particles and to reject $e^+e^-$ pairs from $\gamma$ conversion. It is based on a mature technology developed by LHCb experiment, achieving  a spatial resolution $\le~80~\mu m$. MARS15\cite{MARS15:2016, MARS15:2019} studies indicate that irradiation doses will be similar to those at LHCb.
\item  \textbf{Central Tracker -} 
It relies on the threshold \v{C}erenkov effect to separate slower particles (nuclear interactions) from fast leptons associated with New Physics. Fast timing is needed to cope with the large event rate and for input to a Level-1 trigger. Two options are being considered: an Optical-TPC (OTPC)\cite{Oberla_2016} and an LGAD Tracker with $\sim30~ps$ timing\cite{SHI2020164382, CMOS}  surrounded by Quartz bars. An OTPC is blind to most hadronic background while sensitive to faster $\eta$ decay products. An LGAD Tracker has the granularity to reject multi-hadron events via a 4D track reconstruction. The Quartz bars provide fast input to the Level-0 trigger and complement the LGAD detector for  Time (ToF) measurements.
\item  \textbf{5D Calorimeter -}
The proposed integrally active ADRIANO2 calorimeter~\cite{T1015:2011, T1015:2015, Gatto:2016jtz} combines the dual-readout and the PFA techniques. It has an excellent energy and position resolution and PID capabilities (5D calorimeter). The scintillation and the \v{C}erenkov lights are read out by on-tile SiPMs\cite{BLAZEY2009277} or  SPAD\cite{LEMAIRE2020163538, NOLET201829, NOLET2020162891, 9078416, SPAD2018, PET2020}. The SPAD timing resolution ($15~ps$), based on Single-Photon Avalanche Diode arrays, can be exploited for ToF measurements and as input to the Level-0 trigger.
\item \textbf{Muon Polarimeter -}
It  measures the polarization of muons from $\eta/\eta'$ decays. A non-zero measurement would be an indication of BSM Physics\cite{Gan:2020aco}. Several technologies are being considered. The baseline design is composed by a sandwich of fused silica and Si-pixel extending  the technique developed by CALICE\cite{collaboration_2008}.
\item \textbf{Forward Detector for Run II/III -}
The tagging $\ce{^3He^++}$ ions with momentum above $1.24~GeV$ are mostly emitted in a $[3^\circ\div 5^\circ]$ angular range. The very forward region will be instrumented with a fast LGAD pixel tracker and a sandwich of active fused silica and Si-pixel, combining  technologies  developed for ADRIANO2\cite{T1015:2011,Gatto:2016jtz} and CALICE Si/W electromagnetic\cite{collaboration_2008, Kawagoe_2020} calorimeters.
\end{itemize}

\subsection{Accelerator Requirements}
The availability of a slowly extracted proton beam in the $1.8-2.2~GeV$ energy range at the Delivery Ring (DR) would allow to obtain the required $\eta$ yield in Run-I with a relatively low beam intensity ($\approx 30~W$) based on an intra-nuclear baryonic resonance production mechanism. Similarly,  an energy of $3~GeV$ allows REDTOP to run as an $\eta'$ factory using the same accelerator complex.
With the PIP-II facility, currently being under construction at Fermilab, the $800-920~MeV$, high intensity (100KW-1MW), CW proton beam, opens up  a different  production mechanism of the $\eta$-meson and the possibility to tag the latter  trough the nuclear reaction: $p + De \rightarrow 3He+\eta$. ~Besides the tagging, the kinematic of the process is completely closed, opening the opportunity to explore, in Run-II, long-lived particles  escaping detection. 
A similar production mechanism would be available for the $\eta' $ only above $\approx 1.4~GeV$. Therefore, an improved booster is required for Run-III.

\paragraph{Accelerated particles:}
Protons.

\paragraph{Beam Energy:}
For the $\eta$-factory, simulation indicates that $1.8~GeV$ on the REDTOP target will produce $2\times10^{13}~ \eta$/year while for the $\eta'$-factory, the optimal beam energy is $3~GeV$. Both energy and power requirements can be easily met with the Delivery Ring accelerator complex.
For the Run-II, an upgraded version of the REDTOP, t-REDTOP, is required and a gaseous deuterium target replacing the solid Lithium foils. 

\paragraph{Beam intensity:}
Run-I: $30-50 W$ is sufficient to reach the required $\eta/\eta'$ yield. \\
Run-II and Run-III: $>200 KW$.  

\paragraph{Beam time structure:}
CW mode (slow-extraction for Run-I).

\paragraph{Target requirements:}
Ten round foils of low Z (beryllium or lithium), each about 1/3 mm (3/4 mm for Li)  thick and about 1 cm of diameter inside a beam pipe made of either carbon-fiber or beryllium.
Gaseous Deuterium for Run-II and Run-III.

\paragraph{Timescales, R\&D needs, and similar facilities:}
Novel detector techniques need to be developed to cope with the high interaction rate. REDTOP detector will offer the opportunity for a broad physics program, exploiting different beam energies and beam configurations along with the world’s  leading  facility to search for Physics BSM in flavor-conserving  processes.\\
Future High Energy and High Intensity experiments will benefit from the ensuing R\&D.
Studies of the experiment sensitivity to a broad range of Physics BSM have been carried as part of the Snowmass 2021 process. Details and references can be found in the Snowmass 2020 LOI\cite{redtop_loi:2020} and the Snowmass 2021 White Paper\cite{redtop_wp:2022}.

\begin{figure}[!ht]
    \centering
    \includegraphics[width=7cm]{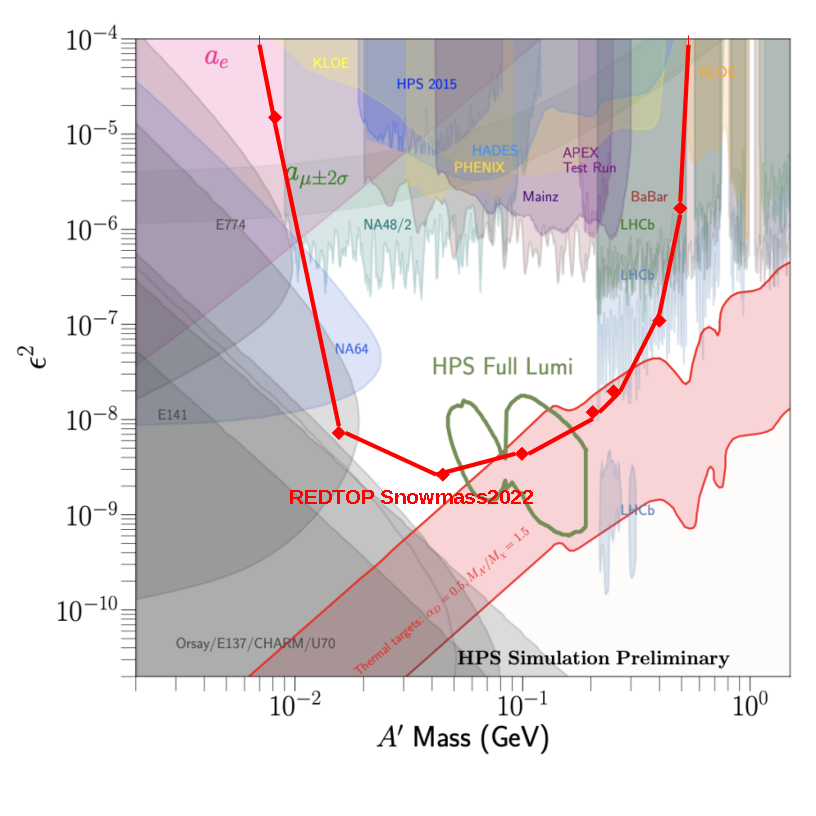}
    \includegraphics[width=7cm]{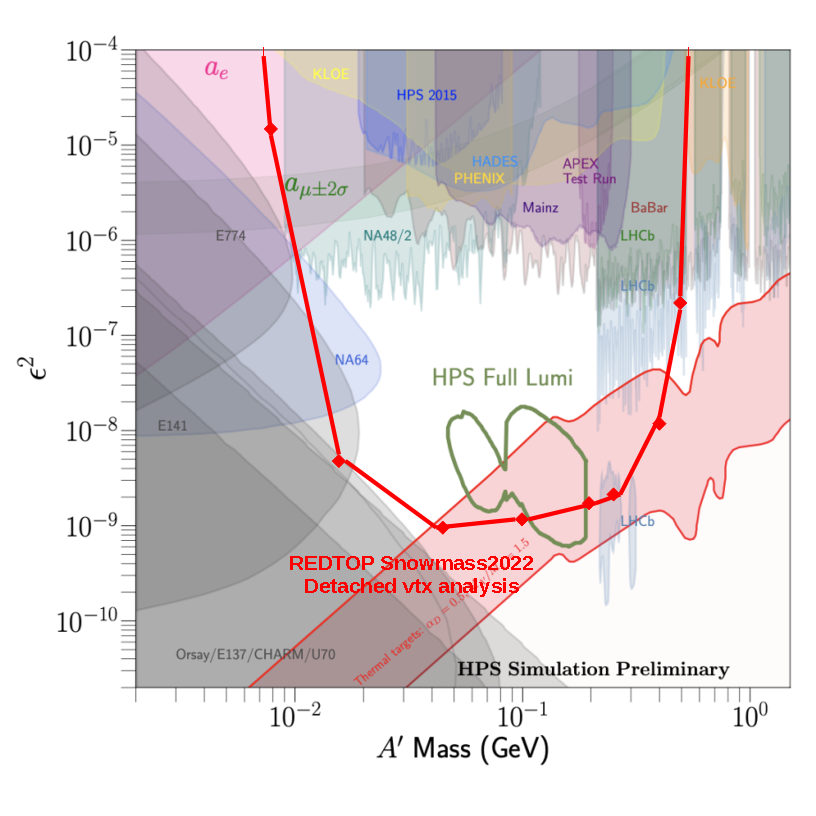}
    \caption{REDTOP sensitivity for dark vectors with Snowmass 2021 detector. Left plot: bump-hunt analysis. Right plot: detached vertex
    analysis.}
        \label{fig:PBC}
\end{figure}

\subsection{Global Context}
The proposed REDTOP experiment at Fermilab, with a yield of the order $10^{14}$($10^{12}$) $\eta(\eta')$
mesons, can breach into decays that violate conservation laws or for searching for new
particles not accessible to other experiments. 
Such an experiment would have enough sensitivity to explore all four portals connecting
the Dark Sector with the Standard Model as well as to probe conservation
laws.
The proposed muon polarimeter (and an optional photon polarimeter) for the REDTOP apparatus will offer additional 
capability to measure the longitudinal polarization of final-state muons (and possibly photons), 
which are not available in most other experiments.

Although the background in REDTOP is expected about several orders of magnitude higher than in the JEF experiment,
this will be compensated for by an enormous $\eta$ yield. 
The recoil detection technique considered for phase II and phase III in the tagged mode will further reduce 
the backgrounds. 
Preliminary sensitivity studies indicated that the background rejection in a tagged $\eta$-factory is at least one 
order of magnitude better than in the untagged production mode.  
Furthermore, the missing 4-momentum technique would make the experiment sensitive to long-lived dark particles, 
similar to B-factories, but with x40,000 the yield.

No similar experiment exists or is currently planned by the international HEP Community.

%% file: NNbar.tex
\section{Cold Neutron source for $n\leftrightarrow \bar n$ Oscillations and  Fundamental Physics }


\paragraph{Authors and proponents:} 
Kent K. Leung, Duke University\footnote{
\href{mailto:kent.leung@duke.edu}{kent.leung@duke.edu}},
Albert R. Young, North Carolina State University\footnote{ \href{mailto:aryoung@ncsu.edu}{aryoung@ncsu.edu}},
Joshua L. Barrow, The University of Tennessee\footnote{ \href{mailto:jbarrow3@vols.utk.edu}{jbarrow3@vols.utk.edu}},
G\"unter Muhrer, European Spallation Source ERIC\footnote{ \href{mailto:gunter.muhrer@ess.eu}{gunter.muhrer@ess.eu}}

\subsection{Physics Goals, Motivation, and Setup}


Both the neutron static electric dipole moment (EDM) and neutron-antineutron oscillations ($n\rightarrow\bar{n}$) probe scenarios tied closely to the Sakharov conditions required to achieve proper baryogenesis \cite{Sakharov:1967dj} and provide some of our most sensitive low energy constraints on BSM physics \cite{Cirigliano:2013,Babu:2020}. Searches for unambiguous \cite{Abe:2011ky,Abi:2020evt,Abi:2020kei,Barrow:2019viz} sources of $CP$ and baryon number violation are thus of great interest to the fundamental high-energy and nuclear physics communities. With deep ties to the US and international particle physics community, locating a future fundamental-physics spallation neutron source at Fermilab could provide the world with complimentary~\citep{Phillips:2014fgb,Addazi:2020nlz} or world-leading sensitivities to $n\rightarrow\bar{n}$~\citep{Kuzmin:1970nx,Kuzmin:1985mm,Kuzmin:1987wn,Mohapatra:1980de,Mohapatra:1980qe} using cold or ultracold neutrons.   

Experimental limits for neutron EDMs and $n\rightarrow\bar{n}$ are typically limited by available densities/fluxes of neutrons. In this subsection, we identify a conceptual approach to a high-intensity ultracold neutron source, suitable for world leading EDM and $n\rightarrow\bar{n}$ experiments. The program of research at such a spallation source could prove remarkably cross-disciplinary, including but not limited to studies of the neutron lifetime, fundamental symmetries and decay correlations, short-ranged interactions, $n\rightarrow n'$~\citep{Berezhiani:2017azg} studies, as well as EDM and $n\rightarrow\bar{n}$. Such a prospective source would greatly alleviate beam-time competition and over-subscription pressures felt by HEP-oriented collaborations at material sciences-focused general-purpose neutron scattering facilities. An $n\rightarrow\bar{n}$ experiment itself would require the involvement of multi-disciplinary experts in neutronics, detectors, and magnetics, creating highly dynamic collaboration across these fields.

\subsection{The source concept}

Ultracold neutrons (UCN) generally move at speeds of $\lesssim8\,$m/s, and can be stored in material or magnetic traps for $\sim100\,$s (up to times approaching the neutron lifetime of $880\,$s)~\cite{golub_UCN_book}. This long storage time makes UCN ideal for many fundamental physics applications, but requires neutrons produced at spallation targets to be moderated through several cooling stages to temperatures of $\sim25\,$K before they are introduced into a cryogenic converter material which can produce UCN from a single scattering. The challenge is to couple extremely low temperature converter materials (in the source proposed here, superfluid ${}^4$He) very closely to a high-power spallation target.  
\begin{figure}[ht]
 \centerline{ 
 \includegraphics[width=0.8\textwidth]{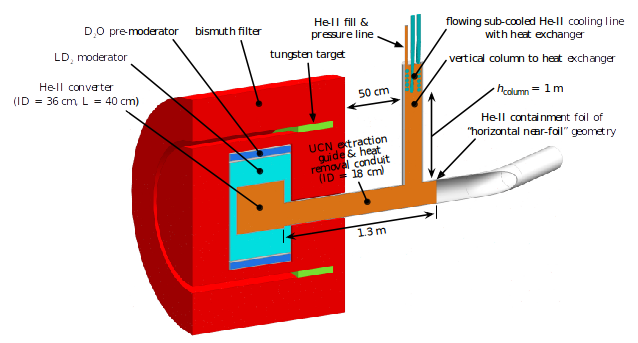}
 }
\caption{Diagram of a high-power, next-generation spallation source of ultracold neutrons (UCN). Up to $\sim1\,$MW of proton beam can be rastered over a cylindrical tungsten target embedded in a bismuth shielding and moderator assembly. UCN are produced within a $40\,$l superfluid ${}^4$He volume held at $1.6\,$K at one atmosphere of pressure, surrounded by an outer D${}_2$O and inner liquid D${}_2$O moderator. UCN are extracted through a guide system behind the source assembly.}
\label{fig:invsource}
\end{figure}

Leung et al.~\cite{Leung:2019} have proposed a solution compatible with $1\,$MW or more of CW proton beam based on utilizing superfluid $^4$He (or He-II) as the converter material and the cryogenic coolant in a sub-cooled He system similarly utilized at CERN~\cite{LeBrun:1994}. The source is depicted in Fig.~\ref{fig:invsource}. In order to simultaneously satisfy the problems of adequate target cooling and close coupling to the source, the target is designed for edge-coupled water cooling with target damage mitigated by rastering the proton beam. The UCN converter is located inside a nested set of shielding and moderating components composed of successively bismuth, a D$_2$O moderator, and a liquid D$_2$O moderator. UCN are produced in the sub-cooled He-II volume (nominally $40\,$L held at one atmosphere of pressure and $1.6\,$K). The impressive cooling power of sub-cooled cryogenic systems permits heat loads of up to $100\,$W on the low temperature components of the source, allowg for continuous proton beam powers in excess of $1\,$MW (at an assumed $800\,$MeV proton beam energy) incident on the spallation target in our optimized design. UCN are delivered to experiments through a $18\,$cm diameter UCN guide exiting the rear of the source assembly, with a high UCN-transmission foil providing the boundary of the He-II system. 

At $1\,$MW of proton beam power, inside the converter a UCN production rate of $2\times 10^9\,$s$^{-1}$ and density of $5\times 10^4\,$UCN/cm$^3$ can be reached. The expected delivery to an external experiment would be a UCN integrated flux of $5\times 10^8\,$UCN/s and a density of $1\times 10^4\,$UCN/cm$^3$. With these parameters, this source would have the highest integrated flux in the world, and ideally suited for an optimized $n\rightarrow\bar{n}$ experiment, for example.


\paragraph{Accelerated particles:}
Protons

\paragraph{Beam Energy:}
$0.8$-$2\,$GeV

\paragraph{Beam intensity:}
Up to $\sim 1\,$MW

\paragraph{Beam time structure:}
A quasi-continuous beam with minimum instantaneous power is optimal for this source. The current design also relies on rapid rastering the proton beam over the front surface of the spallation target. 

\paragraph{Target requirements:}
The target will be thick. Remote handling will be required for target and source service.

\paragraph{Other requirements:}
Water cooling for the spallation target and cryogenic systems are required for the cryogenic components of the source.  

\paragraph{Timescales, R\&D needs, and similar facilities:}

Source and UCN extraction technologies are essentially established, with the possible exception of the foil used to separate the He-II volume from UCN guides under vacuum. UCN guide strategies and more effective moderator geometries using, for example, nanodiamond reflectors, are useful R\&D targets for this project. The proposed source's integrated flux would be superior to other UCN sources currently operating at the Institut Laue-Langevin, the Paul Scherrer Insitute, Los Alamos National Laboratory and the Mainz Triga reactor, and compared to sources under construction at the Petersburg Nuclear Physics Institute and TRIUMF. A conceptual analysis of a $n\rightarrow\bar{n}$ experiment at the PNPI source~\cite{Fomin:2017} found possible improvements of $10$-$40$ times current free neutron experimental limits, mostly depending on the neutron wall reflection model for their experimental geometry and expected source performance.

%% file: proton-storage-ring.tex
\section{Proton Storage Ring: EDM and Axion Searches}

\paragraph{Authors and proponents:} 
William M. Morse and Yannis K. Semertzidis

\subsection{Physics Goals, Motivation, and Setup}

The value of the srEDM experiment is that it can provide substantial insight into the strong CP-problem by improving our sensitivity to qQCD, the P and T-violating parameter in the QCD Lagrangian, by more than three orders of magnitude; can establish the energy scale of the next international collider by probing New Physics at high-mass scales of the order $10^3$ TeV [1-3]; and at $10^{-29}\, e {\rm cm}$ can probe CP-violation with the greatest existing sensitivity, in what could turn out to be the field responsible not only for the generation of lepton masses, but also the matter- antimatter asymmetry of our universe, i.e., the Higgs sector. Like the EDMs of the electron and neutron, it can be the only practical possibility of accessing the very small coupling to first-generation fermions, assuming they do violate CP-symmetry in the $H_{\gamma\gamma}$ coupling interaction. Finally, recent theoretical work on oscillating hadronic EDMs points to a new method of looking for axion dark matter and dark energy, one more-sensitive than the neutron EDM experiments by several orders of magnitude.
Details and references can be found in the related Snowmass LOI~\cite{Yannis}, the paper on the comprehensive treatment of systematic errors and references therein~\cite{Omarov:2020kws}.

\subsection{Accelerator Requirements}

\paragraph{Accelerated particles:}
Protons

\paragraph{Beam momentum:}
$p = 0.7$ GeV/c\,.

\paragraph{Beam intensity:}
$10^{11}$ polarized protons per fill.

\paragraph{Beam time structure:}
Fill the ring every $10^3$ s.

\paragraph{Target requirements:}
N/A

\paragraph{Other requirements:}
Polarization: greater than 80\%, horizontal and vertical rms emittance of about 0.2 mm-mrad, rms momentum spread about $10^{-4}$ and EDM ring with electric bending.

%% file: tau-neutrinos.tex
\newcommand{\nue}{\ensuremath{\nu_{e}}}
\newcommand{\num}{\ensuremath{\nu_{\mu}}}
\newcommand{\nut}{\ensuremath{\nu_{\tau}}}
\newcommand{\nus}{\ensuremath{\nu_{s}}}
\newcommand{\anue}{\ensuremath{\bar{\nu}_{e}}}
\newcommand{\anum}{\ensuremath{\bar{\nu}_{\mu}}}
\newcommand{\anut}{\ensuremath{\bar{\nu}_{\tau}}}
\newcommand{\anus}{\ensuremath{\bar{\nu}_{s}}}

\section{Tau Neutrinos} 

\paragraph{Authors and proponents:} 
Adam Aurisano (University of Cincinnati), Joshua Barrow (University of Tennessee at Knoxville, Fermilab), Andr{\'e} de Gouv{\^e}a (Northwestern University), Laura Fields (Fermilab), 
Elena Gramellini (Fermilab),
Jeremy Hewes (University of Cincinnati),
Thomas Junk (Fermilab),
\textbf{Kevin J. Kelly} (Fermilab \& CERN) [kjkelly@cern.ch],
Pedro Machado (Fermilab),
Ivan Martinez-Soler (Fermilab/Northwestern University),
Irina Mocioiu (Pennsylvania State University),
Yuber F. Perez-Gonzalez (Fermilab/Northwestern University),
Gianluca~Petrillo (SLAC),
Holger Schulz (University of Cincinnati),
Alex Sousa (University of Cincinnati),
Yu-Dai Tsai (Fermilab),
Yun-Tse Tsai (SLAC),
Jessica Turner (Fermilab),
Tingjun Yang (Fermilab)

\subsection{Physics Goals, Motivation, and Setup}

Despite increasing understanding of neutrino oscillations from $\nu_e$ and $\nu_\mu$, there is less direct experimental knowledge of $\nu_\tau$ than any other Standard Model particle. The $\nu_\tau$ was not directly observed until 2000 by the DONuT experiment, which collected nine $\nu_\tau$ candidate events~\cite{DONUT2001, DONUT2007}. The OPERA experiment was designed with high resolution emulsion technology to discover $\nu_\tau$ appearance in a $\nu_\mu$ beam~\cite{OPERAProp}, succeeding in discovering $\nu_\tau$ appearance in 2015~\cite{OPERA2015}; however, the baseline and energy of the experiment was unfavorable, and thus, only ten \nut\ candidates were observed~\cite{OPERA2018}. The Super-Kamiokande experiment has recently developed a method to statistically separate a sample of $\nu_\tau$ events in atmospheric neutrinos to exclude the no-$\nu_\tau$ appearance hypothesis at a 4.6$\sigma$ significance level, and measured the normalization of the $\nu_\tau$ sample relative to expectations to be $1.47\pm 0.32$~\cite{SKNuTau2013, SKNuTau2017}. The IceCube experiment performed a similar analysis using data from the DeepCore detector component. Using CC events only, they were able to exclude the no-$\nu_\tau$ appearance hypothesis at the 2.0$\sigma$ level and measure the $\nu_\tau$ normalization to be $0.57^{+0.36}_{-0.30}$~\cite{DeepCoreNuTau}. In both cases, the limitations of Cherenkov detectors prevented the experiments from improving the purity of their samples beyond 5\%.

Our knowledge of the \nut\ cross-section is inferred from measurements of \num\ assuming lepton universality, such that any cross-section differences are only due to the large mass of the $\tau$ lepton. However, recent data from Belle, BaBar, and LHCb, as combined by the Heavy Flavor Averaging Group, shows that $\mathcal{B}(B \rightarrow D \tau \nut)/\mathcal{B}(B \rightarrow D l \nu_{l}\
)$ and $\mathcal{B}(B \rightarrow D^* \tau \nut)/\mathcal{B}(B \rightarrow D^* l \nu_{l})$ differ from Standard Model predictions by 3.9$\sigma$~\cite{HFAG2017}, assuming lepton universality. Similarly, almost all knowledge of \nut\ mixing-matrix elements comes from assuming unitarity of the mixing matrix. Without assuming unitarity, $|U_{\tau 3}|$ is only known to only 60\%~\cite{ParkeNonUnitarity,Ellis:2020hus}.

Over the next two decades, several currently available and developing sources will allow for direct measurement of $\nu_\tau$. Soon, it is expected that IceCube will be able to use DeepCore data to constrain the normalization of the $\nu_\tau$ sample at the $10\%$ level~\cite{summer_blot_2020_3959546}. Future data from the IceCube upgrade will allow this measurement to be effectively systematically limited. 
The upcoming DsTau/NA65 experiment \cite{Aoki:2019jry} (based at CERN) will directly study tau neutrino production using  a measurement of $D_s\to \tau \, X$ decays following high-energy proton-nucleus interactions.  DsTau aims to  provide an independent $\nu_\tau$ flux prediction for future neutrino beams with
accuracy under 10$\%$ which will reduce the systematic uncertainty of the $\nu_\tau$ CC cross section measurement. FASER$\nu$ will also have capability in measuring this high-energy cross section~\cite{Abreu:2020ddv}. In the Deep Underground Neutrino Experiment (DUNE), significant oscillation of the $\nu_\mu$ beam into $\nu_\tau$ can allow for a precise measurement of the appearance oscillation probability at long baselines and $\mathcal{O}(5\ \mathrm{GeV})$ energy~\cite{NuTauDUNE,Ghoshal:2019pab}. Additionally, DUNE's atmospheric neutrino sample will contain a large number of $\nu_\tau$ events; albeit the identification of the $\tau$ track is unattainable in DUNE, $\nu_\tau$ events can be identified via statistically inference by analyzing the event kinematics
~\cite{ConradAtmNuTau,DUNEtdr2}. Importantly, performing such studies in an environment like DUNE requires new techniques to reduce the other neutrino-related backgrounds from the intense beam~\cite{Machado:2020yxl}.

If a sufficiently high statistics sample can be generated with adequate background rejection and a deep understanding of \nut\-$A$ final state topologies, either directly or through oscillations, detailed studies of the differential \nut-CC cross section could be possible. This could potentially answer questions which cannot be answered using \nue-CC and \num-CC interactions. For example, most formulations of the quasielastic pseudoscalar form factor are calculated in the $Q^2 =0$ limit. Due to the high kinematic threshold for \nut-CC events, most events at threshold will be quasielastic with a large $Q^2$. Similarly, the form factors $F_4$ and $F_5$ are suppressed when the mass of the charged lepton is small compared to the neutrino energy, so they are negligible except in \nut-CC interactions~\cite{SHIPProposal}.

With these upcoming measurements, one will have the ability to better understand oscillations involving $\nu_\tau$ in the standard three-massive-neutrinos paradigm and beyond. If only three neutrinos exist, then the oscillations involving $\nu_\tau$ can be determined perfectly by measuring only oscillations involving $\nu_e$ and $\nu_\mu$. Precision understanding of $\nu_\tau$ can serve as a (relatively weak) cross check of this determination~\cite{NuTauDUNE,Ghoshal:2019pab}. Additionally, in beyond-the-Standard-Model scenarios of neutrino oscillations, $\nu_\tau$ measurements can provide unique information beyond that inferred from $\nu_\mu$ and $\nu_e$ oscillations.

To be a valid description of a physical process, neutrino mixing must be unitary. However, many new physics models predict heavy fermionic gauge singlets, and these states can mix with the familiar neutrino flavors. The mixing of these states is described by an expanded mixing matrix of size $n\times n$ which must be unitary, but the $3\times 3$ sub-matrix describing the mixing of the known states would not be. If these extra states have masses near the GUT scale, they can explain the lightness of the known neutrinos via the seesaw mechanism
~\cite{seesaw,Mohapatra:1979ia}. However, since no known symmetry protects the masses of the extra states, there is no theoretical reason to prefer any mass scale. For very low mass scales, the extra states could be the sterile neutrinos potentially observed at LSND and MiniBooNE. At higher masses, above the kaon mass, the extra states are kinematically inaccessible at neutrino oscillation experiments and can be integrated out.

The effect on oscillations between the known flavors when the heavy states are kinematically inaccessible can be described through a non-unitary modification of the mixing matrix~\cite{PhysRevD.92.053009, NonUnitaritySteriles}.
The effect of apparent non-unitary mixing produces zero-baseline flavor change (effectively a normalization shift at short baselines), and a modification to the matter potential for propagating neutrinos~\cite{NuTauDUNE}. For GUT-scale sterile states, non-unitarity is highly constrained by rare decays like $\mu \rightarrow e\gamma$, but at lower energy scales, these constraints are no longer valid~\cite{ParkeNonUnitarity}. Therefore, searching for non-unitary neutrino mixing provides a way to extend the search for sterile neutrinos to mass scales typically believed to be inaccessible for neutrino oscillation experiments, and one in which \nut\ appearance can make great strides toward.

Beyond the non-unitarity hypothesis, measuring $\nu_\mu \to \nu_\tau$ oscillations in a beam- or atmospheric neutrino context can allow for improved limits (or discovery potential) in searches for sterile neutrinos~\cite{NuTauDUNE,Ghoshal:2019pab}, non-standard neutrino interactions~\cite{NuTauDUNE,Ghoshal:2019pab}, and neutrino decays~\cite{Ghoshal:2020hyo}. Beyond these, one could learn even more with a clean, well-understood, $\nu_{\tau}$-enriched source: measurements of $\nu_\tau$ disappearance, like those performed for $\nu_\mu$ disappearance currently, would provide exciting complementary information to the broader neutrino program. While no such source is currently planned, it it nevertheless useful for the community to consider what can be learned from such experiments.

While there exist obvious challenges in the regime of measuring and identifying $\nu_\tau$ interactions in a neutrino oscillation experiment, the benefits are plenty. Given that new physics is required to explain non-zero neutrino masses, the community should exploit upcoming and future experiments in as many ways as possible to learn about neutrinos, especially those of the \nut\ variety. It is quite possible that $\nu_\tau$ are a unique entry point to uncovering new physics which may be difficult to elucidate in any other way.

The potential for creating a large sample of oscillated $\nu_\tau$ is a unique feature of LBNF/DUNE among accelerator-based neutrino experiments.  Although most of the $\nu_\mu$ neutrinos from LBNF that oscillate will oscillate to $\nu_\tau$, most of these are unobservable because they are created below the 2.5 GeV threshold for charged-current $\nu_\tau$ interactions, or in the 2.5-5 GeV region where the charged-current cross section is very small.  With the CP-optimized LBNF beam, there will be of order 100 $\nu_\tau$ charged current interactions per year\footnote{Assuming 1.2 MW proton beam and a 40 kTon detector}.  

Increasing the LBNF neutrino flux above 5 GeV can improve $\nu_\tau$ appearance rates.  Preliminary studies by the DUNE collaboration indicate that NuMI-style horns can be used to increase the rate to approximately 1000 $\nu_tau$ charged-current interactions per year.  In addition to improving prospects for tau appearance, this high-energy beam extends the L/E range that DUNE can measure, opening a range of phase space that is completely inaccessible to Hyper-Kamiokande, including searches for sterile neutrinos in a detector with excellent $\nu_e$ appearance capabilities, and increased ability to disentangle CP violation and Non-Standard Interaction (NSI) signals.  

\subsection{Accelerator Requirements}

\paragraph{Accelerated particles:} protons

\paragraph{Beam Energy:} All studies so far have assumed 120 GeV protons; $\nu_\tau$ production at other energies is possible, and higher energies may be preferable, since neutrino energies about 5 GeV are optimal for $\nu_\tau$ appearance.  

\paragraph{Beam intensity:} as high as possible.

\paragraph{Beam time structure:} Either pulsed or CW is possible in principle; pulsed is necessary to preserve non-DUNE physics (e.g. supernovae searches).  

\paragraph{Target requirements:} LBNF-like target, requirements likely driven by beam intensity.  Short ( less than 1.5 m) targets are preferable for tau appearance.   

\paragraph{Other requirements:} Tau appearance rates at DUNE are higher with NuMI-style parabolic horns than the CP-optimized LBNF horns; so either NuMI-style or new horns optimized for tau appearance will be needed.  

\paragraph{Timescales, R\&D needs, and similar facilities:}
A tau-optimized LBNF beam would only be needed after the 'standard' DUNE program, so this is a long-term opportunity.  The ability to tune to higher energies is unique to LBNF, so there is no currently planned direct competition, although many experiments will continue to exploit atmospheric $nu\tau$ measurements, as described above.

%% file: proton-irradiation.tex
\section{Proton Irradiation Facility}\label{sec:irrad}

\paragraph{Authors and proponents:} 

Maral Alyari, Artur Apresyan, Doug Berry, Zoltan Gecse, Mandy Kiburg, Thomas Kobilarcik, Ron Lipton, Petra Merkel, Evan Niner, Eduard Pozdeyev (Fermilab)  \footnote{corresponding author: petra@fnal.gov}

\paragraph{Related sections:}
Section~\ref{sec:test}.

\subsection{Physics Goals, Motivation, and Setup}

The goal is to create a high-intensity proton irradiation facility at the Fermilab Booster Replacement to benefit future collider detector development. The current Fermilab Irradiation Test Area (ITA), which is under construction right now at the end of the LINAC, is designed for fluences needed for the HL-LHC detector upgrades. However, for future collider detectors doses up to two orders of magnitude higher are expected. It is paramount that detector elements under development can be tested for radiation hardness to these levels. Currently, there is no facility in the world, that would allow such tests at a reasonable timescale. It would be desirable to reach on the order of 10$^{18}$ protons within a few hours. The exact beam energy is less critical. The DOE program manager for detectors has expressed interest in Fermilab creating such a facility. The Fermilab Booster Replacement seems like an ideal candidate with its high intensity proton beam. The proposal is to build a tangential arm to set up an experimental hall where devices under test can be placed for irradiation under controlled conditions. 

\subsection{Accelerator Requirements}

It would be desirable to reach on the order of 10$^{18}$ protons within a few hours. The exact beam energy is less critical. The beamsize at the device under test (DUT) should be on the order of a few millimeters to two centimeters. A pulsed beam would be preferable for cooling and readout reasons.  

\paragraph{Accelerated particles:}
The main particle type that is needed would be protons, but being able to switch to electrons, or even ions at times, would be beneficial. 

\paragraph{Beam Energy:}
We would take the maximum achievable beam energy. The absolute value is not so relevant, as long as it is stable and known.  

\paragraph{Beam intensity:}
It would be desirable to reach on the order of 10$^{18}$ protons within a few hours.

\paragraph{Beam time structure:}
A pulsed beam would be preferable, or even necessary. A continuous beam would likely overheat the DUT and make cooling very challenging. Even with a pulsed beam, cooling of the DUT will be one of the main challenges. A similar time structure to the current LINAC could be suitable.

\paragraph{Target requirements:}
No target is needed. We would need a stripping station to strip off the electrons from the beam, but the target would be the user provided DUTs. 

\paragraph{Other requirements:}
It would be good to have an adjacent, shielded counting room, where users could sit and operate and monitor their DUTs. This would also require some user infrastructure, such as cables, power supplies, cooling, to be placed either in the experimental hall or the adjacent counting room. Cables would need to run between the two. There should be a cold, dark box for the DUTs available that can be moved in and out of the beam, including beam scans. Beam monitoring data need to be made available to the users as well. The irradiated DUTs would be handled by rad techs. and stored cold until they can be retrieved by the users.

\paragraph{Timescales, R\&D needs, and similar facilities:}
Likely, cooling of the DUTs is the main challenge and some R\&D needs to go into finding a solution for this. Similarly, having robotic control of handling the box that contains the DUTs and that needs to move in and out of the beam would be highly desirable. This could be either addressed by a commercial solution or by in-house development. The timescale by which such a facility is needed would realistically be somewhere in the second half of the 2020s. Before then, detector components aimed to sustain these kinds of radiation levels will likely not be ready for testing.

%% file: test-beam.tex
    \section{Test-beam Facility}\label{sec:test}

\paragraph{Authors and proponents:} 
{
 Mandy Kiburg, Thomas Kobilarcik, Petra Merkel, Evan Niner, (all Fermilab)  \footnote{corresponding author: edniner@fnal.gov}} 

\paragraph{Related sections:}
Section~\ref{sec:irrad}.

\subsection{Physics Goals, Motivation, and Setup}

The goal is to create a multi-purpose facility at the Fermilab Booster Replacement to benefit future detector development. The current Fermilab Test Beam Facility (FTBF) is built at a terminus of the switchyard beam line and has two lines that range from 120 GeV primary protons down to a few hundred MeV mixed tertiary beam.  The facility services several hundred users and about 20 dedicated experiments each year from around the world.  These experiments cover the breadth of collider, muon, and neutrino physics and general detector prototyping.  The switchyard beam lines are very old and take a long, $\approx$ mile, path to the facility which is not needed.  The Test Beam program would also benefit from a purpose-built experimental area to accommodate a broad experimental program.  There are increased requests in the user community for clean low-energy beams of a variety of particles which is not easily achieved in the existing facility.  The proposal is to build a tangential arm to setup an experimental hall with at least two independent beam lines and infrastructure to support short and long term experimental setups.  This would bring the FTBF operations closer to the main campus, reduce maintenance on the beam line length, and establish a set of beam lines suited to the future needs of the R\&D community.  This facility would be ideally located in the same area or building as the proposed proton irradiation facility. 

\subsection{Accelerator Requirements}

At least two beam lines should be available in the Test Beam area.  These lines should be capable of operating simultaneously and at independent energies and intensities.  It is important at any energy that the beam focus to a manageable size, typically several centimeters or less, and have understood backgrounds and particle composition. It is desirable to have at least one beamline dedicated to low energy muons if possible.  

\paragraph{Accelerated particles:}
Protons, muons, electrons, pions are all useful at a variety of energies.  It is desirable to have relatively pure beam compositions or the particle identification capacity to ID events in the devices under test.  

\paragraph{Beam Energy:}
A range of energies is desired, from 120 GeV protons down to a few hundred MeV. High energy 120 GeV proton beams are desired for many collider studies.  At low energies electrons, muons, pions, protons are of interest for a variety of applications including neutrino detector and calorimeter development.  The energy of the beam should be well understood.     

\paragraph{Beam intensity:}
It varies by experiment.  Most detectors are looking at intensities on the testing apparatus in the range of 10 to 100 kHz although higher is desirable in some cases.

\paragraph{Beam time structure:}
A pulsed beam would be preferable.  The current FTBF program operates with a four second spill once per minute.  A higher frequency pulse would increase the experimental yield.

\paragraph{Target requirements:}
The option to insert a target material to produce a secondary or tertiary beam.    

\paragraph{Other requirements:}
The experimental space should have space for dedicated staff and staging equipment as well as electronics and mechanical work spaces.  There should be a control room for monitoring by users. The facility should have multiple beam enclosures capable of supporting several experiments simultaneously.  The facility should be instrumented to support beam monitoring, particle tracking and identification, gas systems, a DAQ, motion controls, and experiment space accessible by crane.  There should be space to support experiments both on the scale of weeks and months.  The floor plan should have the flexibility to accommodate installations of varying size and have a high bay. There should be video conferencing spaces available. 
.

\paragraph{Timescales, R\&D needs, and similar facilities:}
There is strong international demand for increased test beam resources.  FTBF, CERN, DESY, and other test beams are heavily subscribed and subject to various operational limitations.  A new facility at Fermilab would enable robust detector $R\&D$ across all frontiers looking forward.  This project needs community driven input to identify the beam capabilities with the highest impact.  Having facility planning begin now to come online in the later part of the 2020s would
establish a clear course for detector development.

%% file: DUNE-proton-momentum.tex
\section{Appendix: DUNE Physics Performance Versus LBNF Proton Energy}
\label{sec:ebeamdumps}

\paragraph{Author:} 
Laura Fields\footnote{ljf26@fnal.gov} (Fermilab)

The LBNF neutrino beam is being designed to use a proton beam tunable between 60 - 120 GeV.  This energy range is driven by a balance between the need of the experiment to have a somewhat flexible energy spectrum and by the projected PIP-II capabilities of Fermilab's Main Injector (MI).  The MI cannot provide protons above 130 GeV, and while it can provide protons below 60 GeV, beam power drops sharply below 60 GeV, where the MI cycle time reaches its lower limit (0.7 microseconds) and cannot be further reduced to compensate for lower proton momentum.  However, MI power is relatively constant between 60 and 120 GeV, making this an optimum range for DUNE physics.  
This note describes a GLoBES-based study of various DUNE physics metrics versus proton momentum.  

\subsection{Power Assumptions}
\label{sec:power}
All results are shown for two different power scenarios:
\begin{enumerate}
    \item PIP-II power power projections of 1.1e21 protons per year at 120 GeV (1.2 MW), 1.47e21 protons per year at 80 GeV (1.07 MW), and 1.89 protons per year at 60 GeV (1.03 MW).  Between 120 GeV and 60 GeV, the number of protons per year is estimated using a linear interpolation between the above values.  Below 60 GeV, a fixed 1.89 protons per year is assumed.  
    \item Flat power: 1.2 MW, regardless of proton momentum.
\end{enumerate}

These estimates of protons on target per year take into account estimated downtime due to mainetence and annual shutdowns.  Plot labels referring to 'years' should therefore be considered to be calendar years (at 1.2 MW and 40 kTon) rather than continuous years of running.  

\subsection{Software Configurations}
\label{sec:software}
The beam simulation used here is g4lbnf~\cite{g4lbnf} version v3r5p7, which implements the optimized beam design~\cite{optcdr}, with small changes to the horn A to reflect engineering changes made during preliminary, and with a 1.5 m cantilevered cylindrical target.

All studies use a configuration of GLoBES~\cite{Huber:2004ka,Huber:2007ji} that approximately reproduces physics studies reported in the DUNE technical design report~\cite{Abi:2020evt}.  Unless otherwise noted, metrics quoted for a single exposure correspond to 7 years of running at 1.2 MW with a 40 kTon detector with the exposure split evenly between neutrino and antineutrino enriched beams.  

\subsection{Results}

\begin{figure} [htpb] 
\begin{center}
\includegraphics[width=0.49\textwidth]{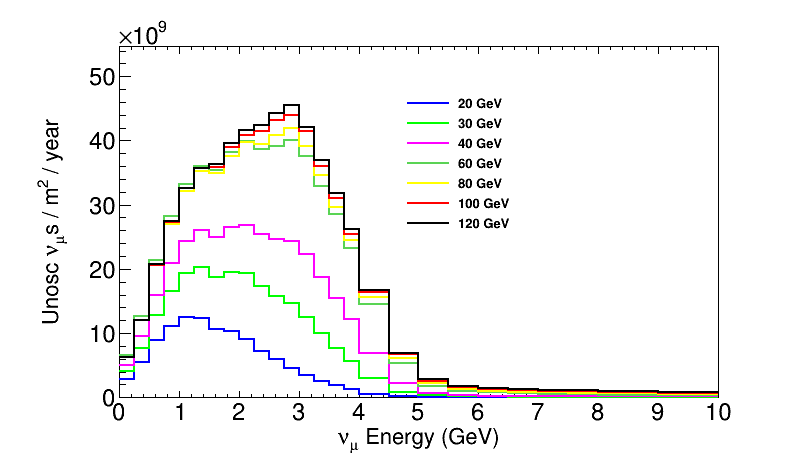}
\includegraphics[width=0.49\textwidth]{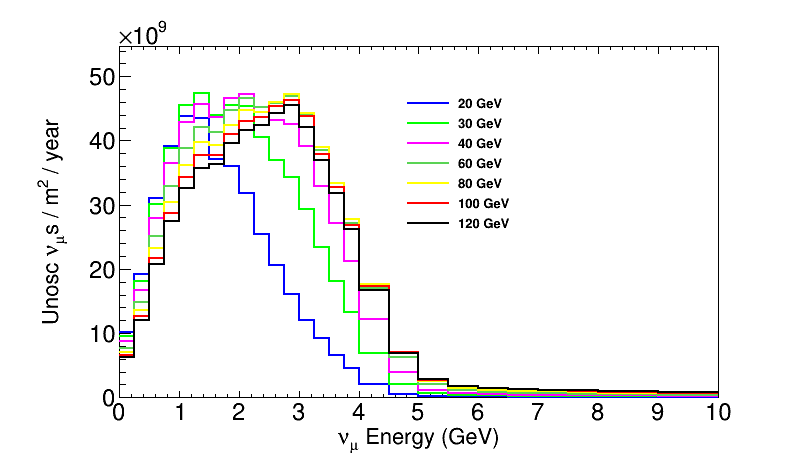}
\includegraphics[width=0.49\textwidth]{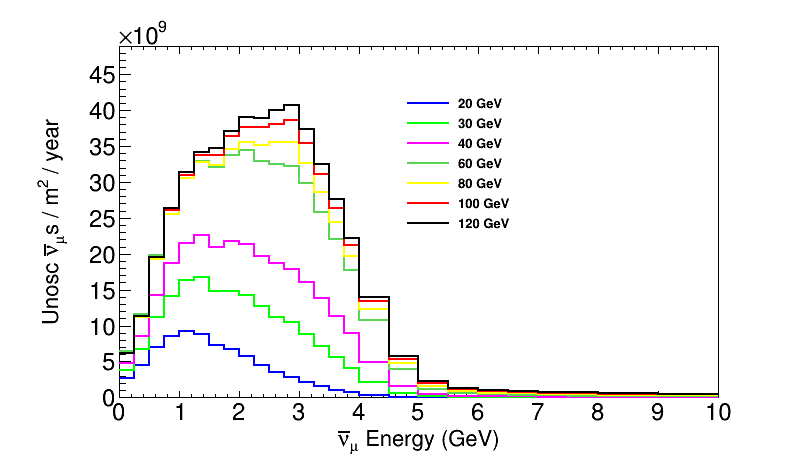}
\includegraphics[width=0.49\textwidth]{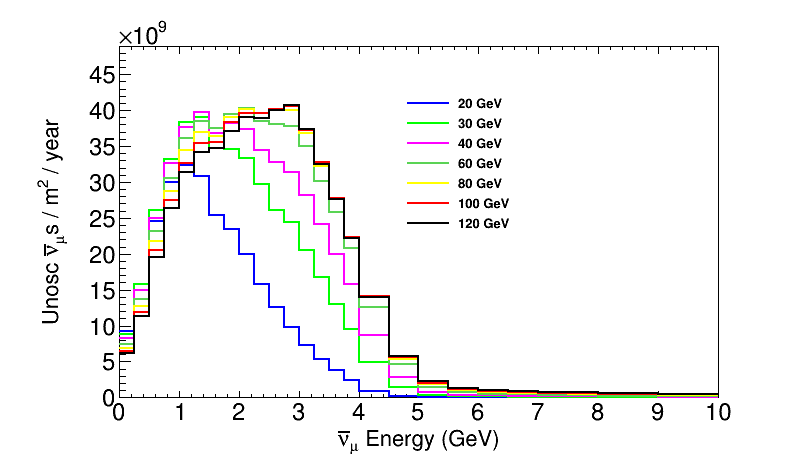}
\includegraphics[width=0.49\textwidth]{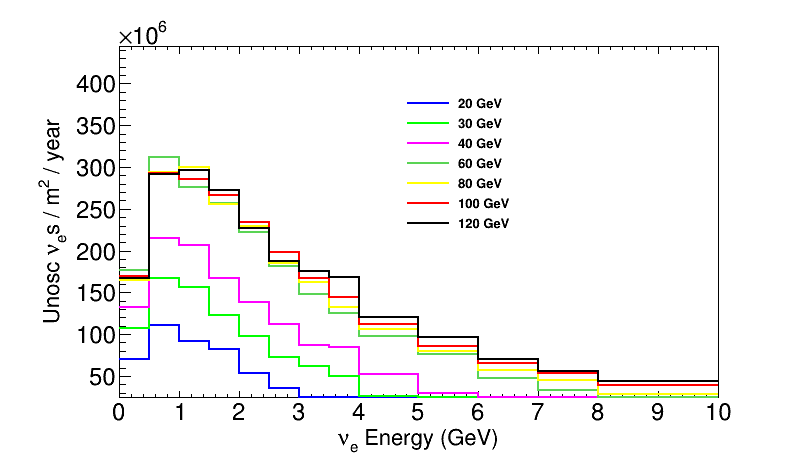}
\includegraphics[width=0.49\textwidth]{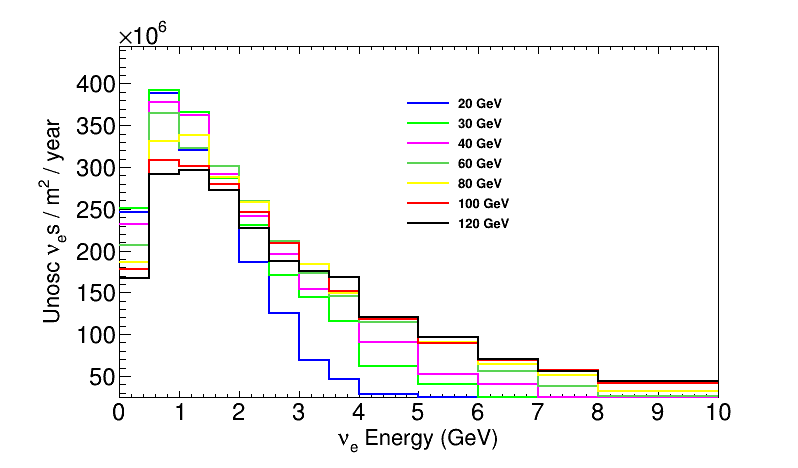}
\includegraphics[width=0.49\textwidth]{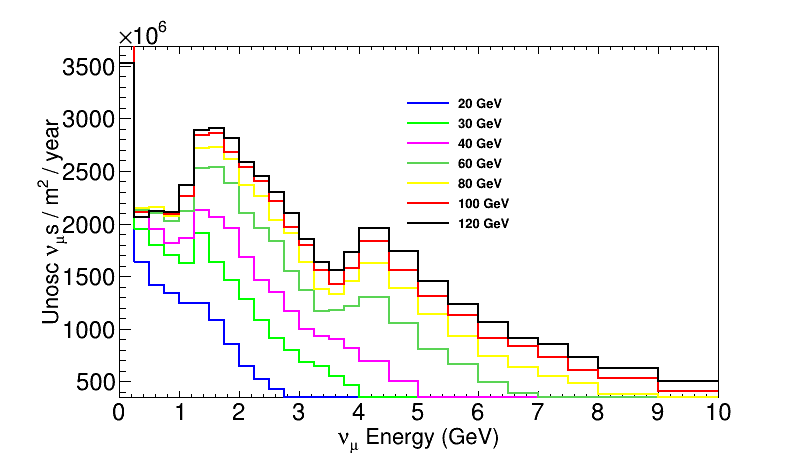}
\includegraphics[width=0.49\textwidth]{Figures-Laura/proton_momentum_flux_antineutrinomode_numu_flatPower.png}

\end{center}
\caption{\label{fig:flux} Neutrino flux versus energy for muon neutrinos in neutrino mode (1st row), muon antineutrinos in antineutrino mode (2nd row), electron neutrinos in neutrino mode (3rd row) and muon neutrinos in antineutrino modes (4th row).   }
\end{figure}

\begin{figure} [htpb] 
\begin{center}
\includegraphics[width=0.49\textwidth]{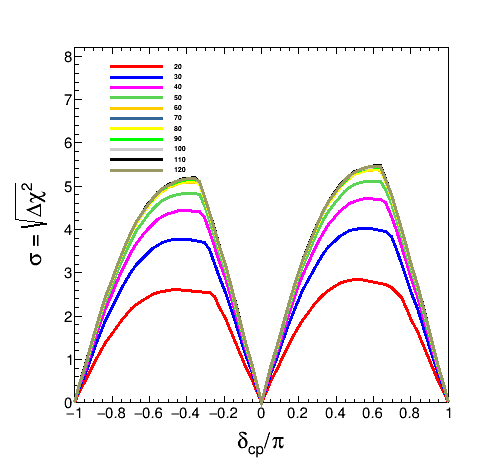}
\includegraphics[width=0.49\textwidth]{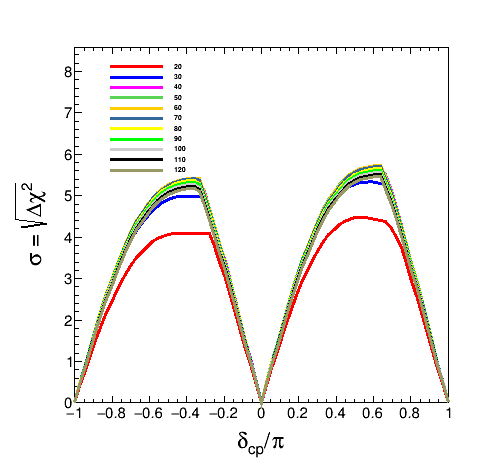}
\includegraphics[width=0.49\textwidth]{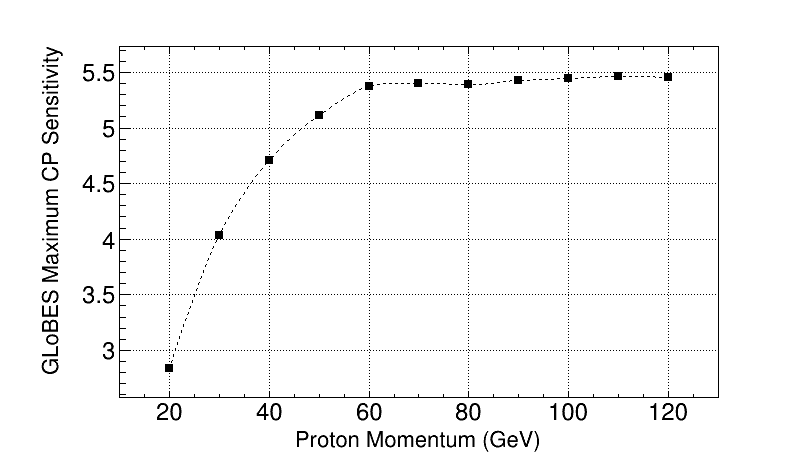}
\includegraphics[width=0.49\textwidth]{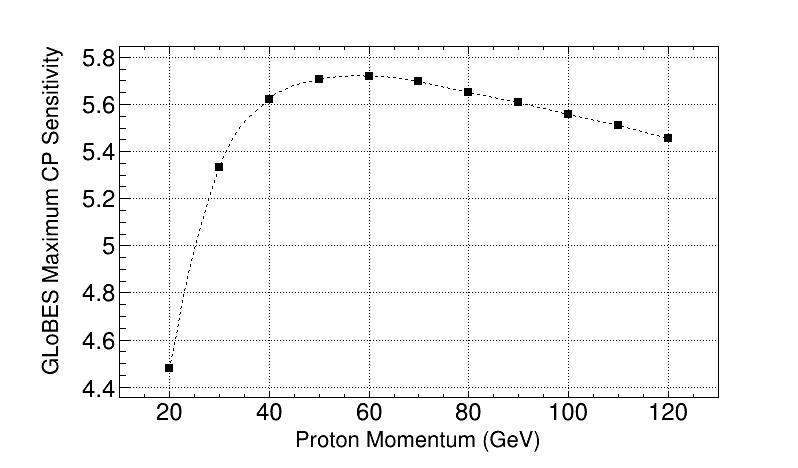}
\includegraphics[width=0.49\textwidth]{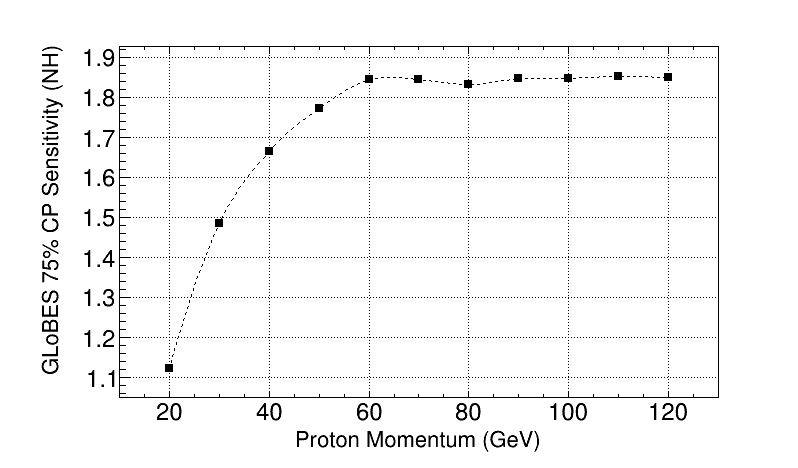}
\includegraphics[width=0.49\textwidth]{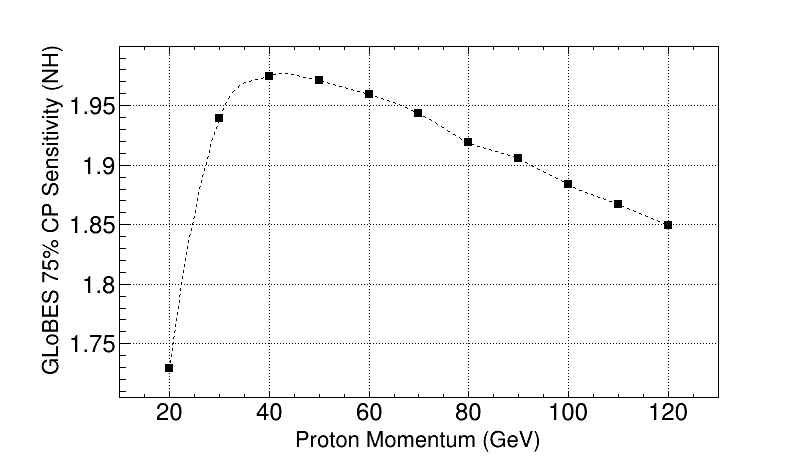}

\end{center}
\caption{\label{fig:cpsens} CP sensitivity versus $\delta_{CP}$ (first row), the minimum CP sensitivity to 75\% of $\delta_{CP}$ phase space (second row), and the maximum CP sensitivity over all of $\delta_{CP}$ phase space, with a 7 year exposure and a 40 kTon detector.  }
\end{figure}

\begin{figure} [htpb] 
\begin{center}
\includegraphics[width=0.49\textwidth]{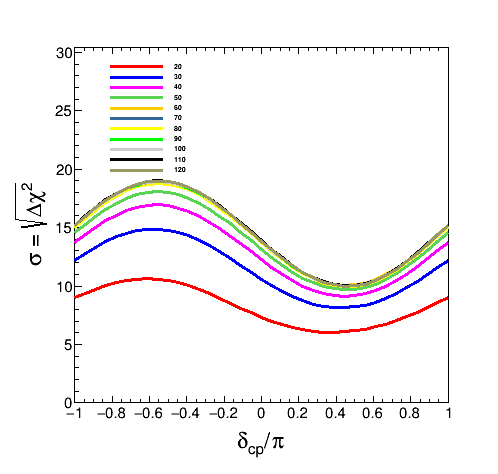}
\includegraphics[width=0.49\textwidth]{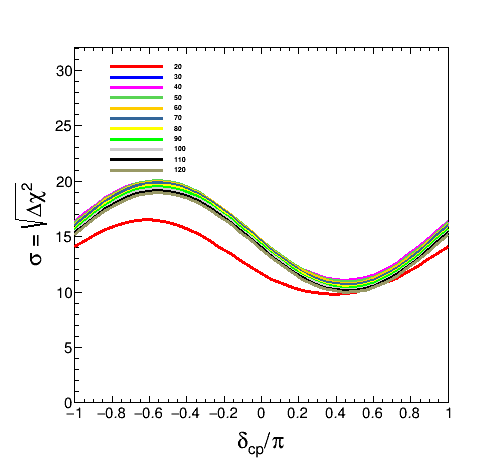}
\includegraphics[width=0.49\textwidth]{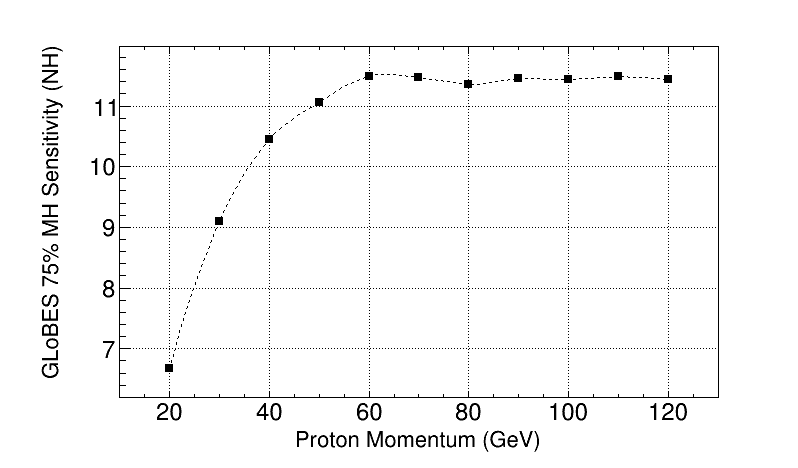}
\includegraphics[width=0.49\textwidth]{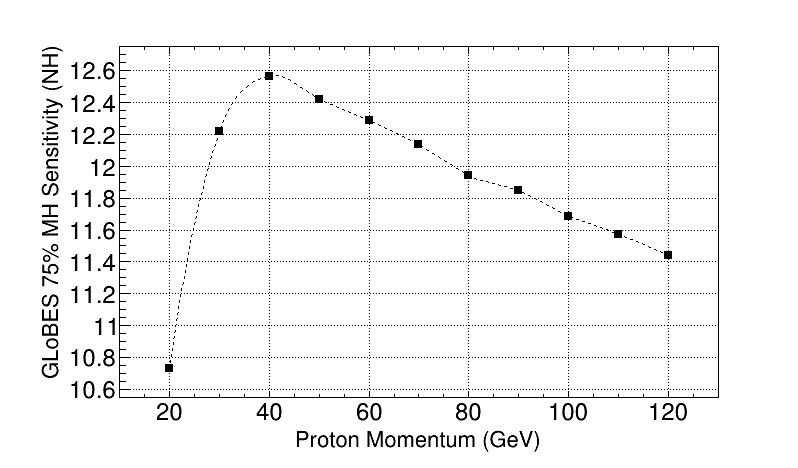}
\includegraphics[width=0.49\textwidth]{Figures-Laura/ProtonMomenta_MH_75.png}
\includegraphics[width=0.49\textwidth]{Figures-Laura/ProtonMomentaFlatPower_MH_75.png}

\end{center}
\caption{\label{fig:mhsens} Mass hierarchy sensitivity versus $\delta_{CP}$ (first row), the minimum mass hierarchy sensitivity to 75\% of $\delta_{CP}$ phase space (second row), and the maximum mass hierarchy sensitivity over all of $\delta_{CP}$ phase space, with a 7 year exposure and a 40 kTon detector.  }
\end{figure}

Neutrino fluxes for several key neutrino mode/flavor combinations are shown in Figure~\ref{fig:flux} for the two power scenarios.  The most critical neutrino fluxes for long-baseline measurements are the so-called "right-sign" fluxes -- muon neutrinos in neutrino mode and muon antineutrinos in antineutrino mode.  In general, the peak neutrino flux is similar for these fluxes in a flat power scenario across many proton momenta, but the focusing peak shifts to lower energy and becomes narrower as proton momentum decreases.  In the PIP-II power scenario, fluxes are similar between 120 GeV and 60 GeV, although peak flux reduces modestly with proton momentum, and flux at lower energies (below 2 GeV) increases slightly as proton momentum decreases.  These fluxes reduce dramatically below 60 GeV in the PIP-II power scenario.  Electron neutrino backgrounds follow similar trends, which is expected given that they arise from the same hadrons as the right-sign fluxes.  Wrong-sign muon neutrino backgrounds (muon antineutrinos in neutrino mode and muon neutrinos in antineutrino mode) decrease substantially with proton momentum, and measurements sensitive to these channels would benefit significantly from lower proton momenta.

Sensitivity to $\delta_{CP}$ and the mass hierarchy for the various momenta and for the two power scenarios was studied using the GLoBES configuration described in Section~\ref{sec:software}.  Results assuming a 7 year exposure are shown in Figures~\ref{fig:cpsens} and~\ref{fig:mhsens}. In the PIP-II power scenario, sensitivities are relatively constant between 60 and 120 GeV, with the optimal value being generally near or slightly below 120 GeV.  In that scenario, physics performance falls sharply below 60 GeV.  In the flat power scenario, sensitivities are more strongly dependant on proton momentum, peaking between 40-60 GeV depending on the metric.  

These Trends in physics performance continue when longer exposures are considered.  Tables~\ref{tab:flat} and~\ref{tab:pipII} show the exposures in kTon MW years required to reach several milestones related to CP sensitivity.  In the PIP-II (flat) power scenario, the smallest exposures are required for proton momenta of 110 GeV (40 GeV).

\begin{table}
\caption{Exposure in kTon MW years required to reach four CP-sensitivity goals in the PIP-II power scenario. \label{tab:pipII}  }
\begin{center}
\begin{tabular}{ |c|cccc| } 
 \hline
  & 3$\sigma$ 50 \% & 5$\sigma$ 50\% & 3$\sigma$ 75\% & 5$\sigma$ 75\% \\ \hline
 20 & 661 & 2570 & 3372 & 8283 \\
 30 & 342 & 1356 & 1921 & 6292 \\
 40 & 259 & 986 &  1489 & 5604 \\
 50 & 223 & 855 &  1284 & 5232 \\
 60 & 203 & 776 & 1174 & 5006 \\
 70 & 202 & 776 & 1175 & 5043 \\
 80 & 205 & 794 & 1205 & 5146 \\
 90 & 201 & 776 & 1181 & 5099 \\
 100 &  201  & 778 & 1187 & 5132 \\
 110 &  199 & 771 &  1180 & 5117 \\
 120 &  200 & 776 & 1188 & 5150 \\
 \hline
\end{tabular}
\end{center}
\end{table}

\begin{table}
\caption{Exposure in kTon MW years required to reach four CP-sensitivity goals in the flat power scenario. \label{tab:flat} }
\begin{center}
\begin{tabular}{ |c|cccc| } 
 \hline
  & 3$\sigma$ 50 \% & 5$\sigma$ 50\% & 3$\sigma$ 75\% & 5$\sigma$ 75\% \\ \hline
 20 & 240 & 932 & 1290 & 4910 \\
 30 & 177 & 701 & 994 & 4448 \\
 40 & 168 & 644 & 959 & 4397 \\
 50 & 169 & 648 & 971 & 4444 \\
 60 & 172 & 660 & 995 & 4552 \\
 70 & 175 & 673 & 1022 & 4656 \\
 80 & 181 & 697 & 1066 & 4795 \\
 90 & 185 & 712 & 1088 & 4863 \\
 100 & 191 & 737 & 1127 & 4982 \\
 110 & 195 & 755 & 1157 & 5060 \\
 120 & 200 & 776 & 1188 & 5150 \\
 \hline
\end{tabular}

\end{center}
\end{table}

\subsection{Summary}

DUNE neutrino fluxes and projected physics performance were compared for various LBNF primary beam proton momenta assuming the expected PIP-II power-versus-momentum profile and assuming flat power versus proton momentum.  For the PIP-II power profile, DUNE physics performance is optimized with a primary proton beam momentum of around 110 GeV, but performance is very similar for proton momenta between 60-120 GeV.  In the flat power scenario, physics performance is a stronger function of proton momentum, with proton momenta near 40 GeV providing the optimal neutrino flux and physics performance.  


%% file: main.bbl
\begin{thebibliography}{100}

\bibitem{ParticlePhysicsProjectPrioritizationPanel(P5):2014pwa}
Steve Ritz et~al.
\newblock {Building for Discovery: Strategic Plan for U.S. Particle Physics in
  the Global Context}.
\newblock 5 2014.

\bibitem{Kronfeld:2013uoa}
Usama Al-Binni et~al.
\newblock {Project X: Physics Opportunities}.
\newblock 6 2013.

\bibitem{Ainsworth:2021ahm}
Robert Ainsworth et~al.
\newblock {An Upgrade Path for the Fermilab Accelerator Complex}.
\newblock 6 2021.

\bibitem{Essig:2013lka}
Rouven Essig et~al.
\newblock {Working Group Report: New Light Weakly Coupled Particles}.
\newblock In {\em {Community Summer Study 2013}: {Snowmass on the
  Mississippi}}, 10 2013.

\bibitem{Alexander:2016aln}
Jim Alexander et~al.
\newblock {Dark Sectors 2016 Workshop: Community Report}.
\newblock 8 2016.

\bibitem{BRN}
{Summary of the High Energy Physics Workshop on Basic Research Needs for Dark
  Matter Small Projects New Initiatives}.
\newblock In {\em {Basic Research Needs For Dark Matter Small Projects New
  Initiatives}}, 10 2018.

\bibitem{Battaglieri:2017aum}
Marco Battaglieri et~al.
\newblock {US Cosmic Visions: New Ideas in Dark Matter 2017: Community Report}.
\newblock In {\em {U.S. Cosmic Visions: New Ideas in Dark Matter}}, 7 2017.

\bibitem{Lee:1977ua}
Benjamin~W. Lee and Steven Weinberg.
\newblock {Cosmological Lower Bound on Heavy Neutrino Masses}.
\newblock {\em Phys. Rev. Lett.}, 39:165--168, 1977.

\bibitem{Boehm:2003hm}
C.~Boehm and Pierre Fayet.
\newblock {Scalar dark matter candidates}.
\newblock {\em Nucl. Phys. B}, 683:219--263, 2004.

\bibitem{Slatyer:2009yq}
Tracy~R. Slatyer, Nikhil Padmanabhan, and Douglas~P. Finkbeiner.
\newblock {CMB Constraints on WIMP Annihilation: Energy Absorption During the
  Recombination Epoch}.
\newblock {\em Phys. Rev. D}, 80:043526, 2009.

\bibitem{Berlin:2018bsc}
Asher Berlin, Nikita Blinov, Gordan Krnjaic, Philip Schuster, and Natalia Toro.
\newblock {Dark Matter, Millicharges, Axion and Scalar Particles, Gauge Bosons,
  and Other New Physics with LDMX}.
\newblock {\em Phys. Rev. D}, 99(7):075001, 2019.

\bibitem{Hochberg:2014dra}
Yonit Hochberg, Eric Kuflik, Tomer Volansky, and Jay~G. Wacker.
\newblock {Mechanism for Thermal Relic Dark Matter of Strongly Interacting
  Massive Particles}.
\newblock {\em Phys. Rev. Lett.}, 113:171301, 2014.

\bibitem{Fitzpatrick:2020vba}
Patrick~J. Fitzpatrick, Hongwan Liu, Tracy~R. Slatyer, and Yu-Dai Tsai.
\newblock {New Pathways to the Relic Abundance of Vector-Portal Dark Matter}.
\newblock 11 2020.

\bibitem{Hochberg:2014kqa}
Yonit Hochberg, Eric Kuflik, Hitoshi Murayama, Tomer Volansky, and Jay~G.
  Wacker.
\newblock {Model for Thermal Relic Dark Matter of Strongly Interacting Massive
  Particles}.
\newblock {\em Phys. Rev. Lett.}, 115(2):021301, 2015.

\bibitem{Berlin:2018tvf}
Asher Berlin, Nikita Blinov, Stefania Gori, Philip Schuster, and Natalia Toro.
\newblock {Cosmology and Accelerator Tests of Strongly Interacting Dark
  Matter}.
\newblock {\em Phys. Rev.}, D97(5):055033, 2018.

\bibitem{Abi2021}
B.~Abi et~al.
\newblock Measurement of the positive muon anomalous magnetic moment to
  0.46~ppm.
\newblock {\em Physical Review Letters}, 126(14), April 2021.

\bibitem{Kuno:1999jp}
Yoshitaka Kuno and Yasuhiro Okada.
\newblock {Muon decay and physics beyond the standard model}.
\newblock {\em Rev. Mod. Phys.}, 73:151--202, 2001.

\bibitem{Raidal:2008jk}
M.~Raidal et~al.
\newblock {Flavour physics of leptons and dipole moments}.
\newblock {\em Eur. Phys. J. C}, 57:13--182, 2008.

\bibitem{deGouvea:2013zba}
Andre de~Gouvea and Petr Vogel.
\newblock {Lepton Flavor and Number Conservation, and Physics Beyond the
  Standard Model}.
\newblock {\em Prog. Part. Nucl. Phys.}, 71:75--92, 2013.

\bibitem{Calibbi:2017uvl}
Lorenzo Calibbi and Giovanni Signorelli.
\newblock {Charged Lepton Flavour Violation: An Experimental and Theoretical
  Introduction}.
\newblock {\em Riv. Nuovo Cim.}, 41(2):71--174, 2018.

\bibitem{Kahn:2018cqs}
Yonatan Kahn, Gordan Krnjaic, Nhan Tran, and Andrew Whitbeck.
\newblock {M$^{3}$: a new muon missing momentum experiment to probe
  $(g-2)_{\mu}$ and dark matter at Fermilab}.
\newblock {\em JHEP}, 09:153, 2018.

\bibitem{ali2021muon}
Hind~Al Ali, Nima Arkani-Hamed, Ian Banta, Sean Benevedes, Dario Buttazzo,
  Tianji Cai, Junyi Cheng, Timothy Cohen, Nathaniel Craig, Majid Ekhterachian,
  JiJi Fan, Matthew Forslund, Isabel~Garcia Garcia, Samuel Homiller, Seth
  Koren, Giacomo Koszegi, Zhen Liu, Qianshu Lu, Kun-Feng Lyu, Alberto Mariotti,
  Amara McCune, Patrick Meade, Isobel Ojalvo, Umut Oktem, Diego Redigolo,
  Matthew Reece, Filippo Sala, Raman Sundrum, Dave Sutherland, Andrea Tesi,
  Timothy Trott, Chris Tully, Lian-Tao Wang, and Menghang Wang.
\newblock The muon smasher's guide, 2021.

\bibitem{Neuffer:2018yof}
D.~Neuffer and V.~Shiltsev.
\newblock {On the feasibility of a pulsed 14 TeV c.m.e. muon collider in the
  LHC tunnel}.
\newblock {\em JINST}, 13(10):T10003, 2018.

\bibitem{Delahaye:2019omf}
Jean~Pierre Delahaye, Marcella Diemoz, Ken Long, Bruno Mansouli\'e, Nadia
  Pastrone, Lenny Rivkin, Daniel Schulte, Alexander Skrinsky, and Andrea
  Wulzer.
\newblock {Muon Colliders}.
\newblock 1 2019.

\bibitem{Han:2020uak}
Tao Han, Zhen Liu, Lian-Tao Wang, and Xing Wang.
\newblock {WIMPs at High Energy Muon Colliders}.
\newblock {\em Phys. Rev. D}, 103(7):075004, 2021.

\bibitem{Rodolfo2022}
Rodolfo Capdevilla, David Curtin, Yonatan Kahn, and Gordan Krnjaic.
\newblock No-lose theorem for discovering the new physics of at muon colliders.
\newblock {\em Physical Review D}, 105(1), Jan 2022.

\bibitem{Buttazzo:2020ibd}
Dario Buttazzo and Paride Paradisi.
\newblock {Probing the muon $g-2$ anomaly with the Higgs boson at a muon
  collider}.
\newblock {\em Phys. Rev. D}, 104(7):075021, 2021.

\bibitem{RK-2022}
Guo-yuan Huang, Sudip Jana, Farinaldo~S. Queiroz, and Werner Rodejohann.
\newblock Probing the rk(*) anomaly at a muon collider.
\newblock {\em Physical Review D}, 105(1), Jan 2022.

\bibitem{franceschini2021higgs}
Roberto Franceschini and Mario Greco.
\newblock Higgs and bsm physics at the future muon collider, 2021.

\bibitem{maltoni-VBF}
Antonio Costantini, Federico De~Lillo, Fabio Maltoni, Luca Mantani, Olivier
  Mattelaer, Richard Ruiz, and Xiaoran Zhao.
\newblock Vector boson fusion at multi-tev muon colliders.
\newblock {\em Journal of High Energy Physics}, 2020(9), Sep 2020.

\bibitem{Abusalma:2018xem}
F.~Abusalma et~al.
\newblock {Expression of Interest for Evolution of the Mu2e Experiment}.
\newblock 2 2018.

\bibitem{Barlow:2011zza}
R.J. Barlow.
\newblock {The PRISM/PRIME project}.
\newblock {\em Nucl. Phys. B Proc. Suppl.}, 218:44--49, 2011.

\bibitem{Alekou:2013eta}
A.~Alekou et~al.
\newblock {Accelerator system for the PRISM based muon to electron conversion
  experiment}.
\newblock In {\em {Community Summer Study 2013}: {Snowmass on the
  Mississippi}}, 10 2013.

\bibitem{Kuno:1997dr}
Y.~Kuno.
\newblock {Lepton flavor violating rare muon decays and future prospects}.
\newblock {\em AIP Conf. Proc.}, 435(1):261--273, 1998.

\bibitem{Kuno:2000kd}
Y.~Kuno.
\newblock {PRISM}.
\newblock {\em AIP Conf. Proc.}, 542(1):220--225, 2000.

\bibitem{2003-PRISM-LOI}
{PRISM} working group.
\newblock The {PRISM} project. {A} letter of intent to the {J-PARC 50 GeV}
  proton synchrotron experiments, 2003.
\newblock {(LOI-24)}.

\bibitem{2019-KEK-roadmap}
KEK Science~Advisory Committee.
\newblock Report: the first meeting of the {KEK} science advisory committee.
  {March 23-24}, 2019.

\bibitem{Baldini:2013ke}
A.M. Baldini et~al.
\newblock {MEG Upgrade Proposal}.
\newblock 1 2013.

\bibitem{Blondel:2013ia}
A.~Blondel et~al.
\newblock {Research Proposal for an Experiment to Search for the Decay $\mu \to
  eee$}.
\newblock 1 2013.

\bibitem{Cavoto:2017kub}
G.~Cavoto, A.~Papa, F.~Renga, E.~Ripiccini, and C.~Voena.
\newblock {The quest for $\mu \rightarrow e \gamma $ and its experimental
  limiting factors at future high intensity muon beams}.
\newblock {\em Eur. Phys. J. C}, 78(1):37, 2018.

\bibitem{Iwai:2020jye}
Ryoto Iwai et~al.
\newblock {Development of next generation muon beams at the Paul Scherrer
  Institute}.
\newblock {\em PoS}, NuFact2019:125, 2020.

\bibitem{Freedman:1973yd}
D.~Z. Freedman.
\newblock {Coherent Neutrino Nucleus Scattering as a Probe of the Weak Neutral
  Current}.
\newblock {\em Phys. Rev.}, D9:1389--1392, 1974.

\bibitem{Kopeliovich:1974mv}
V.~B. Kopeliovich and L.~L. Frankfurt.
\newblock {Isotopic and chiral structure of neutral current}.
\newblock {\em JETP Lett.}, 19:145--147, 1974.
\newblock [Pisma Zh. Eksp. Teor. Fiz.19,236(1974)].

\bibitem{Akimov:2017ade}
D.~Akimov et~al.
\newblock {Observation of Coherent Elastic Neutrino-Nucleus Scattering}.
\newblock {\em Science}, 357(6356):1123--1126, 2017.

\bibitem{deNiverville:2015mwa}
Patrick deNiverville, Maxim Pospelov, and Adam Ritz.
\newblock {Light new physics in coherent neutrino-nucleus scattering
  experiments}.
\newblock {\em Phys. Rev. D}, 92(9):095005, 2015.

\bibitem{Anderson:2012pn}
A.J. Anderson, J.M. Conrad, E.~Figueroa-Feliciano, C.~Ignarra, G.~Karagiorgi,
  K.~Scholberg, M.H. Shaevitz, and J.~Spitz.
\newblock {Measuring Active-to-Sterile Neutrino Oscillations with Neutral
  Current Coherent Neutrino-Nucleus Scattering}.
\newblock {\em Phys. Rev. D}, 86:013004, 2012.

\bibitem{Pellico}
William Pellico et~al.
\newblock {FNAL Booster Storage Ring}.
\newblock In {\em {Snowmass 2021 LOI AF-40}}, 8 2020.

\bibitem{Batell:2018fqo}
Brian Batell, Ayres Freitas, Ahmed Ismail, and David Mckeen.
\newblock {Probing Light Dark Matter with a Hadrophilic Scalar Mediator}.
\newblock {\em Phys. Rev. D}, 100(9):095020, 2019.

\bibitem{Batell:2009di}
Brian Batell, Maxim Pospelov, and Adam Ritz.
\newblock {Exploring Portals to a Hidden Sector Through Fixed Targets}.
\newblock {\em Phys. Rev. D}, 80:095024, 2009.

\bibitem{deNiverville:2011it}
Patrick deNiverville, Maxim Pospelov, and Adam Ritz.
\newblock {Observing a light dark matter beam with neutrino experiments}.
\newblock {\em Phys. Rev. D}, 84:075020, 2011.

\bibitem{Kahn:2014sra}
Yonatan Kahn, Gordan Krnjaic, Jesse Thaler, and Matthew Toups.
\newblock {DAE$\delta$ALUS and dark matter detection}.
\newblock {\em Phys. Rev. D}, 91(5):055006, 2015.

\bibitem{Auerbach:2001wg}
L.B. Auerbach et~al.
\newblock {Measurement of electron - neutrino - electron elastic scattering}.
\newblock {\em Phys. Rev. D}, 63:112001, 2001.

\bibitem{Jordan:2018gcd}
Johnathon~R. Jordan, Yonatan Kahn, Gordan Krnjaic, Matthew Moschella, and
  Joshua Spitz.
\newblock {Signatures of Pseudo-Dirac Dark Matter at High-Intensity Neutrino
  Experiments}.
\newblock {\em Phys. Rev. D}, 98(7):075020, 2018.

\bibitem{Akimov:2020pdx}
D.~Akimov et~al.
\newblock {First Detection of Coherent Elastic Neutrino-Nucleus Scattering on
  Argon}.
\newblock 3 2020.

\bibitem{Akimov:2019xdj}
D.~Akimov et~al.
\newblock {Sensitivity of the COHERENT Experiment to Accelerator-Produced Dark
  Matter}.
\newblock 11 2019.

\bibitem{deNiverville:2016rqh}
Patrick deNiverville, Chien-Yi Chen, Maxim Pospelov, and Adam Ritz.
\newblock {Light dark matter in neutrino beams: production modelling and
  scattering signatures at MiniBooNE, T2K and SHiP}.
\newblock {\em Phys. Rev. D}, 95(3):035006, 2017.

\bibitem{Aartsen:2020iky}
M.G. Aartsen et~al.
\newblock {An eV-scale sterile neutrino search using eight years of atmospheric
  muon neutrino data from the IceCube Neutrino Observatory}.
\newblock 5 2020.

\bibitem{Diaz:2019fwt}
A.~Diaz, C.A. Argüelles, G.H. Collin, J.M. Conrad, and M.H. Shaevitz.
\newblock {Where Are We With Light Sterile Neutrinos?}
\newblock 5 2019.

\bibitem{Aguilar:2001ty}
A.~Aguilar-Arevalo et~al.
\newblock {Evidence for neutrino oscillations from the observation of
  $\bar{\nu}_e$ appearance in a $\bar{\nu}_\mu$ beam}.
\newblock {\em Phys. Rev. D}, 64:112007, 2001.

\bibitem{Elnimr:2013wfa}
M.~Elnimr et~al.
\newblock {The OscSNS White Paper}.
\newblock In {\em {Community Summer Study 2013}: {Snowmass on the
  Mississippi}}, 7 2013.

\bibitem{Aguilar-Arevalo:2018wea}
A.A. Aguilar-Arevalo et~al.
\newblock {Dark Matter Search in Nucleon, Pion, and Electron Channels from a
  Proton Beam Dump with MiniBooNE}.
\newblock {\em Phys. Rev. D}, 98(11):112004, 2018.

\bibitem{Machado:2019oxb}
Pedro~AN Machado, Ornella Palamara, and David~W Schmitz.
\newblock {The Short-Baseline Neutrino Program at Fermilab}.
\newblock {\em Ann. Rev. Nucl. Part. Sci.}, 69:363--387, 2019.

\bibitem{Gardner:2015wea}
S.~Gardner, R.~J. Holt, and A.~S. Tadepalli.
\newblock {New Prospects in Fixed Target Searches for Dark Forces with the
  SeaQuest Experiment at Fermilab}.
\newblock {\em Phys. Rev. D}, 93(11):115015, 2016.

\bibitem{Berlin:2018pwi}
Asher Berlin, Stefania Gori, Philip Schuster, and Natalia Toro.
\newblock {Dark Sectors at the Fermilab SeaQuest Experiment}.
\newblock {\em Phys. Rev.}, D98(3):035011, 2018.

\bibitem{Tsai:2019buq}
Yu-Dai Tsai, Patrick deNiverville, and Ming~Xiong Liu.
\newblock {Dark Photon and Muon $g-2$ Inspired Inelastic Dark Matter Models at
  the High-Energy Intensity Frontier}.
\newblock {\em Phys. Rev. Lett.}, 126(18):181801, 2021.

\bibitem{Batell:2020vqn}
Brian Batell, Jared~A. Evans, Stefania Gori, and Mudit Rai.
\newblock {Dark Scalars and Heavy Neutral Leptons at DarkQuest}.
\newblock {\em JHEP}, 05:049, 2021.

\bibitem{Blinov:2021say}
Nikita Blinov, Elizabeth Kowalczyk, and Margaret Wynne.
\newblock {Axion-like particle searches at DarkQuest}.
\newblock {\em JHEP}, 02:036, 2022.

\bibitem{Akesson:2018vlm}
Torsten \r{A}kesson et~al.
\newblock {Light Dark Matter eXperiment (LDMX)}.
\newblock 8 2018.

\bibitem{Izaguirre:2014bca}
Eder Izaguirre, Gordan Krnjaic, Philip Schuster, and Natalia Toro.
\newblock {Testing GeV-Scale Dark Matter with Fixed-Target Missing Momentum
  Experiments}.
\newblock {\em Phys. Rev. D}, 91(9):094026, 2015.

\bibitem{Ankowski_2020}
Artur~M. Ankowski, Alexander Friedland, Shirley~Weishi Li, Omar Moreno, Philip
  Schuster, Natalia Toro, and Nhan Tran.
\newblock Lepton-nucleus cross section measurements for dune with the ldmx
  detector.
\newblock {\em Physical Review D}, 101(5), Mar 2020.

\bibitem{Bernauer:2013tpr}
J.C. Bernauer et~al.
\newblock {Electric and magnetic form factors of the proton}.
\newblock {\em Phys. Rev. C}, 90(1):015206, 2014.

\bibitem{Lee:2015jqa}
Gabriel Lee, John~R. Arrington, and Richard~J. Hill.
\newblock {Extraction of the proton radius from electron-proton scattering
  data}.
\newblock {\em Phys. Rev. D}, 92(1):013013, 2015.

\bibitem{Sick:2012zz}
Ingo Sick.
\newblock {Problems with proton radii}.
\newblock {\em Prog. Part. Nucl. Phys.}, 67:473--478, 2012.

\bibitem{Snowmass2021LoInuHD}
Alvarez-Ruso Luis et~al.
\newblock {Neutrino Scattering Measurements on Hydrogen and Deuterium}.
\newblock In {\em {Snowmass 2021 LOI: Neutrino Scattering Measurements on
  Hydrogen and Deuterium}}, 8 2020.

\bibitem{Gilman:2017hdr}
R.~Gilman et~al.
\newblock {Technical Design Report for the Paul Scherrer Institute Experiment
  R-12-01.1: Studying the Proton "Radius" Puzzle with $\mu p$ Elastic
  Scattering}.
\newblock 9 2017.

\bibitem{Pohl:2016tqq}
Randolf Pohl.
\newblock {Laser Spectroscopy of Muonic Hydrogen and the Puzzling Proton}.
\newblock {\em J. Phys. Soc. Jap.}, 85(9):091003, 2016.

\bibitem{Dupays:2003zz}
Arnaud Dupays, Alberto Beswick, Bruno Lepetit, Carlo Rizzo, and Dimitar
  Bakalov.
\newblock {Proton Zemach radius from measurements of the hyperfine splitting of
  hydrogen and muonic hydrogen}.
\newblock {\em Phys. Rev. A}, 68:052503, 2003.

\bibitem{Adamczak:2016pdb}
A.~Adamczak et~al.
\newblock {Steps towards the hyperfine splitting measurement of the muonic
  hydrogen ground state: pulsed muon beam and detection system
  characterization}.
\newblock {\em JINST}, 11(05):P05007, 2016.

\bibitem{Ma:2016etb}
Y.~Ma et~al.
\newblock {New Precision Measurement for Proton Zemach Radius with Laser
  Spectroscopy}.
\newblock {\em Int. J. Mod. Phys. Conf. Ser.}, 40:1660046, 2016.

\bibitem{Tomalak:2017npu}
Oleksandr Tomalak.
\newblock {Two-photon exchange correction to the hyperfine splitting in muonic
  hydrogen}.
\newblock {\em Eur. Phys. J. C}, 77(12):858, 2017.

\bibitem{Pohl:2010zza}
Randolf Pohl et~al.
\newblock {The size of the proton}.
\newblock {\em Nature}, 466:213--216, 2010.

\bibitem{Antognini:1900ns}
Aldo Antognini et~al.
\newblock {Proton Structure from the Measurement of $2S-2P$ Transition
  Frequencies of Muonic Hydrogen}.
\newblock {\em Science}, 339:417--420, 2013.

\bibitem{Carlson:2011zd}
Carl~E. Carlson and Marc Vanderhaeghen.
\newblock {Higher order proton structure corrections to the Lamb shift in
  muonic hydrogen}.
\newblock {\em Phys. Rev. A}, 84:020102, 2011.

\bibitem{Parthey:2010aya}
Christian~G. Parthey, Arthur Matveev, Janis Alnis, Randolf Pohl, Thomas Udem,
  Ulrich~D. Jentschura, Nikolai Kolachevsky, and Theodor~W. Hänsch.
\newblock {Precision Measurement of the Hydrogen-Deuterium 1$S$-2$S$ Isotope
  Shift}.
\newblock {\em Phys. Rev. Lett.}, 104:233001, 2010.

\bibitem{Parthey:2011lfa}
Christian~G. Parthey et~al.
\newblock {Improved Measurement of the Hydrogen 1S - 2S Transition Frequency}.
\newblock {\em Phys. Rev. Lett.}, 107:203001, 2011.

\bibitem{Tomalak:2018uhr}
Oleksandr Tomalak.
\newblock {Two-Photon Exchange Correction to the Lamb Shift and Hyperfine
  Splitting of S Levels}.
\newblock {\em Eur. Phys. J. A}, 55(5):64, 2019.

\bibitem{Ye:2017gyb}
Zhihong Ye, John Arrington, Richard~J. Hill, and Gabriel Lee.
\newblock {Proton and Neutron Electromagnetic Form Factors and Uncertainties}.
\newblock {\em Phys. Lett. B}, 777:8--15, 2018.

\bibitem{Borah:2020gte}
Kaushik Borah, Richard~J. Hill, Gabriel Lee, and Oleksandr Tomalak.
\newblock {Parameterization and applications of the low-$Q^2$ nucleon vector
  form factors}.
\newblock 3 2020.

\bibitem{Punjabi:2015bba}
V.~Punjabi, C.F. Perdrisat, M.K. Jones, E.J. Brash, and C.E. Carlson.
\newblock {The Structure of the Nucleon: Elastic Electromagnetic Form Factors}.
\newblock {\em Eur. Phys. J. A}, 51:79, 2015.

\bibitem{Pauk:2015oaa}
Vladyslav Pauk and Marc Vanderhaeghen.
\newblock {Lepton universality test in the photoproduction of $e^- e^+$ versus
  $\mu^- \mu^+$ pairs on a proton target}.
\newblock {\em Phys. Rev. Lett.}, 115(22):221804, 2015.

\bibitem{Carlson:2018ksu}
Carl~E. Carlson, Vladyslav Pauk, and Marc Vanderhaeghen.
\newblock {Dilepton photoproduction on a deuteron target}.
\newblock {\em Phys. Lett. B}, 797:134872, 2019.

\bibitem{Grieser:2018qyq}
S.~Grieser, Daniel Bonaventura, Philipp Brand, Catharina Hargens, Benjamin
  Hetz, Lukas Leßmann, Christina Westphälinger, and Alfons Khoukaz.
\newblock {A Cryogenic Supersonic Jet Target for Electron Scattering
  Experiments at MAGIX@MESA and MAMI}.
\newblock {\em Nucl. Instrum. Meth. A}, 906:120--126, 2018.

\bibitem{Caiazza:2020sda}
S.S. Caiazza et~al.
\newblock {The MAGIX focal plane time projection chamber}.
\newblock {\em J. Phys. Conf. Ser.}, 1498(1):012022, 2020.

\bibitem{Mainz_neutron}
S1 project of sfb 1044.
\newblock \url{https://sfb1044.uni-mainz.de/s1-baryon-form-factors/}.
\newblock Accessed: 2020-06-21.

\bibitem{Xiong:2019umf}
W.~Xiong et~al.
\newblock {A small proton charge radius from an electron--proton scattering
  experiment}.
\newblock {\em Nature}, 575(7781):147--150, 2019.

\bibitem{Faus-Golfe:2019lxi}
Angeles Faus-Golfe, Bowen Bai, Patricia Duchesne, Yanliang Han, Denis Marchand,
  Cynthia Vallerand, and Eric Voutier.
\newblock {The PRORAD Beam Line Design for PRAE}.
\newblock In {\em {10th International Particle Accelerator Conference}}, page
  THPMP003, 2019.

\bibitem{Rachek:2014fam}
I.A. Rachek et~al.
\newblock {Measurement of the two-photon exchange contribution to the elastic
  $e^{\pm}p$ scattering cross sections at the VEPP-3 storage ring}.
\newblock {\em Phys. Rev. Lett.}, 114(6):062005, 2015.

\bibitem{Adikaram:2014ykv}
D.~Adikaram et~al.
\newblock {Towards a resolution of the proton form factor problem: new electron
  and positron scattering data}.
\newblock {\em Phys. Rev. Lett.}, 114:062003, 2015.

\bibitem{Rimal:2016toz}
D.~Rimal et~al.
\newblock {Measurement of two-photon exchange effect by comparing elastic
  $e^\pm p$ cross sections}.
\newblock {\em Phys. Rev. C}, 95(6):065201, 2017.

\bibitem{Henderson:2016dea}
B.S. Henderson et~al.
\newblock {Hard Two-Photon Contribution to Elastic Lepton-Proton Scattering:
  Determined by the OLYMPUS Experiment}.
\newblock {\em Phys. Rev. Lett.}, 118(9):092501, 2017.

\bibitem{Denisov:2018unj}
B.~Adams et~al.
\newblock {Letter of Intent: A New QCD facility at the M2 beam line of the CERN
  SPS (COMPASS++/AMBER)}.
\newblock 8 2018.

\bibitem{Zhan:2011ji}
X.~Zhan et~al.
\newblock {High-Precision Measurement of the Proton Elastic Form Factor Ratio
  $\mu_pG_E/G_M$ at low $Q^2$}.
\newblock {\em Phys. Lett. B}, 705:59--64, 2011.

\bibitem{Batell:2014mga}
Brian Batell, Rouven Essig, and Ze'ev Surujon.
\newblock {Strong Constraints on Sub-GeV Dark Sectors from SLAC Beam Dump
  E137}.
\newblock {\em Phys. Rev. Lett.}, 113(17):171802, 2014.

\bibitem{Dudek:2012vr}
Jozef Dudek et~al.
\newblock {Physics Opportunities with the 12 GeV Upgrade at Jefferson Lab}.
\newblock {\em Eur. Phys. J. A}, 48:187, 2012.

\bibitem{Dalesandro:2017rtx}
Andrew~A. Dalesandro, Joshua Kaluzny, and Arkadiy Klebaner.
\newblock {Thermodynamic Analyses of the LCLS-II Cryogenic Distribution
  System}.
\newblock {\em IEEE Trans. Appl. Supercond.}, 27(4):0500804, 2017.

\bibitem{Raubenheimer:2018mwt}
Tor Raubenheimer, Anthony Beukers, Alan Fry, Carsten Hast, Thomas Markiewicz,
  Yuri Nosochkov, Nan Phinney, Philip Schuster, and Natalia Toro.
\newblock {DASEL: Dark Sector Experiments at LCLS-II}.
\newblock 1 2018.

\bibitem{NA64:2019imj}
D.~Banerjee et~al.
\newblock {Dark matter search in missing energy events with NA64}.
\newblock {\em Phys. Rev. Lett.}, 123(12):121801, 2019.

\bibitem{Achenbach:2008uf}
Patrick Achenbach.
\newblock {The Physics Program at MAMI-C}.
\newblock 2 2008.

\bibitem{Raggi_2014}
Mauro Raggi and Venelin Kozhuharov.
\newblock Proposal to search for a dark photon in positron on target collisions
  at da$\upphi$ne linac.
\newblock {\em Advances in High Energy Physics}, 2014:1–14, 2014.

\bibitem{Battaglieri:2016ggd}
M.~Battaglieri et~al.
\newblock {Dark Matter Search in a Beam-Dump eXperiment (BDX) at Jefferson
  Lab}.
\newblock 7 2016.

\bibitem{frankenthal2019searching}
Andre Frankenthal.
\newblock Searching for dark photons with padme, 2019.

\bibitem{accardi2020ejlab}
A.~Accardi, A.~Afanasev, I.~Albayrak, S.~F. Ali, M.~Amaryan, J.~R.~M. Annand,
  J.~Arrington, A.~Asaturyan, H.~Avakian, T.~Averett, C.~Ayerbe Gayoso,
  L.~Barion, M.~Battaglieri, V.~Bellini, F.~Benmokhtar, V.~Berdnikov, J.~C.
  Bernauer, A.~Bianconi, A.~Biselli, M.~Boer, M.~Bondì, K.~T. Brinkmann, W.~J.
  Briscoe, V.~Burkert, T.~Cao, A.~Camsonne, R.~Capobianco, L.~Cardman,
  M.~Carmignotto, M.~Caudron, L.~Causse, A.~Celentano, P.~Chatagnon, T.~Chetry,
  G.~Ciullo, E.~Cline, P.~L. Cole, M.~Contalbrigo, G.~Costantini, A.~D'Angelo,
  D.~Day, M.~Defurne, M.~De Napoli, A.~Deur, R.~De Vita, N.~D'Hose, S.~Diehl,
  M.~Diefenthaler, B.~Dongwi, R.~Dupré, D.~Dutta, M.~Ehrhart, L.~El-Fassi,
  L.~Elouadrhiri, R.~Ent, J.~Erler, I.~P. Fernando, A.~Filippi, D.~Flay,
  T.~Forest, E.~Fuchey, S.~Fucini, Y.~Furletova, H.~Gao, D.~Gaskell,
  A.~Gasparian, T.~Gautam, F.~X. Girod, J.~Grames, P.~Gueye, M.~Guidal,
  S.~Habet, D.~J. Hamilton, O.~Hansen, D.~Hasell, M.~Hattawy, D.~W.
  Higinbotham, A.~Hobart, T.~Horn, C.~E. Hyde, H.~Ibrahim, A.~Italiano, K.~Joo,
  S.~J. Joosten, N.~Kalantarians, G.~Kalicy, D.~Keller, C.~Keppel, M.~Kerver,
  A.~Kim, J.~Kim, P.~M. King, E.~Kinney, V.~Klimenko, H.~S. Ko, M.~Kohl,
  V.~Kozhuharov, V.~Kubarovsky, T.~Kutz, L.~Lanza, M.~Leali, P.~Lenisa,
  N.~Liyanage, Q.~Liu, S.~Liuti, J.~Mammei, S.~Mantry, D.~Marchand,
  P.~Markowitz, L.~Marsicano, V.~Mascagna, M.~Mazouz, M.~McCaughan,
  B.~McKinnon, D.~McNulty, W.~Melnitchouk, Z.~E. Meziani, M.~Mihovilovič,
  R.~Milner, A.~Mkrtchyan, H.~Mkrtchyan, A.~Movsisyan, M.~Muhoza, C.~Muñoz
  Camacho, J.~Murphy, P.~Nadel-Turoński, J.~Nazeer, S.~Niccolai, G.~Niculescu,
  R.~Novotny, M.~Paolone, L.~Pappalardo, R.~Paremuzyan, E.~Pasyuk, T.~Patel,
  I.~Pegg, C.~Peng, D.~Perera, M.~Poelker, K.~Price, A.~J.~R. Puckett,
  M.~Raggi, N.~Randazzo, M.~N.~H. Rashad, M.~Rathnayake, B.~Raue, P.~E. Reimer,
  M.~Rinaldi, A.~Rizzo, J.~Roche, O.~Rondon-Aramayo, G.~Salmè, E.~Santopinto,
  R.~Santos Estrada, B.~Sawatzky, A.~Schmidt, P.~Schweitzer, S.~Scopetta,
  V.~Sergeyeva, M.~Shabestari, A.~Shahinyan, Y.~Sharabian, S.~Širca, E.~Smith,
  D.~Sokhan, A.~Somov, N.~Sparveris, M.~Spata, S.~Stepanyan, P.~Stoler,
  I.~Strakovsky, R.~Suleiman, M.~Suresh, H.~Szumila-Vance, V.~Tadevosyan, A.~S.
  Tadepalli, M.~Tiefenback, R.~Trotta, M.~Ungaro, P.~Valente, L.~Venturelli,
  H.~Voskanyan, E.~Voutier, B.~Wojtsekhowski, S.~Wood, J.~Xie, Z.~Ye, M.~Yurov,
  H.~G. Zaunick, S.~Zhamkochyan, J.~Zhang, S.~Zhang, S.~Zhao, Z.~W. Zhao,
  X.~Zheng, and C.~Zorn.
\newblock e$^+$@jlab white paper: An experimental program with positron beams
  at jefferson lab, 2020.

\bibitem{Accardi:2012qut}
A.~Accardi et~al.
\newblock {Electron Ion Collider: The Next QCD Frontier}: {Understanding the
  glue that binds us all}.
\newblock {\em Eur. Phys. J. A}, 52(9):268, 2016.

\bibitem{Sakaki:2020mqb}
Yasuhito Sakaki and Daiki Ueda.
\newblock {Search for new light particles at ILC main beam dump}.
\newblock 9 2020.

\bibitem{Chen:2017awl}
Chien-Yi Chen, Maxim Pospelov, and Yi-Ming Zhong.
\newblock {Muon Beam Experiments to Probe the Dark Sector}.
\newblock {\em Phys. Rev. D}, 95(11):115005, 2017.

\bibitem{Chapelain:2017syu}
Antoine Chapelain.
\newblock {The Muon g-2 experiment at Fermilab}.
\newblock {\em EPJ Web Conf.}, 137:08001, 2017.

\bibitem{GORRINGE201573}
T.P. Gorringe and D.W. Hertzog.
\newblock Precision muon physics.
\newblock {\em Progress in Particle and Nuclear Physics}, 84:73 -- 123, 2015.

\bibitem{KANDA2014212}
Sohtaro Kanda.
\newblock Muonium production target for the muon g-2/edm experiment at j-parc.
\newblock {\em Nuclear Physics B - Proceedings Supplements}, 253-255:212 --
  213, 2014.
\newblock The Twelfth International Workshop on Tau-Lepton Physics (TAU2012).

\bibitem{Antognini:2018nhb}
Aldo Antognini, Daniel~M. Kaplan, Klaus Kirch, Andreas Knecht, Derrick~C.
  Mancini, James~D. Phillips, Thomas~J. Phillips, Robert~D. Reasenberg,
  Thomas~J. Roberts, and Anna Soter.
\newblock {Studying Antimatter Gravity with Muonium}.
\newblock {\em Atoms}, 6(2):17, 2018.

\bibitem{TAQQU2011216}
D.~Taqqu.
\newblock Ultraslow muonium for a muon beam of ultra high quality.
\newblock {\em Physics Procedia}, 17:216 -- 223, 2011.
\newblock 2nd International Workshop on the Physics of fundamental Symmetries
  and Interactions - PSI2010.

\bibitem{Bungau:2013hd}
Adriana Bungau, Robert Cywinski, Cristian Bungau, Philip King, and James Lord.
\newblock {Simulations of surface muon production in graphite targets}.
\newblock {\em Phys. Rev. ST Accel. Beams}, 16(1):014701, 2013.

\bibitem{Antognini:2020uyp}
A.~Antognini et~al.
\newblock {Demonstration of Muon-Beam Transverse Phase-Space Compression}.
\newblock 3 2020.

\bibitem{PhysRevLett.97.194801}
D.~Taqqu.
\newblock Compression and extraction of stopped muons.
\newblock {\em Phys. Rev. Lett.}, 97:194801, Nov 2006.

\bibitem{Snowmass2021LoILEMu}
Robert H.~Bernstein et~al.
\newblock Letter of {Interest for an Upgraded Low-Energy Muon Facility at
  Fermilab}.
\newblock
  \url{https://www.snowmass21.org/docs/files/summaries/RF/SNOWMASS21-RF0-AF0-007.pdf},
  2020.
\newblock Submitted to {Snowmass 2021 Workshop}.

\bibitem{Gardner_2020}
Susan Gardner and Jun Shi.
\newblock {Patterns of CP violation from mirror symmetry breaking in the
  $\eta\rightarrow \pi^+\pi^-\pi^0$ Dalitz plot}.
\newblock {\em Physical Review D}, 101(11), Jun 2020.

\bibitem{GardnerShi:2020}
Susan Gardner and Jun Shi.
\newblock {Leading-Dimension Operators with CP and C or P Violation in Standard
  Model Effective Field Theory}.
\newblock (In preparation).

\bibitem{Gan:2020aco}
Liping {Gan}, Bastian {Kubis}, Emilie {Passemar}, and Sean {Tulin}.
\newblock {Precision tests of fundamental physics with $\eta$ and $\eta^\prime$
  mesons}.
\newblock {\em arXiv e-prints}, page arXiv:2007.00664, July 2020.

\bibitem{Beacham:2019nyx}
J.~Beacham et~al.
\newblock {Physics Beyond Colliders at CERN: Beyond the Standard Model Working
  Group Report}.
\newblock {\em J. Phys.}, G47(1):010501, 2020.

\bibitem{gatto2019redtop}
Corrado {Gatto}.
\newblock {The REDTOP experiment}.
\newblock {\em arXiv e-prints}, page arXiv:1910.08505, October 2019.

\bibitem{GENIEHad:2020}
Corrado Gatto.
\newblock {The GenieHad Event Generation Framework}.
\newblock
  \url{https://redtop.fnal.gov/the-geniehadevent-generation-framework/}.

\bibitem{MARS15:2016}
N.V. Mokhov and C.C. James.
\newblock {The MARS Code System User’s Guide, Version 15 (2016)}.
\newblock
  \url{https://lss.fnal.gov/archive/test-fn/1000/fermilab-fn-1058-apc.pdf}.

\bibitem{MARS15:2019}
N.V.~Mokhov et~al.
\newblock {The MARS Code System and The MARS Code System User’s Guide,
  Version 15 (2016)}.
\newblock \url{https://mars.fnal.gov}.

\bibitem{Oberla_2016}
Eric Oberla and Henry~J. Frisch.
\newblock {The design and performance of a prototype water Cherenkov optical
  time-projection chamber}.
\newblock {\em Nuclear Instruments and Methods in Physics Research Section A:
  Accelerators, Spectrometers, Detectors and Associated Equipment},
  814:19–32, Apr 2016.

\bibitem{SHI2020164382}
X.~Shi, M.K. Ayoub, J.~Barreiro~Guimarães {da Costa}, H.~Cui, R.~Kiuchi,
  Y.~Fan, S.~Han, Y.~Huang, M.~Jing, Z.~Liang, B.~Liu, J.~Liu, F.~Lyu, B.~Qi,
  K.~Ran, L.~Shan, L.~Shi, Y.~Tan, K.~Wu, S.~Xiao, T.~Yang, Y.~Yang, C.~Yu,
  M.~Zhao, X.~Zhuang, L.~Castillo García, E.L. Gkougkousis, C.~Grieco,
  S.~Grinstein, M.~Leite, G.T. Saito, A.~Howard, V.~Cindro, G.~Kramberger,
  I.~Mandić, M.~Mikuž, G.~d’Amen, G.~Giacomini, E.~Rossi, A.~Tricoli,
  H.~Chen, J.~Ge, C.~Li, H.~Liang, X.~Yang, L.~Zhao, Z.~Zhao, X.~Zheng,
  N.~Atanov, Y.~Davydov, J.~Grosse-Knetter, J.~Lange, A.~Quadt, M.~Schwickardi,
  S.~Alderweireldt, A.S.C. Ferreira, S.~Guindon, E.~Kuwertz, C.~Rizzi,
  S.~Christie, Z.~Galloway, C.~Gee, Y.~Jin, C.~Labitan, M.~Lockerby, S.M.
  Mazza, F.~Martinez-Mckinney, R.~Padilla, H.~Ren, H.F.-W. Sadrozinski,
  B.~Schumm, A.~Seiden, M.~Wilder, W.~Wyatt, Y.~Zhao, D.~Han, and X.~Zhang.
\newblock {Radiation campaign of HPK prototype LGAD sensors for the
  High-Granularity Timing Detector (HGTD)}.
\newblock {\em Nuclear Instruments and Methods in Physics Research Section A:
  Accelerators, Spectrometers, Detectors and Associated Equipment}, 979:164382,
  2020.

\bibitem{CMOS}
Moritz Kiehn, Francesco Armando Di~Bello, Mathieu Benoit, Raimon Mohr, Hucheng
  Chen, Kai Chen, Sultan D.M.S., Felix Ehrler, Didier Ferrere, Dylan Frizell,
  and Jessica Metcalfe.
\newblock {Performance of CMOS pixel sensor prototypes in ams H35 and aH18
  technology for the ATLAS ITk upgrade}.
\newblock {\em Nuclear Instruments and Methods in Physics Research Section A:
  Accelerators, Spectrometers, Detectors and Associated Equipment}, 924:104 --
  107, Apr 2019.
\newblock 10.1016/j.nima.2018.07.061.

\bibitem{T1015:2011}
Corrado~Gatto et~al.
\newblock {Dual Readout Calorimetry with Glasses}.
\newblock
  \url{https://web.fnal.gov/experiment/FTBF/TSW\%20Library/T1015_mou.pdf}.

\bibitem{T1015:2015}
Corrado Gatto, Vito Di~Benedetto, Anna Mazzacane, and T1015 Collaboration.
\newblock {Status of ADRIANO $R\&D$ in T1015 collaboration}.
\newblock {\em Journal of Physics: Conference Series}, 587, 02 2015.

\bibitem{Gatto:2016jtz}
Corrado Gatto, Vito Di~Benedetto, Eileen Hahn, and Anna Mazzacane.
\newblock {Status of Dual-readout R\&D for a Linear Collider in T1015
  Collaboration}.
\newblock In {\em {International Workshop on Future Linear Colliders}}, 3 2016.

\bibitem{BLAZEY2009277}
G.~Blazey, D.~Chakraborty, A.~Dyshkant, K.~Francis, D.~Hedin, J.~Hill, G.~Lima,
  J.~Powell, P.~Salcido, V.~Zutshi, M.~Demarteau, P.~Rubinov, and N.~Pohlman.
\newblock {Directly coupled tiles as elements of a scintillator calorimeter
  with MPPC readout}.
\newblock {\em Nuclear Instruments and Methods in Physics Research Section A:
  Accelerators, Spectrometers, Detectors and Associated Equipment}, 605(3):277
  -- 281, 2009.

\bibitem{LEMAIRE2020163538}
William Lemaire, Frédéric Nolet, Frédérik Dubois, Audrey~C. Therrien,
  Jean-François Pratte, and Réjean Fontaine.
\newblock {Embedded time of arrival estimation for digital silicon
  photomultipliers with in-pixel TDCs}.
\newblock {\em Nuclear Instruments and Methods in Physics Research Section A:
  Accelerators, Spectrometers, Detectors and Associated Equipment}, 959:163538,
  2020.

\bibitem{NOLET201829}
Frédéric Nolet, Frédérik Dubois, Nicolas Roy, Samuel Parent, William
  Lemaire, Alexandre Massie-Godon, Serge~A. Charlebois, Réjean Fontaine, and
  Jean-Francois Pratte.
\newblock {Digital SiPM channel integrated in CMOS 65 nm with 17.5 ps FWHM
  single photon timing resolution}.
\newblock {\em Nuclear Instruments and Methods in Physics Research Section A:
  Accelerators, Spectrometers, Detectors and Associated Equipment}, 912:29 --
  32, 2018.
\newblock New Developments In Photodetection 2017.

\bibitem{NOLET2020162891}
Frédéric Nolet, William Lemaire, Frédérik Dubois, Nicolas Roy, Simon
  Carrier, Arnaud Samson, Serge~A. Charlebois, Réjean Fontaine, and
  Jean-Francois Pratte.
\newblock {A 256 Pixelated SPAD readout ASIC with in-Pixel TDC and embedded
  digital signal processing for uniformity and skew correction}.
\newblock {\em Nuclear Instruments and Methods in Physics Research Section A:
  Accelerators, Spectrometers, Detectors and Associated Equipment}, 949:162891,
  2020.

\bibitem{9078416}
F.~{Nolet}, N.~{Roy}, S.~{Carrier}, J.~{Bouchard}, R.~{Fontaine}, S.~A.
  {Charlebois}, and J.~{Pratte}.
\newblock {22 $\mu$W, 5.1 ps LSB, 5.5 ps RMS jitter Vernier time-to-digital
  converter in CMOS 65 nm for single photon avalanche diode array}.
\newblock {\em Electronics Letters}, 56(9):424--426, 2020.

\bibitem{SPAD2018}
Frederic Nolet, Samuel Parent, Nicolas Roy, Marc-Olivier Mercier, Serge
  Charlebois, Rejean Fontaine, and Jean-Francois Pratte.
\newblock {Quenching Circuit and SPAD Integrated in CMOS 65 nm with 7.8 ps FWHM
  Single Photon Timing Resolution}.
\newblock {\em Instruments}, 2:19, 09 2018.

\bibitem{PET2020}
Paul Lecoq, Christian Morel, John Prior, Dimitris Visvikis, Stefan Gundacker,
  Etiennette Auffray, Peter Krizan, Rosana~Martinez Turtos, Dominique Thers,
  Edoardo Charbon, Joao Varela, Christophe de~La~Taille, Angelo Rivetti,
  Dominique Breton, Jean-Francois Pratte, Johan Nuyts, Suleman Surti, Stefaan
  Vandenberghe, Paul~K Marsden, Katia Parodi, José~Maria Benlloch, and Mathieu
  Benoit.
\newblock {Roadmap toward the 10 ps time-of-flight PET challenge}.
\newblock {\em Physics in Medicine $\&$ Biology}, 2020.

\bibitem{collaboration_2008}
The~CALICE collaboration, J~Repond, J~Yu, C~M Hawkes, Y~Mikami, O~Miller, N~K
  Watson, J~A Wilson, G~Mavromanolakis, M~A Thomson, and et~al.
\newblock {Design and electronics commissioning of the physics prototype of a
  Si-W electromagnetic calorimeter for the International Linear Collider}.
\newblock {\em Journal of Instrumentation}, 3(08):P08001–P08001, Aug 2008.

\bibitem{Kawagoe_2020}
K.~Kawagoe, Y.~Miura, I.~Sekiya, T.~Suehara, T.~Yoshioka, S.~Bilokin, J.~Bonis,
  P.~Cornebise, A.~Gallas, and A.~et~al. Irles.
\newblock {Beam test performance of the highly granular SiW-ECAL technological
  prototype for the ILC}.
\newblock {\em Nuclear Instruments and Methods in Physics Research Section A:
  Accelerators, Spectrometers, Detectors and Associated Equipment}, 950:162969,
  Jan 2020.

\bibitem{redtop_loi:2020}
REDTOP Collaboration.
\newblock {The REDTOP experiment: an Eta/Eta' factory}.
\newblock
  \url{https://www.snowmass21.org/docs/files/summaries/RF/SNOWMASS21-RF2_RF6-IF6_IF3_REDTOP_Collaboration_-_new-083.pdf}.

\bibitem{redtop_wp:2022}
REDTOP Collaboration.
\newblock {The REDTOP experiment: Rare Eta/Eta' Decays To Observe Physics
  Beyond the Standard Model}.
\newblock {TBP}.

\bibitem{Sakharov:1967dj}
A.~D. Sakharov.
\newblock {Violation of CP Invariance, C asymmetry, and baryon asymmetry of the
  universe}.
\newblock {\em Pisma Zh. Eksp. Teor. Fiz.}, 5:32--35, 1967.
\newblock [Usp. Fiz. Nauk161,no.5,61(1991)].

\bibitem{Cirigliano:2013}
V.~Cirigliano and M.~J. Ramsey-Musolf.
\newblock Low energy probes of physics beyond the standard model.
\newblock {\em Progress in Particle and Nuclear Physics}, 71:2--20, 2013.

\bibitem{Babu:2020}
K.~S. Babu et~al.
\newblock {$\left | \Delta B \right | = 2$: A State of the Field, and Looking
  Forward}.
\newblock 10 2020.

\bibitem{Abe:2011ky}
K.~Abe et~al.
\newblock {The Search for $n-\bar{n}$ oscillation in Super-Kamiokande I}.
\newblock {\em Phys. Rev.}, D91:072006, 2015.

\bibitem{Abi:2020evt}
Babak Abi et~al.
\newblock {Deep Underground Neutrino Experiment (DUNE), Far Detector Technical
  Design Report, Volume II DUNE Physics}.
\newblock 2 2020.

\bibitem{Abi:2020kei}
B.~Abi et~al.
\newblock {Prospects for Beyond the Standard Model Physics Searches at the Deep
  Underground Neutrino Experiment}.
\newblock 8 2020.

\bibitem{Barrow:2019viz}
Joshua~L. Barrow, Elena~S. Golubeva, Eduard Paryev, and Jean-Marc Richard.
\newblock {Progress and simulations for intranuclear neutron-antineutron
  transformations in ${}^{40}_{18} Ar$}.
\newblock {\em Phys. Rev. D}, 101(3):036008, 2020.

\bibitem{Phillips:2014fgb}
D.~G. Phillips, II et~al.
\newblock {Neutron-Antineutron Oscillations: Theoretical Status and
  Experimental Prospects}.
\newblock {\em Phys. Rept.}, 612:1--45, 2016.

\bibitem{Addazi:2020nlz}
A.~Addazi et~al.
\newblock {New high-sensitivity searches for neutrons converting into
  antineutrons and/or sterile neutrons at the European Spallation Source}.
\newblock 6 2020.

\bibitem{Kuzmin:1970nx}
V.~A. Kuzmin.
\newblock {Cp violation and baryon asymmetry of the universe}.
\newblock {\em Pisma Zh. Eksp. Teor. Fiz.}, 12:335--337, 1970.

\bibitem{Kuzmin:1985mm}
V.~A. Kuzmin, V.~A. Rubakov, and M.~E. Shaposhnikov.
\newblock {On the Anomalous Electroweak Baryon Number Nonconservation in the
  Early Universe}.
\newblock {\em Phys. Lett.}, 155B:36, 1985.

\bibitem{Kuzmin:1987wn}
V.~A. Kuzmin, V.~A. Rubakov, and M.~E. Shaposhnikov.
\newblock {Anomalous Electroweak Baryon Number Nonconservation and GUT
  Mechanism for Baryogenesis}.
\newblock {\em Phys. Lett.}, B191:171--173, 1987.

\bibitem{Mohapatra:1980de}
Rabindra~N. Mohapatra and R.~E. Marshak.
\newblock {Phenomenology of neutron oscillations}.
\newblock {\em Phys. Lett.}, 94B:183, 1980.
\newblock [Erratum: Phys. Lett.96B,444(1980)].

\bibitem{Mohapatra:1980qe}
Rabindra~N. Mohapatra and R.E. Marshak.
\newblock {Local $B-L$ Symmetry of Electroweak Interactions, Majorana Neutrinos
  and Neutron Oscillations}.
\newblock {\em Phys. Rev. Lett.}, 44:1316--1319, 1980.
\newblock [Erratum: Phys.Rev.Lett. 44, 1643 (1980)].

\bibitem{Berezhiani:2017azg}
Zurab Berezhiani, Matthew Frost, Yuri Kamyshkov, Ben Rybolt, and Louis
  Varriano.
\newblock {Neutron Disappearance and Regeneration from Mirror State}.
\newblock {\em Phys. Rev.}, D96(3):035039, 2017.

\bibitem{golub_UCN_book}
R.~Golub, D.~Richardson, and S.~K. Lamoreaux.
\newblock {\em Ultra-Cold Neutrons}.
\newblock Taylor \& Francis, New York, NY, 1991.

\bibitem{Leung:2019}
K.~K.~H. Leung, G.~Muhrer, T.~Hügle, T.~M. Ito, E.~M. Lutz, M.~Makela, C.~L.
  Morris, R.~W. Pattie, A.~Saunders, and A.~R. Young.
\newblock A next-generation inverse-geometry spallation-driven ultracold
  neutron source.
\newblock {\em Journal of Applied Physics}, 126(22):224901, 2019.

\bibitem{LeBrun:1994}
P.~LeBrun.
\newblock {Superfluid Helium Cryogenics for the Large Hadron Collider Project
  at CERN}.
\newblock in Proceedings of the 15th International Cryogenic Engineering
  Conference, ICEC 15, June 7-10, Genova, Italy, 1994.

\bibitem{Yannis}
William~M. Morse and Yannis~K. Semertzidis.
\newblock The proton storage ring edm experiment (sredm).
\newblock
  \url{https://www.snowmass21.org/docs/files/summaries/RF/SNOWMASS21-RF3_RF0-AF5_AF0_Yannis_K._Semertzidis-032.pdf}.

\bibitem{Omarov:2020kws}
Zhanibek Omarov, Hooman Davoudiasl, Selcuk Haciomeroglu, Valeri Lebedev,
  William~M. Morse, Yannis~K. Semertzidis, Alexander~J. Silenko, Edward~J.
  Stephenson, and Riad Suleiman.
\newblock {Comprehensive symmetric-hybrid ring design for a proton EDM
  experiment at below 10-29e\textperiodcentered{}cm}.
\newblock {\em Phys. Rev. D}, 105(3):032001, 2022.

\bibitem{DONUT2001}
K.~Kodama et~al.
\newblock {\em Phys. Lett.}, B504:218, 2001.

\bibitem{DONUT2007}
K.~Kodama et~al.
\newblock {\em Phys. Rev.}, D78:052002, 2008.

\bibitem{OPERAProp}
M.~Guler et~al.
\newblock Opera: An appearance experiment to search for $\nu_{\mu} \rightarrow
  \nu_{\tau}$ oscillations in the cngs beam. experimental proposal.
\newblock Technical Report CERN-SPSC-2000-028, CERN, 2000.

\bibitem{OPERA2015}
N.~Agafonova et~al.
\newblock Discovery of tau neutrino appearance in the cngs neutrino beam with
  the opera experiment.
\newblock {\em Phys. Rev. Lett.}, 115:121802, 2015.

\bibitem{OPERA2018}
N.~Agafonova et~al.
\newblock Final results of the opera experiment on $\nu_{\tau}$ appearance in
  the cngs neutrino beam.
\newblock {\em Phys. Rev. Lett.}, 120:211801, 2018.

\bibitem{SKNuTau2013}
K.~Abe et~al.
\newblock Evidence for the appearance of atmospheric tau neutrinos in
  super-kamiokande.
\newblock {\em Phys. Rev. Lett.}, 110:181802, 2013.

\bibitem{SKNuTau2017}
Z.~Li et~al.
\newblock Measurement of the tau neutrino cross section in atmospheric neutrino
  oscillations with super-kamiokande.
\newblock {\em Phys. Rev. D}, 98:052006, 2018.

\bibitem{DeepCoreNuTau}
M.~G. Aartsen et~al.
\newblock Measurement of atmospheric tau neutrino appearance with icecube
  deepcore.
\newblock {\em Phys. Rev. D}, 99:032007, 2019.

\bibitem{HFAG2017}
Y.~Amhis et~al.
\newblock Averages of b-hadron, c-hadron, and $\tau$-lepton properties as of
  summer 2016.
\newblock {\em Eur. Phys. J. C}, 77:895, 2017.

\bibitem{ParkeNonUnitarity}
Stephen Parke and Mark Ross-Lonergan.
\newblock Unitarity and the three flavor neutrino mixing matrix.
\newblock {\em Phys. Rev. D}, 93:113009, 2016.

\bibitem{Ellis:2020hus}
Sebastian Alfonso~Richard Ellis, Kevin~James Kelly, and Shirley~Weishi Li.
\newblock {Current and Future Neutrino Oscillation Constraints on Leptonic
  Unitarity}.
\newblock 8 2020.

\bibitem{summer_blot_2020_3959546}
Summer Blot.
\newblock {N}eutrino oscillation measurements with {I}ce{C}ube, June 2020.

\bibitem{Aoki:2019jry}
Shigeki Aoki et~al.
\newblock {DsTau: Study of tau neutrino production with 400 GeV protons from
  the CERN-SPS}.
\newblock {\em JHEP}, 01:033, 2020.

\bibitem{Abreu:2020ddv}
Henso Abreu et~al.
\newblock {Technical Proposal: FASERnu}.
\newblock 1 2020.

\bibitem{NuTauDUNE}
A.~de~Gouv\^ea, K.~J. Kelly, G.~V. Stenico, and P.~Pasquini.
\newblock {Physics with Beam Tau-Neutrino Appearance at DUNE}.
\newblock {\em Phys. Rev. D}, 100:016004, 2019.

\bibitem{Ghoshal:2019pab}
Anish Ghoshal, Alessio Giarnetti, and Davide Meloni.
\newblock {On the role of the $\nu_{\tau}$ appearance in DUNE in constraining
  standard neutrino physics and beyond}.
\newblock {\em JHEP}, 12:126, 2019.

\bibitem{ConradAtmNuTau}
Janet Conrad, Andr\'e de~Gouv\^ea, Shashank Shalgar, and Joshua Spitz.
\newblock Atmospheric tau neutrinos in a multi-kiloton liquid argon detector.
\newblock {\em Phys. Rev. D}, 82:093012, 2010.

\bibitem{DUNEtdr2}
B.~Abi et~al.
\newblock Deep underground neutrino experiment (dune), far detector technical
  design report, volume ii, dune physics, 2020.

\bibitem{Machado:2020yxl}
Pedro Machado, Holger Schulz, and Jessica Turner.
\newblock {Tau neutrinos at DUNE: new strategies, new opportunities}.
\newblock 6 2020.

\bibitem{SHIPProposal}
Sergey Alekhin et~al.
\newblock A facility to search for hidden particles at the cern sps: the ship
  physics case.
\newblock {\em Rep. Prog. Phys}, 79:124201, 2016.

\bibitem{seesaw}
P~Minkowski.
\newblock $\mu \rightarrow e\gamma$ at a rate of one out of $10^9$ muon decays?
\newblock {\em Phys. Lett. B}, 67:421, 1977.

\bibitem{Mohapatra:1979ia}
Rabindra~N. Mohapatra and Goran Senjanovic.
\newblock {Neutrino Mass and Spontaneous Parity Nonconservation}.
\newblock {\em Phys. Rev. Lett.}, 44:912, 1980.

\bibitem{PhysRevD.92.053009}
F.~J. Escrihuela, D.~V. Forero, O.~G. Miranda, M.~T\'ortola, and J.~W.~F.
  Valle.
\newblock On the description of nonunitary neutrino mixing.
\newblock {\em Phys. Rev. D}, 92:053009, 2015.
\newblock Erratum Phys. Rev. D {\textbf 93}, 119905 (2016).

\bibitem{NonUnitaritySteriles}
M.~Blennow, P.~Coloma, E.~Fernandez-Martinez, et~al.
\newblock Non-unitarity, sterile neutrinos, and non-standard neutrino
  interactions.
\newblock {\em J. High Energy. Phys.}, 2017:153, 2017.

\bibitem{Ghoshal:2020hyo}
A.~Ghoshal, A.~Giarnetti, and D.~Meloni.
\newblock {Neutrino Invisible Decay at DUNE: a multi-channel analysis}.
\newblock 3 2020.

\bibitem{g4lbnf}
\url{https://cdcvs.fnal.gov/redmine/projects/lbne-beamsim/wiki}.

\bibitem{optcdr}
\url{https://docs.dunescience.org/cgi-bin/ShowDocument?docid=4559}.

\bibitem{Huber:2004ka}
Patrick Huber, M.~Lindner, and W.~Winter.
\newblock {Simulation of long-baseline neutrino oscillation experiments with
  GLoBES (General Long Baseline Experiment Simulator)}.
\newblock {\em Comput. Phys. Commun.}, 167:195, 2005.

\bibitem{Huber:2007ji}
Patrick Huber, Joachim Kopp, Manfred Lindner, Mark Rolinec, and Walter Winter.
\newblock {New features in the simulation of neutrino oscillation experiments
  with GLoBES 3.0: General Long Baseline Experiment Simulator}.
\newblock {\em Comput. Phys. Commun.}, 177:432--438, 2007.

\end{thebibliography}
